%% file: main.tex
\documentclass[twocolumn]{aastex631}
\usepackage{appendix}
\pdfoutput=1 
\usepackage{amsmath}
\usepackage{courier}
\usepackage{booktabs}
\usepackage{upgreek}

\newcommand{\zsys}{z$_{sys}$}
\newcommand{\N}{$N$}

\newcommand{\cf}{$C_f$}
\newcommand{\bd}{$b$}
\newcommand{\ta}{$\tau_0$}
\newcommand{\tmeas}{$\tau_0$(\ion{Mg}{2})$\rm_{meas}$}
\newcommand{\tinf}{$\tau_0$(\ion{Mg}{2})$\rm_{inf}$}
\newcommand{\kmps}{\ensuremath{\mathrm{km~s}^{-1}}}
\newcommand{\cm}{\ensuremath{\mathrm{cm}^{-2}}}
\newcommand{\mgii}{\ion{Mg}{2}}
\newcommand{\mgo}{\ion{Mg}{2}$\lambda$2796}
\newcommand{\mgd}{\ion{Mg}{2}$\lambda$2803}
\newcommand{\mgb}{\ion{Mg}{2}$\lambda\lambda$2796, 2803}
\newcommand{\mn}{\ion{Mn}{2}}
\newcommand{\mgi}{\ion{Mg}{1}}
\newcommand{\feii}{\ion{Fe}{2}}
\newcommand{\oii}{\ion{[O}{2]}}

\shorttitle{Extreme Starburst Galactic Outflows}
\shortauthors{Perrotta et al.}

\graphicspath{{./}{figures/}}

\begin{document}

\title{Kinematics, Structure, and Mass Outflow Rates of Extreme Starburst Galactic Outflows}

\correspondingauthor{Serena Perrotta}
\email{s2perrotta@ucsd.edu}

\author{Serena Perrotta}
\affiliation{Department of Astronomy, University of California, San Diego, CA 92092, USA}

\author{Alison L. Coil} 
\affiliation{Department of Astronomy, University of California, San Diego, CA 92092, USA}

\author{David S.~N. Rupke}
\affiliation{Department of Physics, Rhodes College, Memphis, TN, 38112, USA}

\author{Christy A. Tremonti}
\affiliation{University of Wisconsin-Madison, 475 N. Charter St. Madison, WI 53706}

\author{Julie D. Davis}
\affiliation{University of Wisconsin-Madison, 475 N. Charter St. Madison, WI 53706}

\author{Aleksandar M. Diamond-Stanic}
\affiliation{Department of Physics and Astronomy, Bates College, Lewiston, ME, 04240, USA}

\author{James E. Geach}
\affiliation{Centre for Astrophysics Research, University of Hertfordshire, Hatfield, Hertfordshire AL10 9AB, UK}

\author{Ryan C. Hickox}
\affiliation{Department of Physics and Astronomy, Dartmouth College, Hanover, NH 03755, USA}

\author{John Moustakas}
\affiliation{Department of Physics and Astronomy, Siena College, Loudonville, NY 12211, USA}

\author{Gregory H. Rudnick} 
\affiliation{Department of Physics and Astronomy, University of Kansas, Lawrence, KS 66045, USA}

\author{Paul H. Sell}
\affiliation{Department of Astronomy, University of Florida, Gainesville, FL, 32611 USA}

\author{Cameren N. Swiggum} 
\affiliation{University of Wisconsin-Madison, 475 N. Charter St. Madison, WI 53706}
\affiliation{Department of Astrophysics, University of Vienna, Türkenschanzstrasse 17, 1180 Wien, Austria}

\author{Kelly E. Whalen}
\affiliation{Department of Physics and Astronomy, Dartmouth College, Hanover, NH 03755, USA}

\begin{abstract}

We present results on the properties of extreme gas outflows in massive ($\rm M_* \sim$10$^{11} \ \rm M_{\odot}$), compact, starburst ($\rm SFR \sim$$200 \, \rm M_{\odot} \ yr^{-1}$) galaxies at z = $0.4-0.7$ with very high star formation surface densities ($\rm \Sigma_{SFR} \sim$$2000 \,\rm M_{\odot} \ yr^{-1} \ kpc^{-2}$).
Using optical Keck/HIRES spectroscopy of 14 HizEA starburst galaxies we identify outflows with maximum velocities of $820 - 2860$ \kmps. High-resolution spectroscopy allows us to measure precise column densities and covering fractions as a function of outflow velocity and characterize the kinematics and structure of the cool gas outflow phase (T $\sim$10$^4$ K). 
We find substantial variation in the absorption profiles, which likely reflects the complex morphology of inhomogeneously-distributed, clumpy gas and the intricacy of the turbulent mixing layers between the cold and hot outflow phases. There is not a straightforward correlation between the bursts in the galaxies' star formation histories and their wind absorption line profiles, as might naively be expected for starburst-driven winds. 
The lack of strong \mgii \ absorption at the systemic velocity is likely an orientation
effect, where the observations are down the axis of a blowout. We infer high mass outflow rates of $\rm \sim$50 $-$ 2200 $\rm M_{\odot} \, yr^{-1}$, assuming a fiducial outflow size of 5 kpc, and mass loading factors of $\eta\sim$5 for most of the sample. 
While these values have high uncertainties, they suggest that starburst galaxies are capable of ejecting very large amounts of cool gas that will substantially impact their future evolution.
\end{abstract}


\section{Introduction} \label{sec:intro}

In the last decade, the potential impact of galaxy-scale outflows in galactic evolution has become widely recognized \citep[e.g.,][]{kor13, som15, vei20}. Outflows provide a mechanism that can regulate and possibly quench star formation activity in the galaxy by blowing away the gas that feeds star formation and supermassive black hole (SMBH) growth, enriching the large-scale galactic environment with metals. Theoretical studies and simulations show that this ``feedback'' offers a natural explanation for a variety of observations, e.g., the chemical enrichment of the circumgalactic and intergalactic medium (CGM, IGM), the self-regulation of the growth of the SMBH and of the galactic bulge, the relative dearth of both low and high mass galaxies in the stellar mass function, and the existence of the red sequence of massive, passive galaxies \citep[e.g.,][]{sil98,tho05,ker05, opp10, mur10,tum17,dav19,nel19}. While galactic outflows appear to be crucial to rapidly shutting down star formation, the physical drivers of this ejective feedback are still unclear. In particular, the relative role of feedback from stars versus SMBHs in quenching star formation in massive galaxies remains widely debated. 

Our team has been studying a sample of massive (M$_* \sim$10$^{11}$ M$_{\odot}$) galaxies at z = 0.4$-$0.8, originally selected from the Sloan Digital Sky Survey \citep[SDSS;][]{yor00} Data Release 8 \citep[DR8;][]{Aihara11} to have typical signatures of young post-starburst galaxies. We refer to these galaxies as the HizEA\footnote{HizEA was originally coined as shorthand for High-$z$ E+A, meaning high redshift post-starburst. However, subsequent work has shown that many of the galaxies host on-going starbursts (see Section~\ref{subsection:properties}).} sample. Their spectra exhibit both strong stellar Balmer absorption from A- and B-stars and weak or absent nebular emission lines, implying minimal ongoing star formation. These galaxies are driving extremely fast ionized gas outflows, as revealed by highly blueshifted \mgii \ absorption in $\sim$90\% (Davis et al., submitted) of their optical spectra, with velocities of $\sim$1000$-$2500 \kmps, an order of magnitude larger than typical z $\sim$1 star-forming galaxies \citep[e.g.,][]{wei09, mar12}.
These results initially led our team to conclude that these galaxies are post-starburst systems with powerful outflows that may have played a critical role in quenching their star formation \citep{tre07}. 

However, several of these galaxies are detected in the Wide-field Infrared Survey Explorer \citep[WISE;][]{wri10}, and their ultraviolet (UV) to near-IR spectral energy distributions (SEDs) indicate a high level of heavily obscured star formation ($>$ 50 M$_{\odot}$ yr$^{-1}$; \citealp{dia12}). 
Furthermore, Hubble Space Telescope (HST) imaging of 29 of these galaxies reveals they are in the late stages of highly dissipational mergers with exceptionally compact central starforming regions ($R_e \sim$few 100 pc; \citealp{dia12, dia21, sel14}). Combining star formation rate (SFR) estimates from WISE rest-frame mid-IR luminosities with physical size estimates from HST imaging, we derive extraordinarily high SFR surface densities $\Sigma_{SFR}$ $\sim$10$^{3}$ M$_{\odot}$ yr$^{-1}$ kpc$^{-2}$ \citep{dia12}, approaching the maximum allowed by stellar radiation pressure feedback models \citep{leh96, meu97, mur05, tho05}.
These findings paint a different picture where our targets are starburst galaxies with dense, dusty star-forming cores \citep{per21}, and a substantial portion of their gas and dust is blown away by powerful outflows. Millimeter observations for two galaxies in our sample suggest that the reservoir of molecular gas is being consumed by the starburst with remarkable efficiency \citep{gea13}, and ejected in a spatially extended molecular outflow \citep{gea14}, leading to rapid gas depletion times. Notably, we find little evidence of ongoing AGN activity in these systems based on X-ray, IR, radio, and spectral line diagnostics \citep{sel14, per21}.
These results agree with models in which stellar feedback is the primary driver of the observed outflows \citep[e.g.][]{hop12, hop14}. 
Our sample of compact starburst galaxies is characterized by the highest $\Sigma_{SFR}$ and the fastest outflows ($>$1000 \kmps) observed among star-forming galaxies at any redshift; therefore, they are an ideal laboratory to test the limits of stellar feedback and determine whether stellar processes alone can drive extreme outflows, without the need to invoke feedback from SMBH. They could represent a short but relatively common phase of massive galaxy evolution \citep{wha22}.

Low and medium-resolution observations have shown that large-scale galactic outflows are a common feature of star-forming galaxies across a wide range of masses and redshifts \citep[e.g.,][]{mar98, hec00, wei09, mar09, mar12, rub10, rub14,hec15,hec16,mcq19}. Thanks to these studies we have a fair understanding of the typical outflow kinematics, and how their main properties scale with the fundamental properties of their galaxy hosts.
While absorption-line spectroscopy at low and medium resolution is sufficient to measure velocities and the total amount of absorption within outflows, robust determinations of key physical properties such as the column density, covering fraction, and mass outflow rate require high spectral resolution. In this paper, we present new optical Keck/HIRES observations for 14 of the most well-studied starburst galaxies in our sample. These high-resolution spectra allow us to directly measure accurate column densities and covering fractions as a function of velocity for a suite of Mg and Fe absorption lines. We utilize this to probe the small scale structures of the extreme galactic outflows observed in our sample and investigate the potential impact of these outflows on the evolution of their host galaxies.

The paper is organized as follows: Section \ref{sec:data} describes the sample selection, observations, and data reduction; Section \ref{sec:spec_analysis} illustrates our line profile fitting method and measurements of the absorption line column densities and kinematics; Section \ref{section:discussion} presents our main results in comparison to theoretical models and other galaxy samples, and discusses the more comprehensive implications of our analysis. Our conclusions are reviewed in Section \ref{sec:Conclusions}. 

Oscillator strengths and vacuum wavelengths are taken
from \cite{mor91, mor03}. Throughout the paper, we assume a standard $\Lambda$CDM cosmology, with H$_0$ = 70 \kmps Mpc$^{-1}$, $\Omega_m$ = 0.3, and $\Omega_{\Lambda}$ = 0.7. All spectra are converted to vacuum wavelengths and corrected for heliocentricity.

\section{Sample and Data Reduction} \label{sec:data}

The parent sample for this analysis was selected from the Sloan Digital Sky Survey I (SDSS-I \citealp{yor00}) Data Release 8 (DR8; \citealp{Aihara11}) and is described in \citet{tre07, dia12, sel14, dia21} and Davis et al., (submitted). Briefly, as mentioned above, the original goal was to identify a sample of intermediate redshift (z = $0.35 - 1$) post-starburst galaxies to investigate their star formation quenching mechanisms. 
We focused on redshift $>$ 0.35 so that the \mgb \ doublet, an ISM line broadly used to probe galactic winds, was easily observable with optical spectrographs. Since the SDSS main galaxy sample \citep{strauss02} is magnitude limited (r $<$ 17.7) and does not contain many galaxies at z $>$ 0.25, we concentrated on objects targeted for spectroscopy as quasars because they probe fainter magnitudes (i $<$ 20.4) and higher redshifts. We selected all objects classified by the SDSS spectroscopic pipeline as galaxies with redshifts between z = $0.35 - 1$ and either g $<$ 20 or i $<$ 19 mag (296 galaxies). To select galaxies with a recent ($<$ 1 Gyr) burst of star formation, but little ongoing activity, we required them to show strong Balmer-absorption (suggestive of a burst $50 - 1000$ Myr ago) and moderately weak nebular emission (suggestive of a low SFR in the past 10 Myr). Removing 3\% of the objects by visual examination (blazars or odd quasars with wrong redshifts) yielded a sample of 121 galaxies. More details about the sample selection can be found in Tremonti et al., (in prep.) and Davis et al., (submitted).

We have undertaken extensive follow-up observations on a subsample of 50 of these galaxies. In choosing this subsample, we prioritized galaxies with the largest g-band fluxes and the youngest stellar populations, but we included galaxies with older burst ages for comparison. The sample comprises galaxies with mean stellar ages in the range of 4 - 400 Myr. Notably, we did not prioritize galaxies with \ion{Mg}{2} absorption lines. 

We obtained ground-based spectroscopy of the 50/121 galaxies with the MMT/Blue Channel, Magellan/MagE, Keck/LRIS, Keck/HIRES, and/or Keck/KCWI \citep{tre07, dia12, sel14, rup19}, X-ray imaging with Chandra for 12/50 targets \citep{sel14}, radio continuum data with the NSF’s Karl G. Jansky Very Large Array (JVLA/VLA) for 20/50 objects \citep{pet20}, millimeter data (ALMA) for 2/50 targets \citep{gea14, gea18}, and optical imaging with HST for 29/50 galaxies \citep[``HST sample''][]{dia12, sel14, dia21}. For the HST observations, we first targeted the twelve galaxies that were most likely to have AGN. We observed three galaxies with broad \mgii \ emission (expected to be type I AGN) and nine galaxies with strong \ion{[O}{3]} emission, which could be indicative of an obscured (type II) AGN. 
We subsequently observed an additional seventeen galaxies that have the youngest estimated post-burst ages (t$_{burst}$ $<$ 300 Myr). We also obtained multi-band HST imaging to study the physical conditions at the centers of the 12/29 galaxies with the largest SFR surface densities estimated by \citet{dia12}, (30 M${\odot}$ yr$^{-1}$ kpc$^{-2}$ $<$ $\Sigma_{SFR}$ $<$ 2000 M${\odot}$ yr$^{-1}$ kpc$^{-2}$), and investigated the young compact starburst component that makes them so extreme \citep{dia21}. 

In this paper, we focus on 14 galaxies from the HST sample (6 from the 12/29 most AGN-like galaxies, and 8 from the 17/29 with the youngest post-burst ages). The galaxies and their basic properties are listed in Table~\ref{table1}.

\subsection{Data Acquisition and Reduction}\label{subsection:reduction}
We obtained high–resolution spectra of 14 starburst galaxies spanning emission redshifts 0.4 $<$ z $<$ 0.8, using the Keck High Resolution Echelle Spectrometer (HIRES; \citealp{vogt94}) on the Keck I telescope over the nights of $12-13$ August 2009, $10-11$ December 2009, and $13-14$ May 2012. 
HIRES was configured using the 1.148 $\times$ 7 arcsec C5 decker with no filters and 2x2 binning.
The 1.148 arcsec slit width disperses the light with R = $\lambda$/$\Delta\lambda$ = 37,000 ($\simeq$ 8 \kmps) projected to $\simeq$ 3 pixels per resolution element, $\Delta\lambda$. Individual exposures were 3600 seconds, with total integration times of $2 - 3$ hours per object.

This yielded an observed wavelength coverage from $\sim$3150 to 5800 \AA\, in $>$30 orders, which for our $z = 0.4 - 0.8$ galaxies corresponds to a rest-frame coverage from $\lambda_{rest} \sim$1800$-$3250 to $2300-4100 $~\AA\, depending on redshift. We concentrate our analysis on
the \mgb \ doublet, the \mgi$\lambda$2852 transition, and the \feii$\lambda$2344, 2374, 2582, 2587, and 2600 multiplet. 

\input{Table1.tex}

The data were reduced using the XIDL HIRES Redux code\footnote{Available at \url{https://www.ucolick.org/~xavier/HIRedux/index.html}}. We used this reduction pipeline to perform flat-field corrections, determine wavelength solutions using thorium-argon (ThAr) arcs, 
measure slit profiles for each order, perform sky subtraction, and remove cosmic rays. The output of the spectral extraction is an order-by-order spectrum of counts versus vacuum wavelength. 
For a more comprehensive description of this procedure, we refer the reader to \citet{omea15}. As automated flux calibration of HIRES spectra is unreliable \citep[e.g.,][]{suz05}, each echelle order was examined by eye to perform continuum fitting. A Legendre polynomial was fit to each order and adjusted using anchor points in absorption-free regions with a custom IDL code.

\begin{figure*}[htp!]
 \centering
 \includegraphics[width=0.65\textwidth]{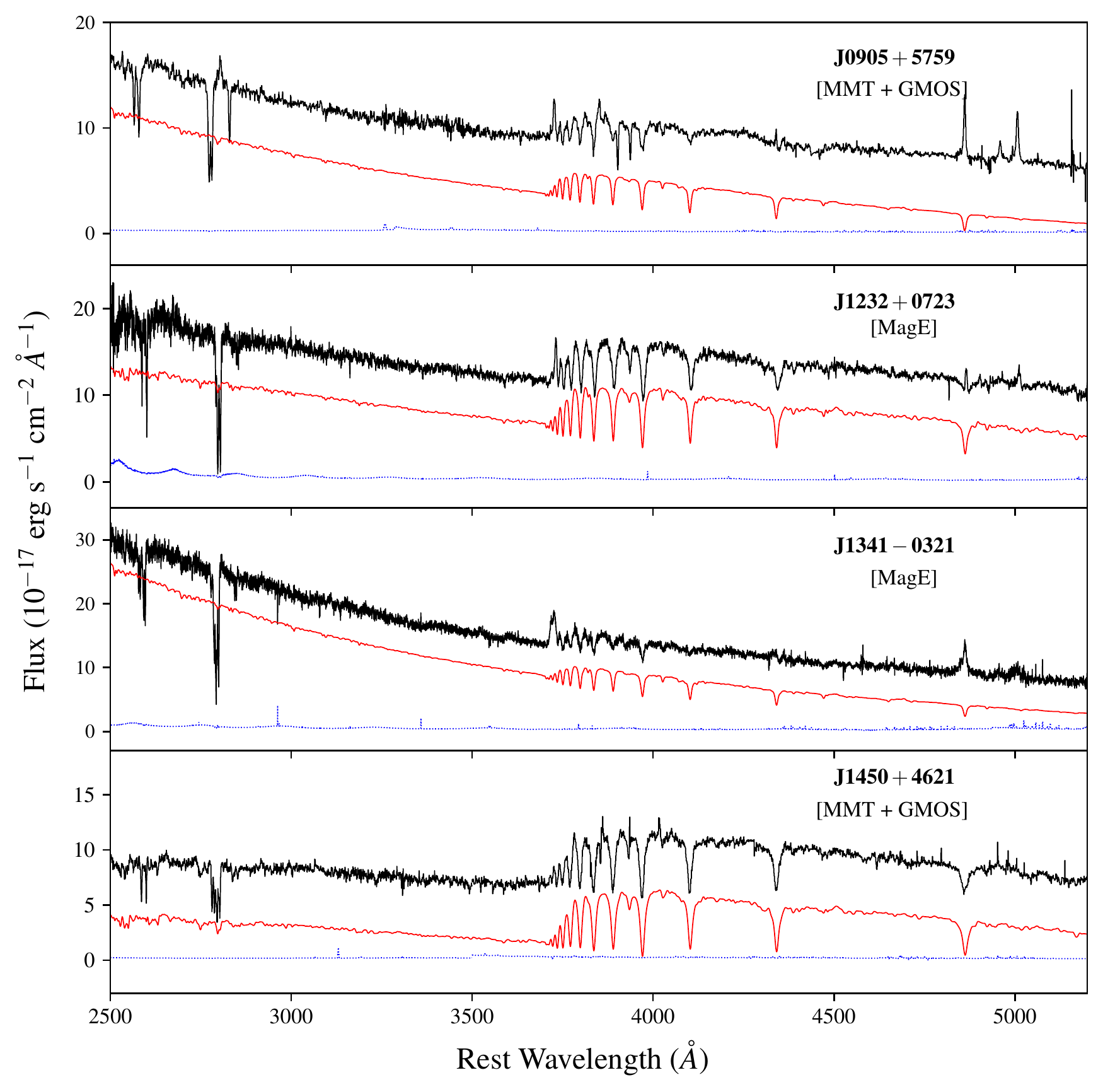}
 \caption{Rest-frame near-UV and optical spectra of 4 representative galaxies in our sample. The black line shows the combined MMT + GMOS or MagE spectra as marked in each panel, and the corresponding errors are shown in blue. The red line shows the continuum model fit, offset in the vertical direction for clarity. The spectra are dominated by the light from a young stellar population but have relatively weak nebular emission lines and strong \ion{Mg}{2} $\lambda \lambda$2796,2803 absorption, originating from the outflows.}
 \label{fig:cont}
\end{figure*}

\subsection{Galaxy properties}\label{subsection:properties}

We report in Table~\ref{table1} relevant derived galaxy properties for our sample. Accurate systemic redshifts are necessary in order to determine outflow velocities. For this purpose, we use stellar continuum fits as derived in Davis et al. (submitted). We have collected low/medium resolution (R = $600-4000$) optical spectra of our parent sample (50 galaxies) using three instruments on $6 - 10$-m class telescopes (MMT/Blue Channel, Magellan/MaGE, and Keck/LRIS; see Davis et al. (submitted) and Tremonti et al. in prep. for details on these data sets). We required our spectra to extend to at least 4000 \AA\, in the restframe to cover strong stellar or nebular features to aid in measuring the galaxy systemic redshift (e.g. \oii, low order Balmer absorption lines). We fit the spectra with a combination of SSP models and a \citet{salim18} reddening law derived from local galaxies that are analogs of massive, high redshift galaxies. We used the Flexible Stellar Population Synthesis code \citep{con09, con10} to generate SSPs with Padova 2008 isochrones \citep{marigo08}, a \citet{sal55} initial mass function (IMF), and the theoretical stellar library C3K (Conroy et al., in prep.) with a resolution of R $\sim$10,000. We utilize solar metallicity SSP templates with 43 ages spanning 1 Myr $ - $ 8.9 Gyr and perform the fit with the Penalized Pixel-Fitting (pPXF) software \citep{cap04, cap17}. We mask the region around \mgii \ and the forbidden emission lines during the fit and implement two separate templates for broad and narrow Balmer emission lines, assuming Case B recombination line ratios. Both the lines and stellar continuum are attenuated by the same amount of dust in the pPXF fit; we find that the results are insensitive to this assumption.
By fitting Balmer emission and absorption lines simultaneously we can take into account the potential infill of the absorption line cores. One of the outputs of our pPXF fit is the galaxy systemic redshift (\zsys), which we list in Table~\ref{table1}. 

The fit produces a stellar continuum model without a nebular emission component (see Figure~\ref{fig:cont}). Most sources, in addition to having strong Balmer absorption, have very blue continua indicating a recent starburst event ($\sim$1$-$10 Myr) that is not highly dust-obscured. For galaxies with young stellar populations, like those in our sample, the stellar contribution to the Mg absorption lines is minimal. However, we utilize our best fit pPXF continuum model to properly remove the stellar absorption features from each spectrum in our sample.

To estimate stellar masses (M$_*$) and star formation rates (SFR) we fit the combined broad-band UV -- mid-IR photometry and optical spectra using the Bayesian SED code Prospector \citep{leja19, johnson21}, as described in Davis et al. (submitted). Briefly, we incorporate the 3500 - 4200 \AA\ spectral region in the SED fit as it covers many age-sensitive features (e.g., D4000, H$\delta$). We generate simple stellar population (SSP) models employing the Flexible Stellar Populations Synthesis code \citep[FSPS;][]{con09}, adopting a Kroupa IMF \citep{kro01} and the MIST isochrones \citep{cho16} and the C3K stellar theoretical libraries (Conroy et al., in prep.). The stellar models are very similar to the ones described above over the wavelength range of interest for this work. We determined the best-fit parameters and their errors from the 16th, 50th, and 84th percentiles of the marginalized probability distribution function. The combined photometry and spectra are well fit by these models (see Davis et al., submitted for examples of the SED fitting). However, the dust emission properties of our sample are poorly constrained due to the low signal-to-noise ratio (SNR) of the WISE W3 and W4 photometry and the limited infrared coverage of the SED. This yields fairly tight constraints on the M$_*$ ($\pm$0.15 dex) and slightly larger errors on the SFR ($\pm$0.2 dex). M$_*$ represents the present-day stellar mass of the galaxy (after accounting for stellar evolution) and not the integral of the star formation history (i.e. total mass formed). We list in Table~\ref{table1} the SFRs calculated from the resulting star formation history (SFHs; see Fig.~\ref{fig:SFH}), averaged over the last 100 Myr. This is the typical timescale for which UV and IR star formation indicators are sensitive \citep{ken12}. As here we are interested in the most recent SFH, we compute the light-weighted age of the stellar populations younger than 1 Gyr, with the light contribution estimated at 5500 \AA. These $<$1 Gyr light-weighted ages more closely replicate the timescale of the peak SFR than the mass-weighted ages. We use light-weighted ages rather than the age since the burst as the light-weighted ages are more robust to changes in our modeling approach.  However, there may be systematic errors related to the stellar population models we adopt.  In practice uncertainties in the treatment of  Wolf-Rayet stars and high mass binary evolution can have a large impact on the UV spectra of galaxies with young stellar populations \citep[e.g.][]{eld16}. The detailed analysis required to produce a quantitative estimate of the systematic errors on the light-weighted ages is beyond the scope of this work. The light-weighted ages for the galaxies in our sample are reported in Table~\ref{table1}.

The effective radii (r$_e$) measurements for the galaxies in our sample are discussed in \citet{dia12, dia21}. Briefly, for 11 galaxies we use the GALFITM software \citep{haussler13, vika13} to perform S\'ersic fits of joint multi-band HST imaging \citep{dia21} in the UVIS/F475W and UVIS/814W filters. 
To prevent uncertainties due to tidal features, we fit the central region of the galaxy and extrapolate the fit to larger radii to compute r$_e$. 
The shorter-wavelength filter ($\lambda_{rest}$(F475W) $\approx$ 3000\AA) traces the young, unobscured stars, while the longer-wavelength filter ($\lambda_{rest}$(F814W) $\approx$ 5200\AA) is more sensitive to the underlying stellar mass. 
Characteristic errors on the effective radius are of the order of 20\%. For the remaining three galaxies (J1125, J1232, and J1450) we quantify the morphology utilizing optical HST UVIS/F814W image \citep{dia12}. We model the two-dimensional surface brightness profile with a single Sersic component (defined by Sersic index $n$ = 4 and r$_e$) using GALFIT \citep{pen02, pen10}. We adopt an empirical model point-spread function (PSF) produced using moderately bright stars in our science images.

\begin{figure*}[htp!]
 \centering
 \includegraphics[width=0.9\textwidth]{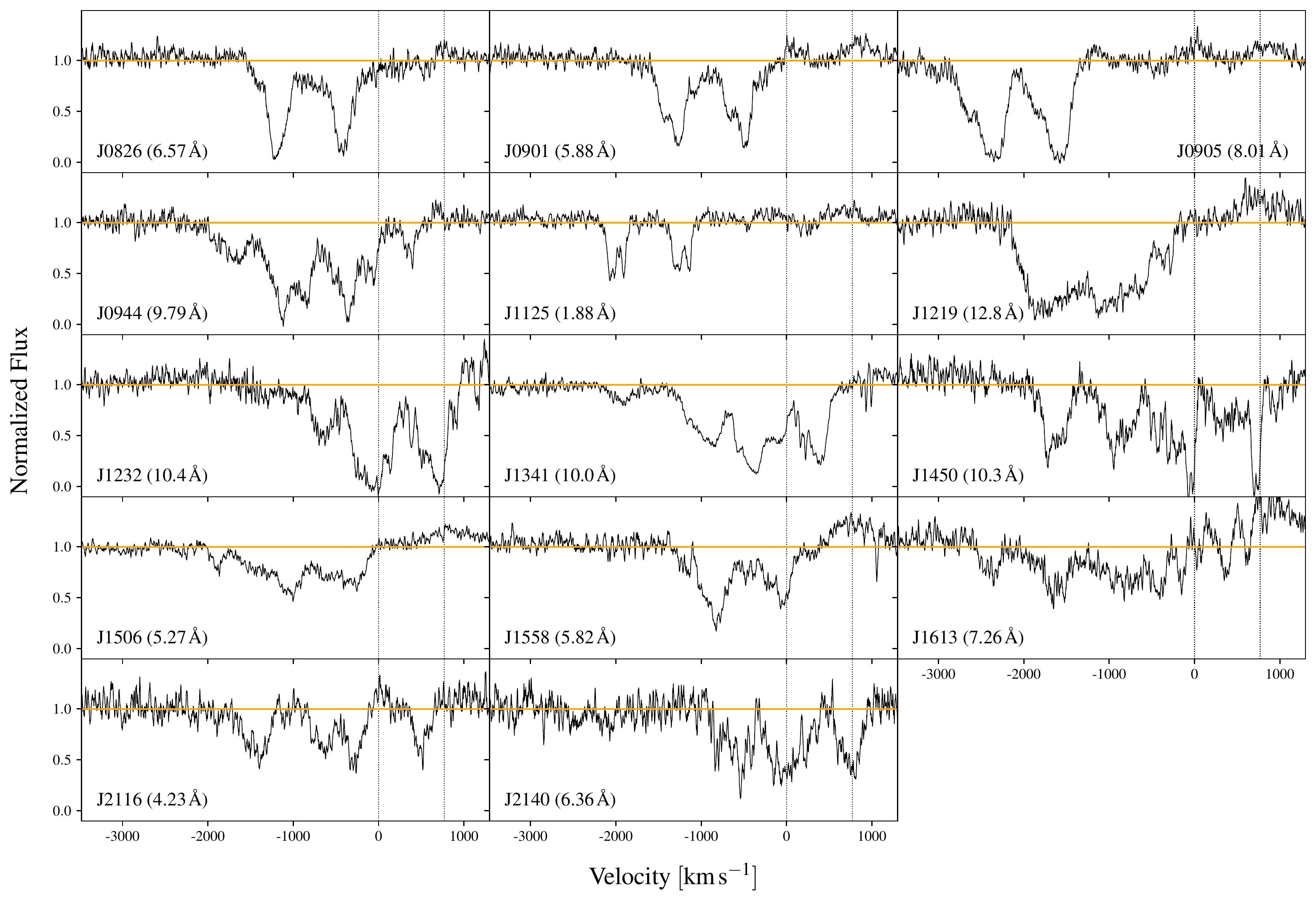}
 \caption{Normalized spectra of the \mgii \ absorption troughs in the galaxies in our sample. The velocities on the horizontal axis refer to the \mgii$\lambda$2796 doublet component, relative to the systemic redshift. The vertical dotted lines mark the \mgii$\lambda$2796 ($v=0$ \kmps) and \mgd \ ($v=+770$ \kmps) position in this velocity space, at the galaxy systemic redshift. We report the total rest frame \mgii \ equivalent width at the bottom of each panel. The spectra here and below have been lightly smoothed by two resolution elements ($\sim$16 \kmps).}
 \label{fig:allfits}
\end{figure*}

\section{Spectral Analysis}\label{sec:spec_analysis}

In this section, we briefly present our data, discuss the method and assumptions adopted for the line profile fitting and describe the techniques used to estimate errors. The line fitting results for each galaxy are listed in Tables~\ref{table2} and ~\ref{table3}.

\subsection{Absorption Lines as Tracers of Galactic Outflows}

Galactic winds are typically recognized through their kinematic signatures. Winds seen in absorption are identified as troughs detected in the foreground of the galaxy stellar continuum, blueshifted with respect to the galaxy systemic velocity \cite[e.g.,][]{mar09, mar12, kor12, rub14, pru21}. In this context the line velocity shift relative to the systemic redshift ($z_{sys}$) at which 98\% ($v_{98}$) or 50\% (v$_{avg}$) of the equivalent width (EW) accumulates moving from red (positive velocities) to blue (negative velocities) across the line profile can be used to parameterize the kinematics of absorption lines.

The galaxies in our sample commonly display extreme kinematics in the \mgii \ absorption lines. Figure~\ref{fig:allfits} shows the continuum-normalized \mgii \ spectral region for each galaxy and illustrates the range of spectral absorption properties in our data. The \mgb \ absorption lines are produced by the blending of multiple velocity components. Additionally, the \ion{Mg}{2}$\lambda$2796 line has blueshifted components that blend with the \mgd \ components in almost all of the galaxies (with the exception of J1125 and J2116). The velocity blueshifts that cause this blending range from several hundred up to over 2,000 \kmps\, in the \ion{Mg}{2} absorbing gas. Given the complex kinematics, we cannot directly integrate the spectrum to measure the EW of the single absorption components. However, we can calculate the total \mgii \ EW by integrating the normalized spectrum over the velocity range for which we detect \mgii \ absorption. The total rest frame EWs for the galaxies in our sample are reported at the bottom of each panel in Figure~\ref{fig:allfits}. The \mgii \ absorption lines are very strong, with EWs from 1.89 to 12.90 \AA, with an average value of 7.49 \AA. (Below in Section~\ref{sec:trends} we compare our galaxies to other samples in the literature.) The $v_{98}$ values in the \mgii \ absorption lines span a range from $-620$ to $-2700 $ \kmps, with an average value of $-1630$. Such large line blueshifts are unambiguous signs of outflowing gas. We note that the potential emission filling effects on the $v_{98}$ estimates are negligible (see Section~\ref{subsection:lack}). In the following section, we construct a model that describes our data in order to quantify the kinematics and strength of the outflows.

\subsection{Line Fitting}\label{sec:fit}

We refer the reader to \citet{rup05} for a complete discussion of the analysis techniques used in absorption line fitting, depending on the type of line profiles studied. Here we summarize the relevant aspects for our sources which have partially covered, blended absorption lines.

The profile of a single absorption trough is parameterized by the distribution of the optical depth ($\tau$) along the line of sight, combined with how the absorbing gas covers the background source. If it completely covers the background source, such that the covering fraction ($C_f$) is unity, then the $\tau$ distribution can be calculated accurately as a function of velocity. This is valid also for a blended doublet, triplet, or higher-order multiplet lines which can be fit by solving a set of linear equations, using a ``regularization'' method \citep{arav99}. However, when the background light source subtends a wide angle on the sky compared to the absorbing clouds, \cf\, is generally less than unity. This latter case may apply to our sample, where the central starburst illuminates gas clouds in the galaxy halo. 

The intensity of an absorption line, where \cf\, and $\tau$ depend on velocity and the continuum level has been normalized to unity, can be described as $I(\lambda) = 1 - C(\lambda) + C(\lambda) e^{-\tau(\lambda)}$. There is a degeneracy when solving for $\tau$ and \cf\, for a single line. This degeneracy can usually be resolved by simultaneously fitting two or more transitions of the same species with known oscillator strengths ($f_0$), as the relative depths of the lines are independent of \cf. However, in the case of doublet or higher-order multiplet lines blended together, we cannot solve for $\tau$ and \cf\, directly, as in the region of overlap the solution is not unique. Nor can we directly integrate the spectrum to estimate the EW of the single absorption troughs. We must thus fit analytic functions to the line profiles. We, therefore, use intensity profiles which are direct functions of physical parameters (i.e., velocity, optical depth, and covering fraction). We assume $\tau$ can be formulated as a Gaussian, $\tau(\lambda) = \tau_0 e^{(\lambda - \lambda_0)^2/(\lambda_0 b/c)^2}$, where $\lambda_0$ and $\tau_0$ are the central wavelength and central optical depth of the line, and $b$ is the velocity width of the line (or Doppler parameter $b =\sqrt{2}\sigma = \text{FWHM}/[2\sqrt{\ln2}]$). The clear advantage of this approach is that the derived profile shapes are readily interpreted in terms of these physical parameters. This method accurately handles the intensity profile at both low and high optical depth in the case of constant \cf\, which is an assumption we make for simplicity. 

To decompose the blended absorption profiles, we need to make assumptions about the geometry of the absorbing gas. Here we refer to ``components'' to describe distinct velocity components of the same transition.

We assume the case of completely overlapping atoms when combining the intensities of two doublet or multiplet lines within a given velocity component. In this context, the atoms at all velocities are placed at the same position in the plane of the sky relative to the background continuum source. This is a simplification, as the broad profiles we observe could be due to the large-scale motions of individual clouds that are not all coincident. The covering fraction in this case is independent of velocity. The expression for the combined intensity of a doublet or multiplet (each with optical depth $\tau_i(\lambda)$ and covering fraction $C_{f,i}(\lambda$)) is then given by

\begin{equation}
I(\lambda) = 1 - C_f + C_f e^{-\tau_1(\lambda)-\tau_2(\lambda)}.
\end{equation}

We assume the case of partially overlapping atoms when combining two different velocity components. The motivation for this is that different components could have different \cf\, if we assume they are spatially distinct. In this scenario, at a given wavelength there is an overlap between the atoms producing the components, and the covering fraction describes the fractional coverage of both the continuum source and the atoms producing the other component. The definition of the intensity is 

\begin{multline}
I(\lambda) = [1 - C_{f,1} + C_{f,1} e^{-\tau_1(\lambda)}] \\ \times [1 - C_{f,2} + C_{f,2} e^{-\tau_2(\lambda)}] = I_1(\lambda)I_2(\lambda).
\end{multline}

We fit our data utilizing a suite of IDL routines from the IFSFIT library \citep{rup14f}. The code combines the two cases illustrated above when simultaneously fitting doublet lines that have multiple blended velocity components.
We describe the normalized line profile intensity for each component as a
function of four parameters: $\lambda_0$, \bd, $C_f$, and column density ($N$). \N can be formulated as
\begin{equation}\label{N}
N[\cm] = \frac{\tau_0 \, b[\kmps]}{1.497 \times 10^{-15} \lambda_0 [\mbox{\normalfont\AA}] f_0}.
\end{equation}

Different transitions of the same ionic species are required to have the same component structure and are fit simultaneously to produce a single solution. However, we do not impose the same kinematics on different ionization states of the same element or on other elements. As we impose no constraints on the decomposition between different ionization states or different ionic species, any qualitative or quantitative correspondences between the fits occur naturally.

A comparison of the doublet (multiplet) line shapes in our spectra often reveals nearly identical intensities. In these cases, the relative intensities of the doublet (multiplet) troughs constrain that transition to be optically thick, setting a lower limit on the optical depth at the line center. We also enforce an upper limit on the optical depth in \mgii \ of $\tau_{2803}=10$ and in \feii\ of $\tau_{2586}=10$. The data are not sensitive to changes in optical depth above these values unless the optical depth is $\gg10$. 

Our best fit is degenerate with fits to larger numbers of components. For example, an absorption line with a single component could be fit with a given \bd\, and \ta, or it could be fit by adding multiple velocity components with narrower \bd\, and larger \ta. Following previous studies \citep{rup05, mar09}, here we adopt the minimum number of velocity components required to describe the doublet (multiplet) absorption troughs.

To compute errors in the best-fit parameters, we follow the Monte Carlo method described in detail in \citet{rup05}. In brief, we first assume the fitted parameters represent the ``real'' parameter values. Then, we add random Poisson noise to our best fit model that we assume to be the ``real'' spectrum, where the extent of the errors is assessed from the data. We re-fit this new spectrum with initial guesses equal to the best-fit parameters in the fit to the real data and record the new best-fit parameters. We repeat this 1000 times. This produces a distribution of fitted parameters for each galaxy. We compute the 1$\sigma$ errors in the best-fit parameters by computing the 34\%\ probability intervals above and below the median in each parameter distribution.

\startlongtable
\input{Table2.tex}

\input{Table3.tex}

\begin{figure*}[htp!]
 \centering
 \includegraphics[width=0.9\textwidth]{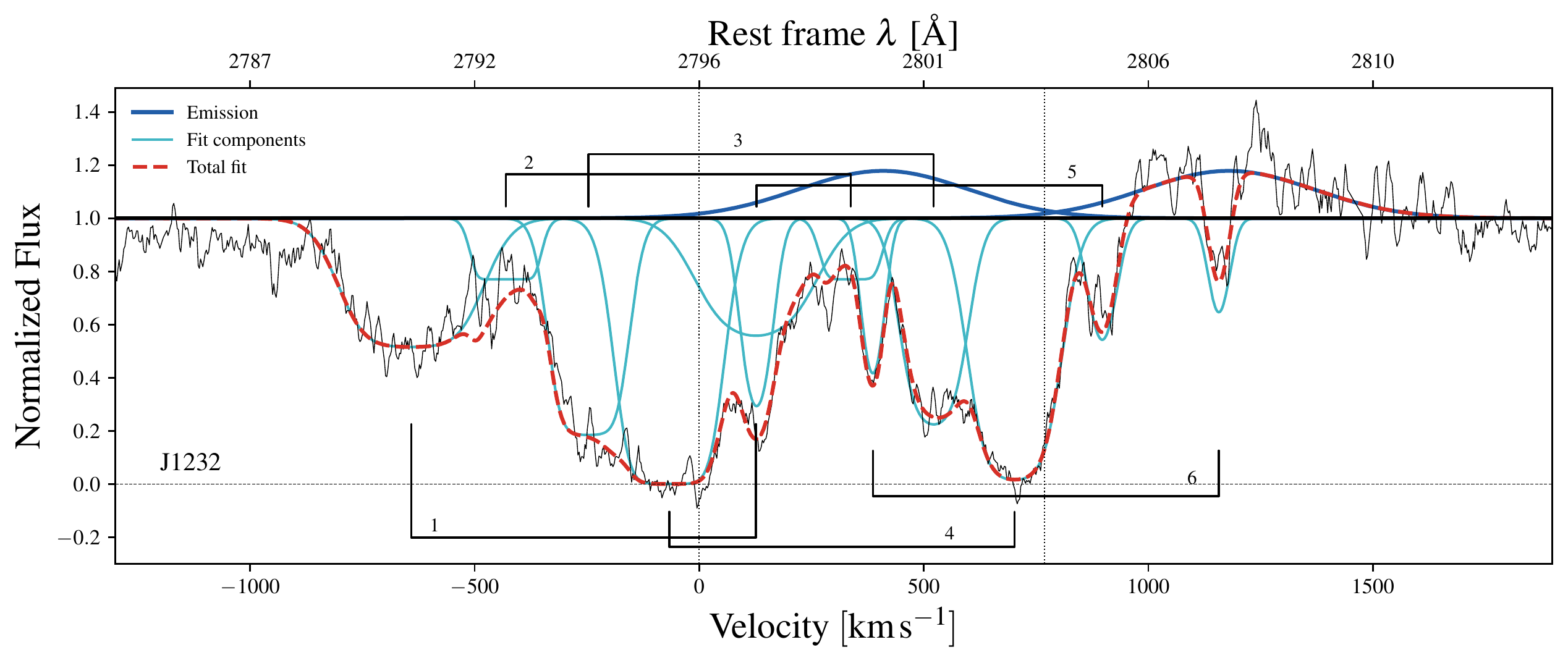}
 \caption{Normalized spectrum of the \mgii \ absorption lines that appear in the multi-component outflow complex in the galaxy J1232. The lower horizontal axis shows the velocity of the \mgii$\lambda$2796 doublet component relative to the galaxy systemic redshift (\zsys). Vertical black dotted lines mark the \mgii$\lambda$2796 and \mgii$\lambda$2803 locations at \zsys, respectively. The upper horizontal axis indicates the rest frame wavelengths. The light and dark blue solid lines show the best fit \mgii \ absorption and emission line profiles, respectively. Brackets mark \mgii \ doublet pairs numbered in decreasing order of blueshift velocity from \zsys. The red dashed line shows the total \mgii \ best fit profile. }
 \label{fig:bestfit_J1232}
\end{figure*}

\begin{figure*}[htp!]
 \centering
 \includegraphics[width=0.8\textwidth]{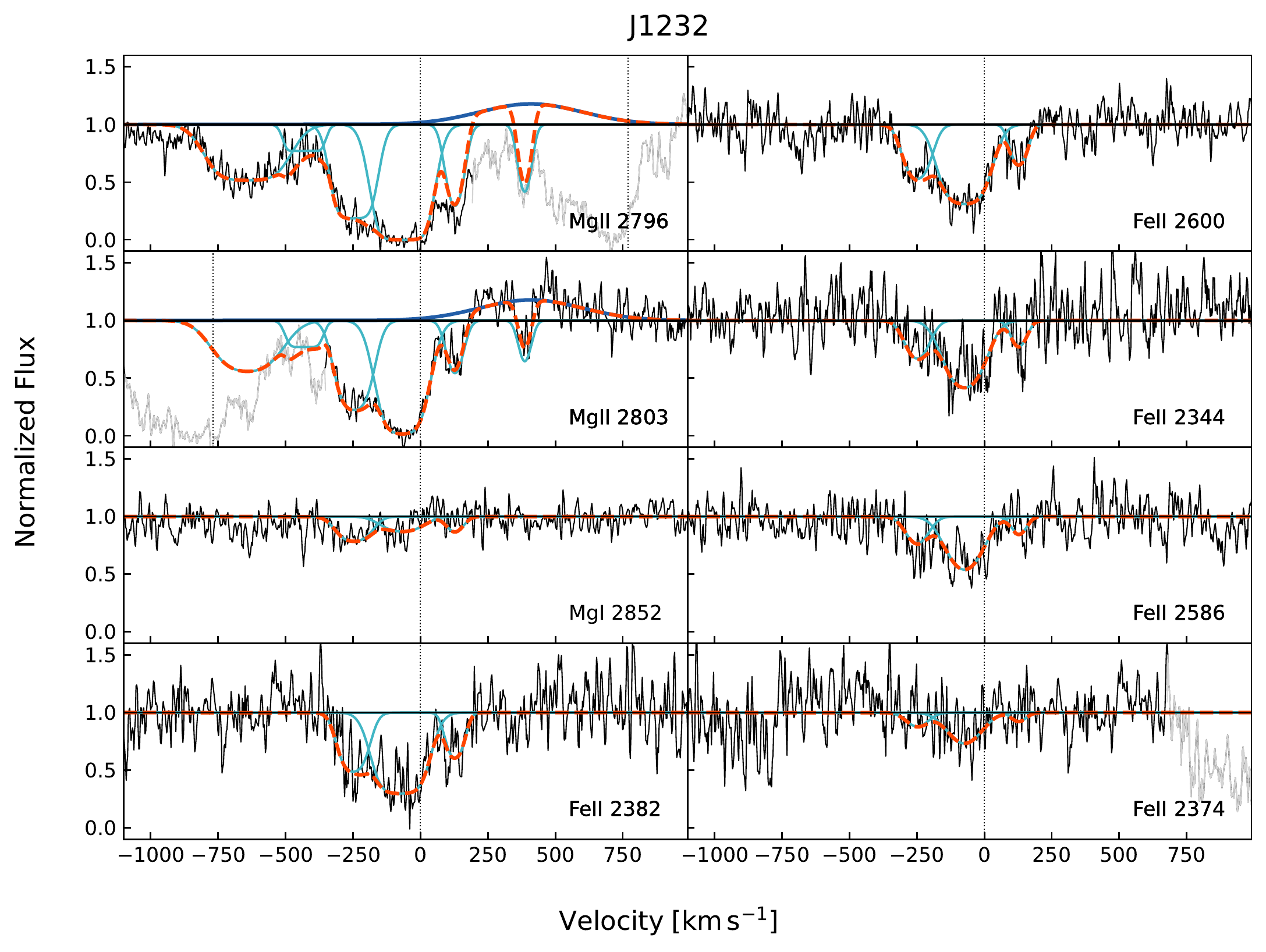}
 \caption{Normalized spectrum of the galaxy J1232 centered on the transitions relevant to this study: \mgii$\lambda\lambda$2796,2803\AA, \mgi$\lambda$2803\AA, \feii$\lambda$2382\AA, \feii$\lambda$2600\AA, \feii$\lambda$2344\AA, \feii$\lambda$2586\AA, and \feii$\lambda$2374\AA. In each panel $v = 0$ corresponds to the transition wavelength at the galaxy systemic redshift. When required, the fit includes emission lines (dark blue solid lines). In each panel, the light blue solid lines represent the individual velocity components fit for the relevant ion, while the dashed orange line shows the total line profile fit. The light grey part of the spectrum shown in some panels indicates a region where multiple transitions overlap in wavelength. The J1232 spectrum exhibits \mgii \ emission within to $+900$ \kmps\, of \zsys. In this galaxy, we see strong absorption from \mgii \ and \feii \ within $-380$ and $+220$ \kmps\, of the systemic redshift. The \mgii \ doublet has three additional absorption components at $v \sim$-640 \kmps, $v \sim$-430 \kmps\, and $v \sim$+385 \kmps. The detected \mgi \ absorption troughs are weak and have velocities within $-380$ and $+220$ \kmps. We show the \mgi$\lambda$2853 best fit profile as derived following the constrained approach (see text for details).}
 \label{fig:j1232_all}
\end{figure*}
\subsubsection{\rm Mg II}\label{mgii}

To quantify the kinematics and absorption strength of \mgii \ lines in our spectra, we fit the line profile shape assuming that the absorption of the continuum emission is due to foreground \mgii \ ions. However, continuum photons can be absorbed by gas both in front of and surrounding the galaxy and successively re-emitted in any direction, such that the excited ions decay straight back to the ground state. This scattering mechanism can produce P Cygni-like line profiles for \mgii \citep{rub11, pro11} which has often been detected in star-forming galaxies at z $\sim$0.3 and $\sim$1 \citep{mar09, wei09, rub10, rub14, rup19, bur21}. We observe \mgii \ emission in 9/14 galaxies in our sample. For 7 of these galaxies, we have KCWI data that confirm the presence of \mgii \ emission (see Section~\ref{subsection:lack}). Another common characteristic in our sample is the absence of strong \mgii \ absorption at the galaxy systemic velocity, which traces absorption in the interstellar medium (ISM) of the galaxy. We note that 11/14 galaxies in our sample have less than four percent of the \mgii \ EW within 200 \kmps of \zsys. We come back to this point and discuss the potential effects of \mgii \ emission filling below in Section~\ref{subsection:lack}.

We quantify the \mgii \ kinematics for each galaxy in our sample by fitting the absorption line profiles as described in Section~\ref{sec:fit}. At a given wavelength the atoms producing the \mgo \ line are separated by 7\AA\, (770 \kmps) from the ones producing the \mgd \ line. However, at a given velocity relative to \zsys, we consider them to have (1) relative $\tau$ defined by atomic physics ($\tau_{2976}$ = 2 $\times$ $\tau_{2803}$; \citealp{mor03}), and (2) equal \bd\, and $C_f$. The best fit parameters are reported in Table~\ref{table2}.

Figure~\ref{fig:bestfit_J1232} shows an example of the best fit to the \mgii \ absorption in the galaxy J1232$+$0723 (hereafter J1232).
The lower horizontal axis shows the velocity of the \mgii$\lambda$2796 doublet component relative to the galaxy systemic redshift (\zsys), and the vertical black dotted lines mark the \mgo \ and \mgd \ location in the velocity space at \zsys, respectively.
The \mgii \ doublet absorption profile is well described by six Gaussian components (shown with light blue solid lines). The brackets in Figure~\ref{fig:bestfit_J1232} mark \mgii \ doublet pairs numbered in decreasing order of blueshift velocity from \zsys. The three most blueshifted components (1, 2, and 3) are characterized by \cf $< 1$. Comparison of the \mgo \ and \mgd \ trough shapes illustrates their nearly identical intensity, but the lines are not black. The remaining three components have \cf $= 1$. The spectrum also clearly includes redshifted ($\sim$410 \kmps) \mgd \ emission observed up to $+900$ \kmps\, relative to \zsys. (The corresponding \mgo \ line is not obvious because of \mgd \ absorption at the same wavelengths.) The inclusion of an emission component in the model significantly improves the fit to the \mgii \ absorption trough. The red dashed line in Figure~\ref{fig:bestfit_J1232} represents the total \mgii \ best fit, including the \mgii \ emission (dark blue solid line).

Figure~\ref{fig:j1232_all} shows an ionic species stack from the spectrum of J1232, centered on the absorption lines included here.
Each panel shows the line decomposition (light blue solid lines) and the total best fit (dashed orange lines) to a different ion absorption trough. The light grey part of the spectrum shown in some panels marks a region where multiple transitions overlap in wavelength. The top two left panels show our best fits to \mgo \ and \mgd, respectively. It is worth noting that where the \mgo \ and \mgd \ velocity components lie on top of each other, then the dashed orange line does not describe the observed spectrum (in black) because it represents the total fit of a single transition only. For example, the high-velocity \mgd \ absorption makes the \mgo \ trough appear to be deeper than the \mgd \ trough between $+190$ and $+330$ \kmps. When the best fit profile of both doublet components are combined they fit the data well, as shown in Figure~\ref{fig:bestfit_J1232} (dashed red line).

\subsubsection{\rm Mg I}\label{fit:mg1}

We also fit \mgi \ absorption in our sample. 
Observations of a single \mgi \ transition do not directly
constrain the \mgi$\lambda$2853 optical depth. To quantify the kinematics and absorption strength of \mgi$\lambda$2853 line profiles in our spectra, therefore, we follow two distinct approaches. First, we perform a fit to the \mgi \ absorption troughs independent from the \mgii \ fit, as described in Section~\ref{sec:fit}. We set \cf\, = 1 and let the other parameters (i.e. \bd, $\lambda_0$, and \ta) vary. This approach assumes that an absorption line that is not black is produced by optically thin gas. If the gas is instead optically thick and the shape of the absorption trough is determined by the \cf\, this method provides a lower limit to its column density. The \mgi \ best fit parameters are listed in Table~\ref{table3} (columns 5, 6, and 7).
 
Our second approach assumes the optical depth of each \mgi \ velocity component to be linked to that of the \mgii \ component. We set \bd\, and $\lambda_0$ for each \mgi \ velocity component to be the same as the corresponding \mgii \ component and estimate what \cf(\mgi) best fits the data. The optical depth in \mgi \ is

\begin{multline}\label{tau_mgi}
\tau_0({\rm MgI_{2853}}) = 6.1 \, \tau_0({\rm MgII_{2803}})\, \times \\ \chi( {\rm MgI})/ \chi({\rm MgII}) \, C_f({\rm MgII})/ C_f({\rm MgI}),
\end{multline}

where $\chi$(\mgi)/$\chi$(\mgii) is the relative ionization correction. For starburst spectral energy distributions (SED), $\chi$(\mgii) $\geq 0.7$ (Martin et al. 2009). We set the neutral fraction to be 30\% (i.e. $\chi$(\mgi) = 0.3). 
This approach is motivated by the fact that the \mgi \ and \mgii \ absorption troughs have similar kinematic structures, however, \mgi \ is seen to be shallower than \mgii. When \ta(\mgii) is large and \cf(\mgii) $< 1$, the neutral Mg fraction must be less than a few percent to produce a shallow \mgi$\lambda$2853 trough with \cf = 1 (i.e. optically thin). In this case, it is more likely that \mgi$\lambda$2853 is optically thick, and the \cf\, determines the shape of the \mgi \ absorption trough. We report the results of this approach in Table~\ref{table3} (columns 2, 3, and 4). We refer to the first approach to the \mgi \ fit as ``independent'', and to the second as ``constrained''.

\begin{figure*}[htp!]
 \centering
 \includegraphics[width=0.8\textwidth]{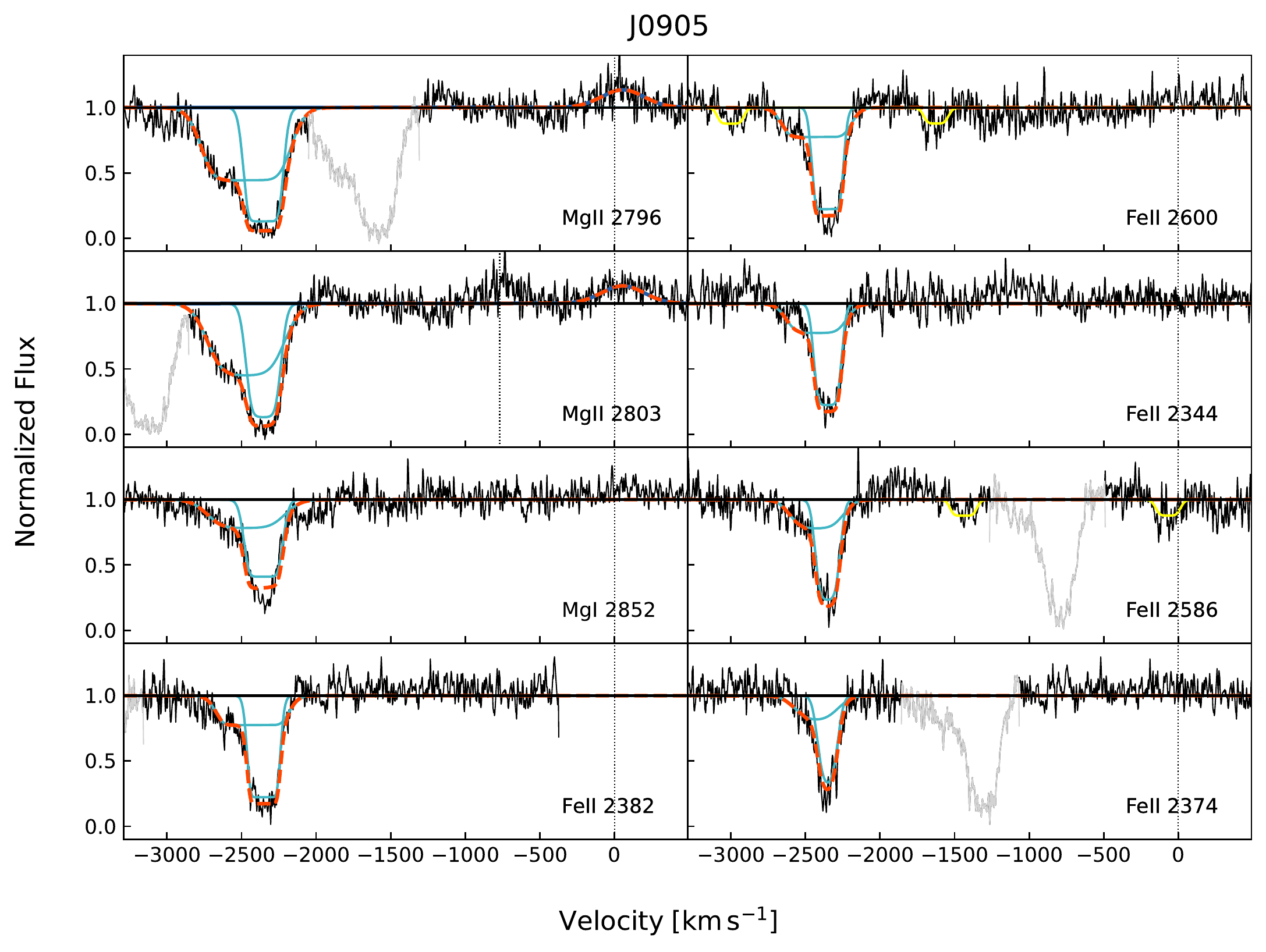}
 \caption{Normalized spectrum of the galaxy J0905 centered on the transitions relevant here. The J0905 spectrum exhibits \mgii \ emission within $-300$ and $+400$ \kmps\, of \zsys. In this galaxy, we see strong absorption from \mgii, \mgi, and \feii \ within $-2900$ and $-2000$ \kmps\, of the systemic redshift. All the detected transitions are well described by two velocity components and show similar kinematics. We also detect the \ion{Mn}{2} $\lambda\lambda\lambda$ 2576, 2594, and 2606 triplet (yellow line).}
 \label{fig:j0905}
\end{figure*}

Figure~\ref{fig:j1232_all} shows an example of our best fit to the \mgi \ lines in the galaxy J1232. The third from the top left panel shows the best fit to \mgi$\lambda$2853. Following the independent approach, we identify three components falling within $-380$ and $+220$ \kmps\, of \zsys. They have good kinematic correspondence to three \mgii \ components, but the \mgi \ profiles are shallower. 
In Figure~\ref{fig:j1232_all} (and for the rest of the sample Figure~\ref{fig:j0905} $-$ \ref{fig:j2140}) we show the \mgi$\lambda$2853 best fit as derived following the constrained approach. Even in this case, we find that three \mgi \ velocity components within $-380$ and $+220$ \kmps\, of \zsys\, provide a good fit to the data. The three \mgi \ components are described by \cf\, values in the range of $16-33\%$ of \cf(\mgii).

\subsubsection{\rm Fe II}

The spectral coverage of our data includes a series of strong \feii \ resonance lines which have some useful advantages over analysis of the \mgii \ doublet. First, the large number of \feii \ transitions (i.e. \feii$\lambda$2344\AA, \feii$\lambda$2374\AA, \feii$\lambda$2382\AA, \feii$\lambda$2586\AA, and \feii$\lambda$2600\AA) span a wide range of oscillator strengths, which makes it possible to place robust bounds on the column density of singly-ionized iron, thereby better constraining the total gas column density.
For example, the \feii$\lambda$2374 oscillator strength is a tenth that of the strongest transition, \feii$\lambda$2382\AA\, (which is nearly equal to that of \mgii$\lambda$2803\AA).
Furthermore, the absorption of \feii \ in several of these transitions is followed by fluorescence (rather than resonance) emission, providing a clearer view of the intrinsic absorption profile \citep{pro11, erb12, mar12}. The \feii$\lambda$2374\AA\, transition is particularly useful as the resonance absorption trough is not filled in by the emission of fluorescent \feii$^*\lambda$2396\AA\, photons. 

We quantify the \ion{Fe}{2} kinematics for each galaxy in our sample by fitting the absorption line profiles as described in Section~\ref{sec:fit}. We model the observed transitions simultaneously to constrain \cf\, and \ta. At a given velocity relative to \zsys, we consider the transitions to have (1) a relative $\tau$ defined by atomic physics (\citealp{mor03}), and (2) equal \bd\, and $C_f$ across the transitions. We do not detect any \feii \ emission. The \feii \ best fit parameters are reported in Table~\ref{table2}. In some spectra we additionally detect and model the \ion{Mn}{2} $\lambda\lambda\lambda$ 2576, 2594, and 2606 triplet which can blend with the \feii$\lambda\lambda$ 2586, 2600 transitions. The \mn \ best fit parameters are also reported in Table~\ref{table4}.

\input{Table4.tex}

As \feii \ provides a more reliable estimate of the central optical depth, for the velocity components that are detected in both \feii \ and \mgii \ we compare \ta(\mgii) derived from the \mgii \ fit (as described in Section~\ref{mgii}) with the optical depth inferred from the \feii \ fit. We assume a solar abundance ratio, which is supported by an ensemble of line ratio diagnostic diagrams used to estimate the metallicities in our sample \citep{per21}. For a solar abundance ratio (log Mg/H = $-4.45$; \citealp{asp21}), and conservatively adopting a comparable ionization correction (e.g. $\chi$(\mgii) $\approx$ $\chi$(\feii)), the optical depth in the weaker magnesium doublet line is 

\begin{multline}\label{tau_mgii}
\tau_0({\rm MgII_{2803}}) = 4.8 \, \tau_0({\rm FeII_{2586}})\, \frac{10^{-4.45}}{10^{\rm log Mg/H}} \, \times\\ \frac{10^{\rm log Fe/H}}{10^{-4.54}} \frac{\rm 10^{-0.5}}{ 10^{d(\rm Mg)}} \frac{10^{\rm d(Fe)}}{\rm 10^{-1.0}}
\end{multline}

While the relative optical depth in \feii$\lambda$2382 and \mgd \ can be similar to each other, the \mgii \ optical depth can be much larger as Fe is more depleted onto dust grains than Mg \citep{sav96, jen09}. 
Since there are not many studies on dust depletion in galactic winds, we adopt dust depletion factors ($d(X)$) consistent with those measured in the Galactic ISM (0.5 dex for Mg and 1.0 dex for Fe; \citealp{jen09}). We note that \citet{tuc18} studied a sample of nine gravitationally lensed $z \approx 2-3$ star-forming galaxies, and inferred that the outflowing medium is characterized by moderate dust depletion $d$(Fe) = $-$0.9 dex, in line with our adopted $d$(Fe).
Assuming a similar dust depletion correction for Mg and Fe, we would infer a \ta(\mgii$_{2803}$) that is systematically lower by a factor of three; this would decrease the inferred \mgii \ column densities (presented below) by roughly a factor of three. Adopting the dust depletion typical of the Galactic halo (0.59 dex for Mg and 0.69 dex for Fe; \citealp{sav96}) would lead to a similar result, with inferred \mgii \ column densities lower by a factor of 2.3.
Hereafter we refer to the \mgd \ central optical depth derived directly from the \mgii \ fit as ``measured'' (\tmeas), and the optical depth deduced using Equation~\ref{tau_mgii} as ``inferred'' (\tinf).

Figure~\ref{fig:j1232_all} shows an example of our best fit to the \feii \ lines in the galaxy J1232. The bottom left and the right column panels show the best fit to the \feii \ transitions observed in this study. The shape of the \feii \ absorption trough is well described by three kinematic components within $-380$ and $+220$ \kmps\, of \zsys. Although we do not force agreement in the kinematics of \feii \ and \mgii, the different velocity components agree remarkably well (with comparable $v$ and \bd\, to within the errors) over the velocity range where we detect \feii. We can therefore directly compare their \cf. The three \feii \ components all have lower \cf\, with values in the range of $46-71\%$ of \cf(\mgii). The \tmeas\, is not consistent with \tinf\, for all the velocity components. From this comparison, we infer that the actual \mgii \ optical depth is $\sim$4 times larger than the minimum required to fit the doublet. The \mgii \ velocity components detected at $-67$ and $-247$ \kmps are saturated and the best fit \ta\, values represent lower limits. For the component detected at 128 \kmps, we ascribe the difference as resulting from uncertainties in the emission filling affecting this component.

\subsection{Individual Systems}\label{sec:systems}

In the sections below we present fits to the \mgii, \mgi, and \feii \ absorption lines detected in other three galaxies in our sample, to illustrate the range of line properties observed in our data. Fits for the remaining galaxies are presented in Appendix~\ref{app:A}. We generally find that the kinematics of \feii \ match those of \mgii \ well, for the components with high EW and SNR.

\subsubsection{J0905+5759}

Figure~\ref{fig:j0905} shows the best fit to the \mgii, \mgi, and \feii \ absorption lines in the galaxy J0905+5759 (hereafter J0905). We present the absorption line fit for this source as an example that exhibits relatively simple kinematics, with few components, and with \mgb \ lines that do not blend together.
There is strong agreement in the component structure of all the absorption lines. The shape of the \mgii, \mgi, and \feii \ troughs are well described by two velocity components spanning $-2900$ to $-2000$ \kmps\, of \zsys. The spectrum also displays slightly redshifted ($+56$ \kmps) \mgb emission spanning $-300$ to $+400$ \kmps. However, the resonance emission does not fill in the high-velocity \mgii \ absorption lines. The comparable intensity at all velocities of the \mgo \ and \mgd \ troughs indicates the \mgii \ is optically thick. As the lines are not black (though the lower velocity component is close to black) the covering fraction (\cf $ < 1$) determines the shape of the absorption troughs. 

The independent \mgi \ fit identifies two components with similar kinematics to the constrained fit within the errors. As seen in J1232 above, here the \mgii \ troughs are deeper than the \mgi \ troughs, which indicates that only a fraction of the cloud volume contains much neutral Mg. It seems more likely that the \mgi \ absorption profile shape is determined by \cf\, rather than optical depth, since the latter (i.e. \cf(\mgi) = 1) would require the Mg neutral fraction to be less than a few percent (see Formula~\ref{tau_mgi}). The covering fraction for \mgi \ is 67 and 40\% that of the corresponding \mgii \ velocity components. Before performing the \feii \ fit, we identify and model the \ion{Mn}{2} $\lambda\lambda\lambda$ 2576, 2594, and 2606 triplet (yellow solid lines). In this case, they do not blend with the \feii$\lambda\lambda$ 2586, 2600 transitions. The \feii \ transitions show roughly equal intensities, suggesting they are also tracing optically thick gas. The two \feii \ velocity components are not black, and we find they are well described by \cf(\feii)$<1$ (90\% and 40\% of \cf(\mgii), respectively). Based on a comparison of \tmeas\, and \tinf, we conclude that the \mgii \ fit provides only a lower limit on N(\mgii) and that the actual central optical depth for the two \mgd \ components is closer to 79 and 67 (rather than 6 and 4 as measured from the \mgii \ fit).

\begin{figure*}[htp!]
 \centering
 \includegraphics[width=0.8\textwidth]{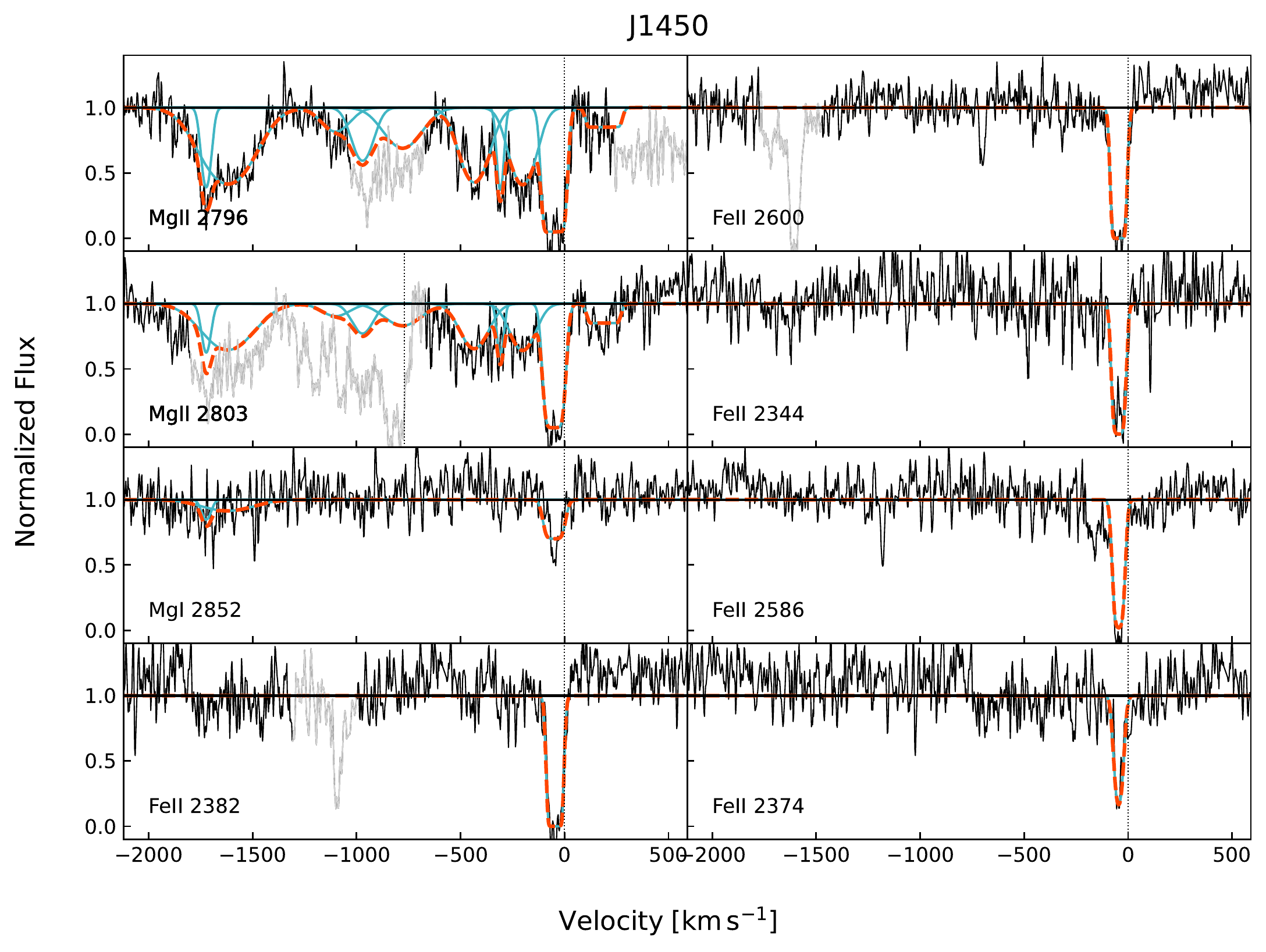}
 \caption{Normalized spectrum of the galaxy J1450 centered on the transitions relevant here. In this galaxy, we see strong absorption from \mgii falling within $-2000$ and $+350$ \kmps\, of the galaxy systemic redshift. The \mgii \ trough is well described by ten velocity components that show extremely complex kinematics. We detect \feii \ and \mgi \ absorption in the deepest velocity component, which is closest to $v=0$ \kmps. In addition \mgi \ shows weak absorption at $\sim$-1700 \kmps.}
 \label{fig:j1450}
\end{figure*}

\subsubsection{J1450+4621}

Figure~\ref{fig:j1450} shows our best fit to the \mgii, \mgi, and \feii \ absorption lines in the galaxy J1450+4621 (hereafter J1450). This spectrum exhibits a strong and complex \mgii \ absorption trough within $-2000$ and $+350$ \kmps\, of \zsys. The \mgo \ and \mgd \ line profiles are well described by ten velocity components. Many components (seven in \mgo, and five in \mgd) blend together. We do not detect \mgii \ emission in this galaxy. We find that two \mgii \ components (at $v=+198$ and $v=-50$ \kmps) are optically thick, and their shape is therefore determined by the covering fraction (\cf = 0.15 and \cf = 0.95). The remaining eight blueshifted components trace optically thin gas with \cf = 1. We detect a strong \feii \ absorption line at $v=-45$ \kmps\, of \zsys, that has good kinematic correspondence to the deepest \mgii \ component. We model this using only one component and find \bd(\feii) to be 26\% narrower than \bd(\mgii). The absorption trough is black for all \feii \ transitions other than \feii$\lambda$2374. This provides a robust constraint on N(\feii). Comparing \tmeas\, and \tinf\, for the only component detected in both \mgii \ and \feii, we infer that the \mgii \ fit provides a lower limit on N(\mgii), and the real \ta(\mgd) is closer to 74 (than 8, as measured from the Mg II fit). We identify one \mgi \ absorption trough close to \zsys (at $v=-38$ \kmps) that is visibly less deep than the corresponding \mgii \ and \feii \ components. The independent and constrained fits agree well in terms of the absorption line kinematics. However, as \ta(\mgii) is at least $\sim$8 (and more likely 19), we conclude that the constrained fit provides only an upper limit to N(\mgi) and that the \mgi \ absorption is tracing optically thick gas (with \cf(\mgi) = 32\% \cf(\mgii)). \mgi \ shows a second weak absorption trough at $v \sim$-1700 \kmps that has a good kinematic correspondence to the two most blueshifted \mgii \ components. We model this using two velocity components and find that the independent fit results in a broader profile than that of the constrained fit. However, the kinematics agree within the errors given the large uncertainties in the independent fit, due to the poor SNR. As the \mgii \ in these two components traces optically thin gas, we conservatively favor the \mgi \ independent fit over the constrained fit.

\begin{figure*}[htp!]
 \centering
 \includegraphics[width=0.8\textwidth]{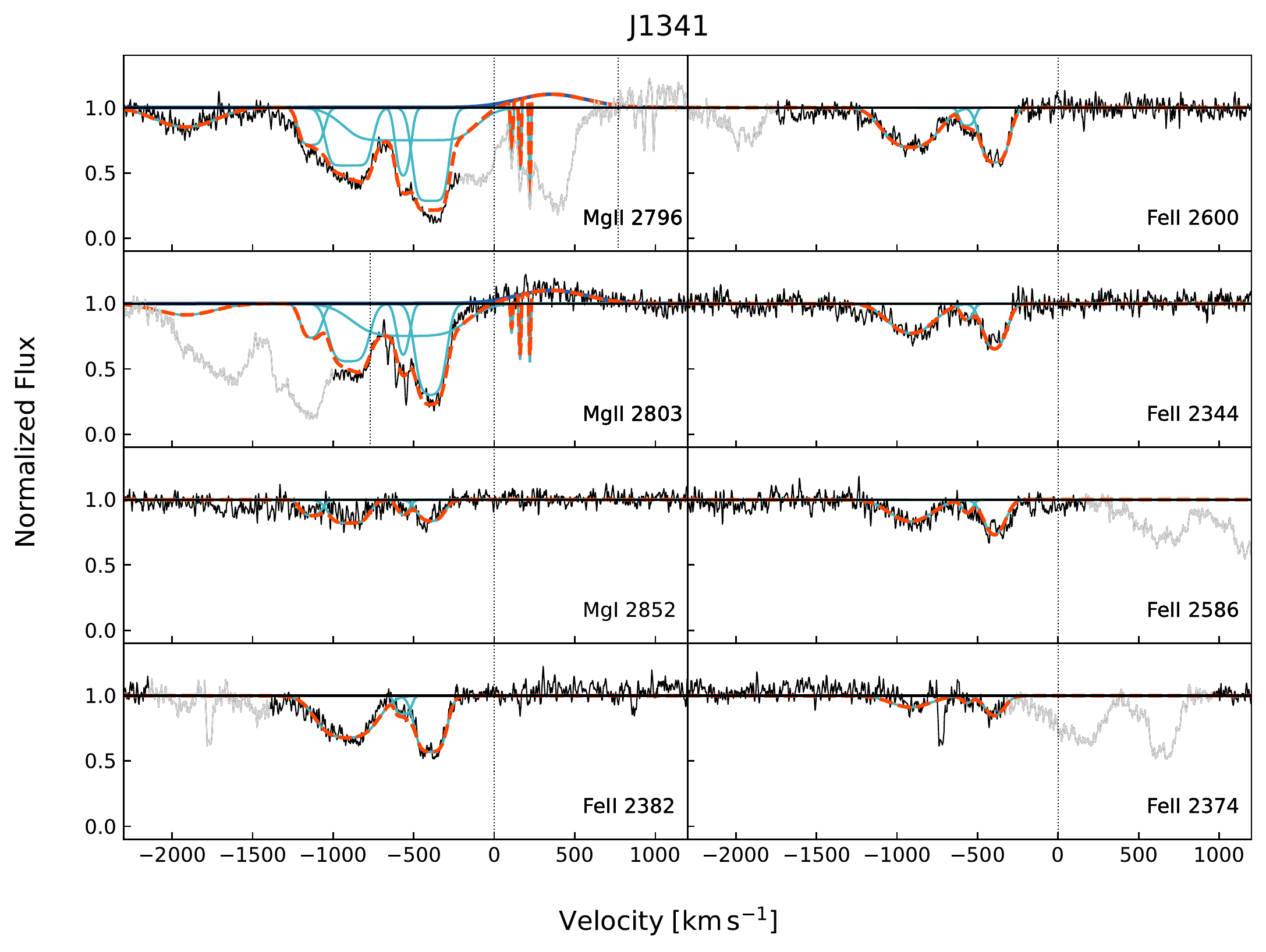}
 \caption{Normalized spectrum of the galaxy J1341 centered on the transitions relevant here. The J1341 spectrum exhibits \mgii \ emission within $-200$ and $+1000$ \kmps\, of \zsys. In this galaxy, we see strong absorption from \mgii falling within $-2300$ and $+250$ \kmps\, of \zsys. The \mgii \ trough is well described by nine velocity components that show complex kinematics, including three extremely narrow redshifted \mgii \ components. These lie on top of the \mgii \ emission and are not spectrally resolved; the best fit \bd\, is only one resolution element ($\sim$8 \kmps). We detect \feii \ and \mgi \ absorption within $-1500$ and $-200$ \kmps\, of \zsys, which we model using four velocity components.}
 \label{fig:j1341}
\end{figure*}

\subsubsection{J1341-0321}
Figure~\ref{fig:j1341} shows the best fit to the \mgii, \mgi, and \feii \ absorption lines in the galaxy J1341-0321 (hereafter J1341).
This spectrum includes \mgii \ redshifted emission ($+355$ \kmps) within $-200$ and $+1000$ \kmps\, of \zsys. We detect strong absorption from \mgii \ falling within $-2300$ and $+250$ \kmps\, of \zsys. We model the \mgo \ and \mgd \ troughs using nine velocity components that are characterized by complex kinematics. Most of the components (eight \mgo, and six \mgd) blend together. The \mgii \ best fit includes a combination of optically thin (in three components) and optically thick (in five components) gas, and all components have \cf $<$ 1. Of note, we identify three extremely narrow redshifted \mgii \ components. These lines lie on top of the \mgii \ emission and are not resolved; their best fit \bd\, parameter is similar to one resolution element ($\sim$8 \kmps).
We detect \mgi \ and \feii \ absorption within $-1500$ and $-200$ \kmps\, of \zsys, which we model using four and three velocity components, respectively. There is strong agreement in the \mgi \ kinematics of the independent and constrained fit solutions (with comparable $v$ and $b$ within the errors). The independent fit results in \cf(\mgi) of $25 - 45$\% of \cf(\mgii). 
There is a remarkable correspondence between \mgii \ and \feii. However, the most blueshifted \feii \ component is $\sim$57\% broader than the corresponding \mgii. 
We ascribe the difference to the lower SNR in the \feii \ spectral region compared to \mgii, which results in fitting \feii \ using one component instead of two as for \mgii. We find \feii \ to have lower \cf\, than \mgii \ ($24-77$ \% of \cf(\mgii)). 
Based on a comparison of \tmeas\, and \tinf, we conclude that the \mgii \ fit provides only a lower limit on N(\mgii) and that the actual central optical depth for the three \mgd \ components is closer to 17, 21, and 12 (rather than 4, 1, and 5 as measured from the \mgii \ fit).

\section{Discussion}\label{section:discussion}

We detect outflows in 14 compact, massive starburst galaxies in absorption from \mgi, \mgii, and \feii \ and show remarkably similar profiles of the absorption troughs in all these transitions. In most of our sample, the velocity dependence of the gas covering fraction (\cf) across components defines the absorption trough profile. Similarities in the profiles suggest these species reside in the same low-ionization gas structures. However, \mgii \ has on average a higher \cf\, than \feii\ at a given velocity, and a higher \cf\, than neutral Mg, implying that the absorbing clouds or filaments are not homogeneous.

We now discuss our results, including the variation of the absorption line profiles and the possible connection with the star formation history of each galaxy (Section~\ref{subsec:variation}). In Section~\ref{subsection:lack} we examine the lack of substantial absorption at the systemic redshift and the potential effect of emission line filling. We then use our results to gain insights into the role of different physical mechanisms in the extreme galactic outflows observed in our sample (Section~\ref{sec:models}). In Section~\ref{sec:trends} we investigate trends (or lack thereof) between the outflow absorption strength and galaxy properties. We conclude by presenting estimates of the mass outflow rates in our sample and discussing the associated uncertainties 
as well as the potential for these extreme outflows to affect the evolution of their host galaxies (Section~\ref{sec:Mdot}).

\begin{figure*}[htp!]
 \centering
 \includegraphics[width=0.9\textwidth]{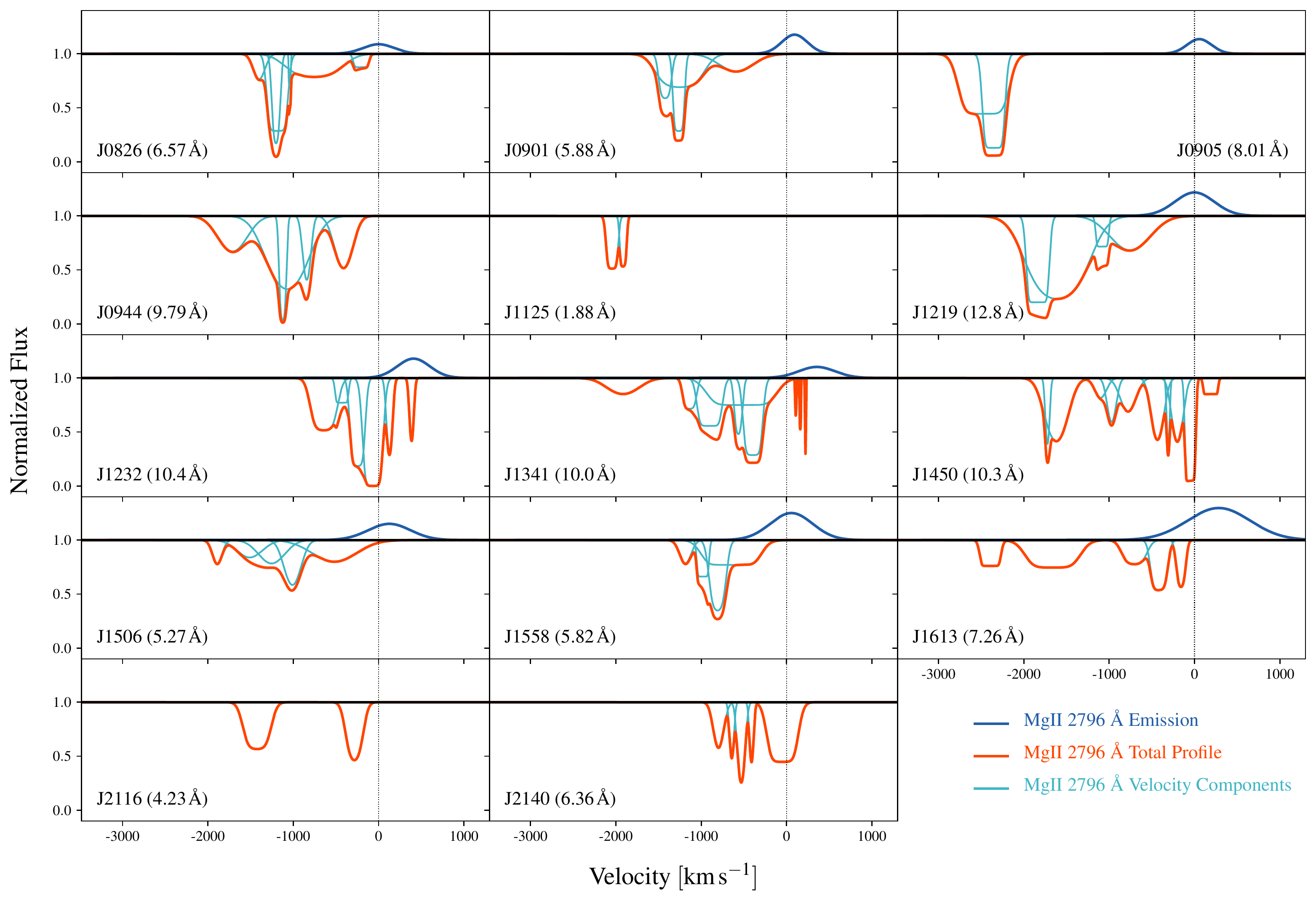}
 \caption{Normalized best fit \mgo \ line profiles for the galaxies in our sample. The x-axis shows the velocity of the \mgo \ doublet component relative to \zsys. Vertical black dotted lines mark \zsys. In each panel, the light blue solid lines show the fitted individual absorption velocity components, while the solid orange line shows the total fitted absorption line profile. When required, the fit includes \mgo \ emission (dark blue solid line). The total rest frame \mgii \ equivalent width is listed at the bottom of each panel. The order of the galaxies in this figure is different from the other figures in the paper and here follows the stellar light weighted ages.}
 \label{fig:Fits}
\end{figure*}

\subsection{Variation of Absorption Line Profiles}\label{subsec:variation}

As shown in Section~\ref{sec:fit}, each of the galaxies in our sample requires multiple components to fit the absorption troughs for the transitions studied here. The \mgii \ absorption troughs delineate the outflow kinematics most cleanly. Fig.~\ref{fig:Fits} shows our best fit to the \mgo \ absorption trough for each galaxy and illustrates the variety of absorption profiles observed in our data. 

First, there is a substantial variation in the
distribution of the \mgii \ absorption troughs in velocity space. Two galaxies (J0905 and J1125) lack any absorption from 0 to $\sim$2000 \kmps. One galaxy (J2116) shows two distinct troughs separated by $\sim$600 \kmps. Five galaxies (J0826, J0901, J1232, J1558, and J2140) exhibit complex kinematics with contiguous blueshifted absorption from $\sim$900$-$1500 \kmps. The remaining six galaxies (J0944, J1219, J1341, J1450, J1506, and J1613) have even more complex absorption profiles with nearly continuous blueshifted absorption out to $\sim$2000 \kmps. 

Second, there is a large variation in the number of velocity components required to describe each profile, as well as their widths (or Doppler parameter, \bd). As discussed in Section~\ref{sec:fit} we fit the blended absorption lines with the minimum number of components needed to characterize the velocity asymmetry. This number varies from two to ten velocity components. Their widths vary from 8 to 344 \kmps, with a mean value of 106 \kmps. Third, there is a large variation of the \mgii\ covering fraction (\cf) in different galaxies. For four galaxies (J0944, J1450, J1506, and J2140) most of their velocity components have \cf\, = 1, such that the shape of their absorption profiles is determined by the optical depth rather than \cf. For the rest of the sample, we find \cf\, as low as 0.12, with a mean value of 0.57. Fourth, three galaxies (J1232, J1341, and J1450) additionally have redshifted velocity components, indicating infalling gas. Lastly, 9 of the 14 galaxies in our sample exhibit clear signs of \mgii\ emission 
with velocity shifts from $z_{sys}$ that vary from 0 \kmps\ to $+450$ \kmps. Overall, there is a remarkable variation in the profiles and kinematics of these galaxies.

\begin{figure*}[htp!]
 \centering
 \includegraphics[width=0.9\textwidth]{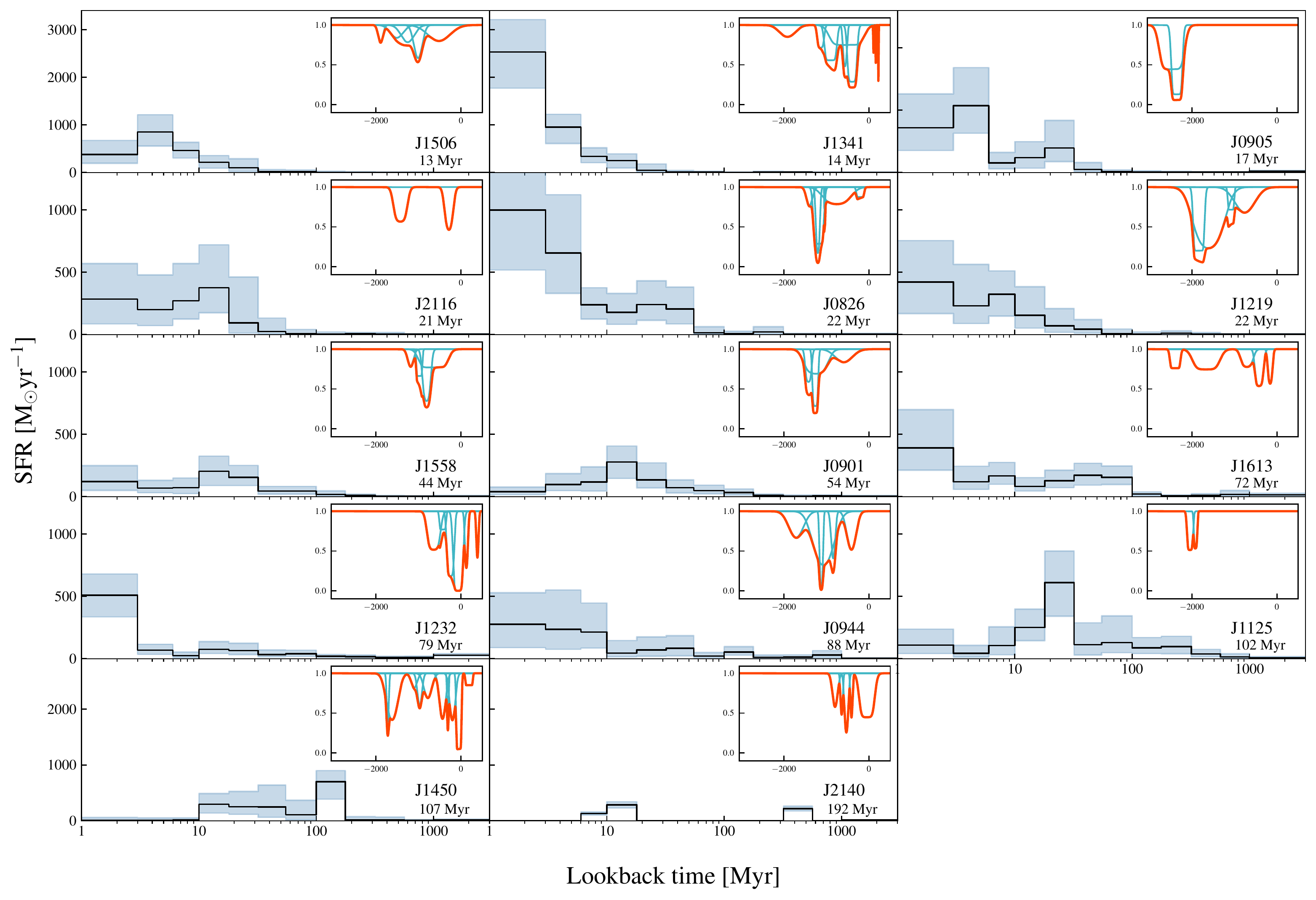}
 \caption{Star formation histories (SFHs) for the galaxies in our sample, derived with Prospector \citep{johnson21}. We note that the SFH for J2140 derives from the pPXF analysis (Davis et al., submitted). A different scale is used for the y-axis of the top and bottom rows to better display the data. Light blue shaded regions show the errors on the SFH. Time is displayed on a logarithmic scale to emphasize the recent SFH. Indeed, on a linear scale, the young bursts are so impulsive they are nearly invisible against the y-axis. The mean light-weighted age of the stellar populations younger than $\sim$1 Gyr is reported at the bottom right of each panel. The \mgo \ absorption profile model is shown for each galaxy in the upper right inset as presented in Fig.~\ref{fig:Fits}. The inset x and y axes are the normalized flux model and velocity of the \mgo \ doublet component relative to the galaxy systemic redshift, respectively.}
 \label{fig:SFH}
\end{figure*}

The galaxies in our sample are characterized by extreme and ``bursty'' star formation episodes that likely drive the powerful outflows observed, which can reach far into the circumgalactic medium (CGM) of the galaxy \citep{rup19}. As mentioned in Section~\ref{sec:intro} our team observed the galaxy Makani with the Keck CosmicWeb Imager (KCWI; Morrissey et al. 2018) and uncovered two distinct outflows traced by [OII] emission: a larger-scale, slower outflow ($\sim$300 \kmps) and a smaller-scale, faster outflow ($\sim$1500 \kmps). The velocities and sizes of the two wind components map exactly to two recent starburst episodes that this galaxy experienced 0.4 Gyr and 7 Myr ago, determined by its star formation history (SFH). To understand if the \mgii \ multiple velocity components seen here in this larger sample are also connected to the highly impulsive ``burstiness'' of the star formation in these galaxies we next investigate potential correlations between the \mgii \ absorption profiles and the SFHs of each source.

Figure~\ref{fig:SFH} shows the SFHs for the galaxies in our sample derived using Prospector \citep{johnson21} as described in Section~\ref{subsection:properties}. The mean light-weighted age of the stellar population younger than $\sim$1 Gyr is reported in the bottom right of each panel. The \mgo \ absorption profile model is shown for each target in the upper right inset as presented in Fig.~\ref{fig:Fits}. We note that the order of the galaxies in this figure is different from the others in the paper and follows their light weighted age.

Unlike Makani which had two clear bursts and two clear outflows at different velocities, we do not find a simple correspondence between the number of bursts in the SFH and the number of \mgii \ absorption features in this study. Galaxies exhibiting similar \mgii \ absorption troughs can have a variety of SFHs and vice-versa. For example, the \mgii \ EWs of J0905 and J1125 are substantially different, 8.01 and 1.88 \AA, respectively. However, their absorption troughs are similar in several regards as they both show no absorption from 0 to $\sim$2000 \kmps and are fit using two velocity components. Their SFHs both display a burst of star formation around 30 Myr ago, but their light-weighted ages are substantially different (17 and 102 Myr, respectively) as J0905's SFH is characterized by a younger burst $\sim$6 Myr ago. As another example, J1341 and J1450 have comparable \mgii \ EWs (10 and 10.3 \AA) and very broad blueshifted troughs ($\sim$2000 \kmps) with extremely complex kinematics fit with a close number of components (nine and ten). Nevertheless, they have SFHs that are substantially different, with inferred light-weighted ages of 14 and 107 Myr, respectively. In particular, J1341 shows a starburst in the last 10 Myr, as opposed to J1450 which has no significant bursts in the same time frame. On the other hand, J0826 has a SFH that is similar to J1341; however J0826 exhibits a remarkably different \mgii \ absorption trough, with a narrower profile ($\sim$1500 \kmps), no absorption at the systemic redshift (\zsys), no redshifted components, and a minimum absorption at a much higher velocity of $\sim$$-1200$ \kmps. Additionally, J0901 has a \mgii \ absorption profile that is comparable to J0826 in shape and EW but has a very different SFH, with older light-weighted age (54 Myr) and a decreasing trend of star formation in the last 10 Myr. 

The galaxies with the fastest velocity components are not necessarily the ones with the most recent bursts, and the galaxies with strong recent bursts do not all have fast outflow components, though some do (e.g. J0826, J0905, J0944, J1219, J1341, J1613).
Furthermore, we do find that eight galaxies (J0826, 
J0901, J0905, J1125, J1450, J1558, J1613, and J2116) have in their SFH an older burst ($>$10 Myr ago), regardless of the presence of a younger burst. Among these 8 galaxies, 6 (all except J0905 and J1125) have \mgii \ components at low velocities that may be tracing slower outflowing gas driven by the older burst of star formation, similarly to Makani.

\subsection{Absorption and Emission at the Systemic Redshift}\label{subsection:lack}

The \mgii \ doublet is the most sensitive transition among the absorption lines covered in this study, however, it is affected by resonantly-scattered wind emission which we expect to fill in the absorption profiles around \zsys\, \citep{mar12, pro11}. As \mgii \ is a resonant transition, the emitted photons are continually reabsorbed due to the absence of fine-structure splitting. This trapping process interferes with the escape of the photons, complicating the interpretation of the origin of \mgii \ emission. In the traditional model of a galactic-scale outflow expanding as a shell, this creates a P-Cygni-like profile for each \mgii \ doublet component, with blueshifted absorption and redshifted emission. The \mgii \ absorption arises from gas moving toward the observer as it absorbs photons in its rest frame. The emission arises from the receding component of the outflow, where photons emitted in the rest-frame, having scattered to escape towards the observer, are redshifted and go through any intervening gas. The emission line profile is centered near the systemic redshift of the galaxy. 
This emission can ``fill in'' the absorption within $\sim$200 \kmps of \zsys, decreasing the EW by up to 50\%, shifting the centroid of the absorption lines by tens of \kmps, and reducing the opacity near systemic. A study of cool gas outflows that ignores this line emission may underestimate the true optical depth and/or incorrectly infer that the wind partially covers the source \citep{pro11}.

In our sample, we detect \mgii\ emission in 9/14 of the galaxies, with velocity shifts of 0 \kmps \ to $+450$ \kmps\ from $z_{sys}$. In J0905 we can observe emission in both \mgii\ lines, with an observed ratio of 1. This ratio is in the optically thick regime \citep[e.g.][]{chi20}, and it agrees with the ratio observed in Makani \citep{rup19}. The \mgii\ trough in J0905 is so blueshifted ($\sim$$-2400$ \kmps) that the \mgii \ absorption profile is not affected by emission filling. For the remaining eight galaxies we adopt a ratio of 1 in our fits, as they clearly show \mgd\ emission but the corresponding \mgo\ line is suppressed by \mgd \ absorption at the same wavelengths. 
If the ratio of the two \mgii\ components is closer to 2 rather than 1 we may underestimate the effect of line-filling. However, the \mgii \ absorption profiles in these galaxies are not substantially affected by emission filling as the resonance emission is not expected to fill in the high velocity components of \mgii \ absorption. 

In our sample the majority of the \mgii \ absorption EW is blueshifted; the minimum intensity of the trough lies near the systemic velocity in only two objects (J1450 and J1232). The intensity minima are blueshifted by $\sim$300 \kmps \ to $\sim$2,000 \kmps \ in the rest of the sample, with little or no absorption at the systemic velocity. We use the \mgd \ absorption profile to quantify the EW at systemic, as it does not suffer from blending with \mgo. We find that 11/14 galaxies in our sample have less than four percent of the \mgii \ EW within 200 \kmps of \zsys. Among the objects that display \mgii \ emission, J1232 is the most potentially affected by emission filling as $\sim$26\% of its \mgii \ EW is within 200 \kmps of \zsys. 
However, emission filling is not a concern for the bulk of our sample of highly blueshifted \mgii \ absorption lines.

The lack of substantial \mgii \ absorption at the systemic velocity is an interesting feature in our sample, as first noted in \citealp{per21} and Davis et al., (submitted). 
One interpretation of the lack of absorption at systemic could be 
that the extreme outflows in these galaxies may have expelled the bulk of the interstellar medium (ISM) in these sources.
However, as many of the galaxies in our sample had a burst of star formation in the last 10 Myr, it is unlikely that these recent winds have entirely removed the ISM on such a short time scale. Additionally, some galaxies likely have ongoing star formation, such that the ISM can not be entirely absent. 

Another possibility is that the lack of absorption at systemic is due to an observational selection effect. 
Our parent sample is characterized by an extremely high outflow detection rate ($\sim$90\%; Davis et al., (submitted)). While it is possible that this high outflow incidence reflects a very wide opening angle of ubiquitous outflows in these galaxies, it could also be that the magnitude and color cuts used to select our sample may have identified galaxies where a powerful outflow has excavated a hole in the ISM, causing the galaxies to appear very bright and blue.
As a consequence, there may be little or no ISM left along the particular lines of sight studied here, while there is still remaining ISM in the galaxies along other sightlines. 

An example of this scenario is provided by the galaxy J0905. \citet{gea14} use the IRAM Plateau de Bure interferometer to study the CO(2-1) emission line in J0905, which is a tracer of the bulk of the cold molecular gas reservoir. They observe a CO emission line at the systemic velocity of the galaxy, which traces $\sim$65\% of the total cold molecular gas in the galaxy with an inferred mass of $M_{H2} = (3.1 \pm 0.6) \times 10^9 M_{\odot}$. This suggests that along some sightlines the amount of gas swept up by the outflow can be large, despite the galaxy retaining a substantial amount of its ISM. 

Another potential explanation for the exceptionally high velocity \mgii \ absorption components and the lack of absorption at \zsys\, is provided by strong radiative cooling \citep[e.g.,][]{bus16, tho16}. It is possible for a cold gas phase to form ``in-situ'' within a large-scale galactic wind via thermal instabilities and condensation of a fast-moving hot wind rather than being entrained and gradually accelerated. The by-product of this mechanism is cold gas at similar high velocities as the hot phase. In this model, the cold clouds accelerated by the hot wind are rapidly destroyed on small scales and at low velocities by hydrodynamical instabilities. As a result, the hot wind with enhanced mass loading and density perturbations can cool radiatively on larger scales forming an extended region of atomic and ionized gas moving at $\sim$10$^3$ \kmps, while the gas at low velocities is not observable. We come back to the origin and formation of cold gas in galactic outflows in the next Section~\ref{sec:models}.

\subsection{Comparison with Theoretical Models}\label{sec:models}
The existence of very fast, cool gas observed in outflowing winds from star-forming galaxies has been a persistent puzzle. In this Section, first, we briefly describe the processes most commonly invoked for cool gas acceleration in winds, then we use the results from this study and our parent sample to understand what insights can be gained into the role of these mechanisms in the extreme galactic outflows observed here.

\subsubsection{Mechanisms of Cool Gas Acceleration}
 The cool gas phase in winds is commonly explained as the acceleration of clouds from the host ISM via ram pressure from the hot phase \citep[e.g.,][]{vei05}. However, several 
simulations have challenged this explanation, demonstrating that ram pressure alone is not effective at accelerating cool gas clouds to the velocities and large scales observed without the clouds being shredded by hydrodynamical instabilities and becoming incorporated into the hot flow \citep{coo09, mcc15, sca15, sch15, sch17, zha17}. Recent work has shown that under the right background conditions and when sufficient large clouds are considered, cool gas {\it can} survive as a result of a mixing and cooling cycle. This may increase the cool gas flux as the hot gas condenses out, effectively growing the clouds rather than destroying them \citep{arm16, gri17, gro18,gro20, fie22}. 

In an alternative model \citep{efs00, sil03, tho16} the cold phase can form ``in-situ'' via thermal instabilities and condensation from the hot wind on large scales, provided it is sufficiently mass-loaded via the destruction of cool gas in the inner regions of the flow \citep[see also][]{loc18}. Additional models for cold cloud acceleration have also been proposed including momentum deposition by supernovae, the radiation pressure of starlight on dust grains \citep{mur05,mur10, mur11, hop12, zha12, kru13, dav14, tho16b}, and cosmic rays \citep{eve08, soc08, uhl12}. It has also been suggested that several of these mechanisms may be taking place simultaneously in order to drive cool outflows efficiently \citep{hop12, vei20}, making it difficult to isolate the different processes potentially at play.

 From an energetic point of view, starburst galaxies with powerful winds, like those in our sample, are ideal candidates for outflows driven by the radiation pressure from Eddington-limited star formation \citep{dia12, gea14, rup19, per21}. Recently, our team obtained results that confirm this hypothesis. Rupke et al., (submitted) present Keck/ESI \citep[Echellette Spectrograph and Imager;][]{she02} long-slit spectra of the two wind episodes observed in the galaxy Makani, drawn from the same parent sample as the galaxies in this study. They infer momentum and energy outflow rates in the inner ($R_\mathrm{II}=0-20$ kpc), recent (7~Myr ago), fast ($\sim$2,000 \kmps) outflow that implies a momentum-driven flow driven by the hot ejecta and radiation pressure from the extreme, possibly Eddington-limited, compact starburst. 

Here, in the Sections below, we focus on the kinematics of low-ionization absorbers tracing the cool, ionized phase of the extreme outflows in our sample to gain insights into the distribution of the outflowing gas and the physical mechanisms that may occur at the interface between the hot and cool wind phases. 

\begin{figure*}[htp!]
 \centering
 \includegraphics[width=0.9\textwidth]{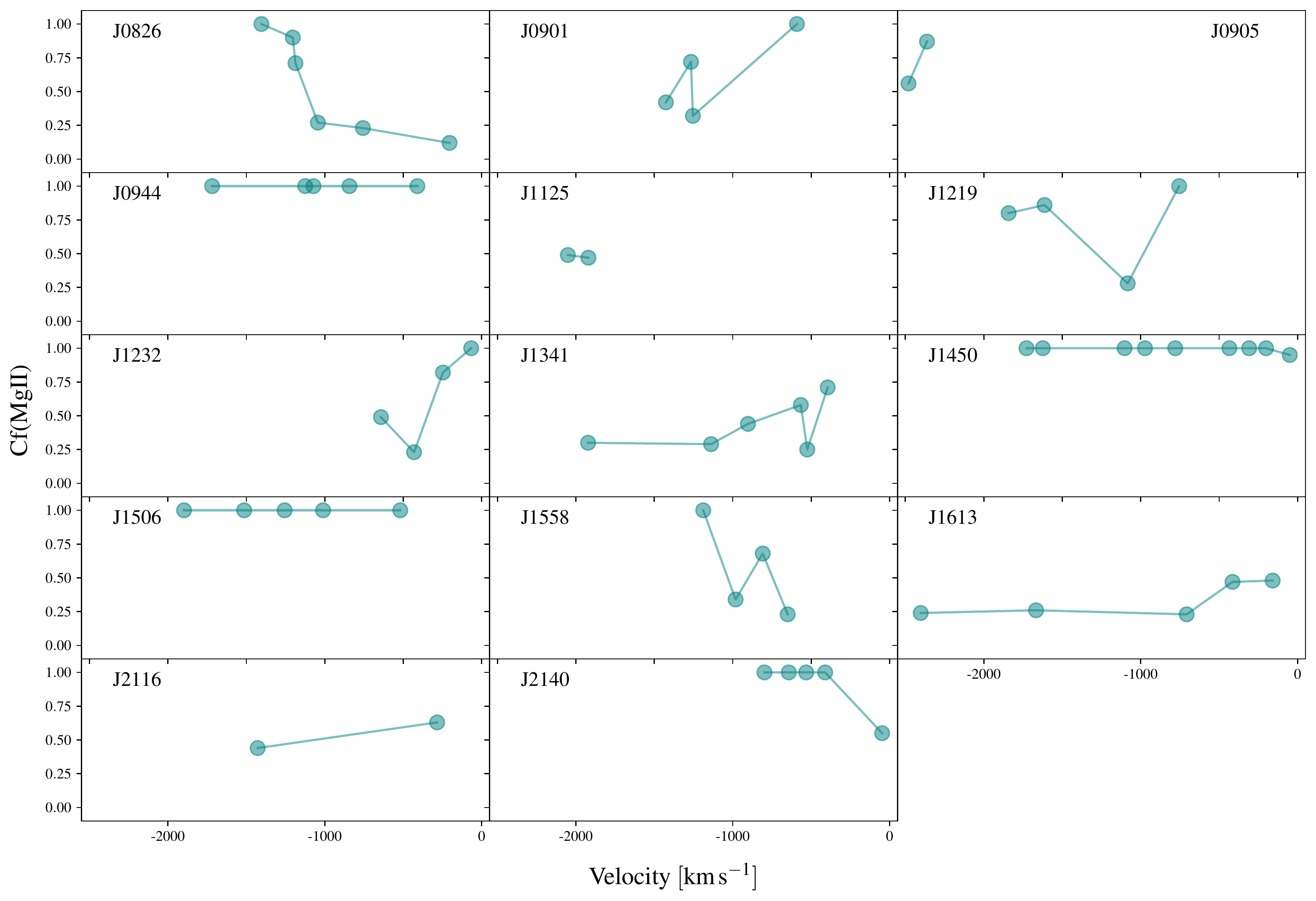}
 \caption{Fitted \mgii \ covering fraction (\cf) as a function of the \mgii \ velocity, for the \mgii \ blueshifted (i.e. outflowing) components. 
 We find that \cf\, determines the shape of most of the absorption troughs in 10/14 galaxies in our sample. In particular, six galaxies (J0905, J1613, J2116, J1341, J1232, and J1125) show deeper troughs at velocities where low-ionization gas covers a higher fraction of the continuum. In the remaining four galaxies (J1450, J0944, J1506, and J2140) the absorption profiles are shaped by the optical depth. Additionally, in our sample, the \cf\, does not show a unique trend with outflow velocity.}
 \label{fig:cf}
\end{figure*}

\subsubsection{Inferred Spatial Distribution of Ionized Gas}\label{subsec:dist}
Simulations indicate that halos of \mgii \ emission are a ubiquitous feature across the galaxy population to at least z = 2 \citep{nel21}.
\mgii \ halos extend from a few to tens of kpc at the highest masses, i.e. far beyond the stellar component of a galaxy. Moreover, they are highly structured, clumpy, and asymmetric. 
\citet{def21} generate mock \mgii \ observations from the TNG50 simulation \citep{nel21} and produce absorption spectra to compare with observed data. Although the mock sight lines are too small to be comparable with observations, they show the non-uniform distribution of the \mgii \ absorption, usually concentrated in discrete clumps. The mock velocity spectra, despite originating from a remarkably small fraction of the total \mgii \ gas in the halo, 
reflect the diversity of the \mgii \ gas distribution and kinematics in halos of similar galaxy mass. On the other hand, halos with similar morphology exhibit similar mock spectra. The sample of star-forming galaxies in \citet{def21} differs from ours as it is characterized by galactic disks, while our galaxies are late-stage mergers with spherical morphologies. However, we also find great diversity in the \mgii \ absorption profiles (see Section~\ref{subsec:variation}). This suggests that the \mgii \ halos of our galaxies, despite spanning only 0.76 dex in stellar mass, may have diverse morphologies.

\mgii \ absorption traces cool $\sim$10$^4$ K gas that is usually found in simulations in high-density clouds or filaments permeating the volume-filling, low-density, hot phase at large radii \citep{sch20, nel21}. The remarkably extended velocity distribution of the \mgii \ absorption profiles in our sample may reflect the filamentary structure of dense outflowing material. 
Additionally, some simulated halos exhibit fountain flows in \mgii \ emitting gas, with signatures of infalling gas clouds in addition to wind-driven outflows \citep{nel21}. We note that we find clear signatures of inflowing gas in the spectra of three galaxies in our sample (J1232, J1341, and J1450), where the components that trace infalling gas are redshifted $\sim$200$-$300 \kmps (from the systemic redshift) and are notably narrower that the blueshifted components that trace outflowing gas. Studies of infalling gas typically utilize surveys of a hundred or more galaxies, as the detection rate of redshifted absorption lines is around 3$-$6\% \citep[e.g.][]{sat09, mar12, rub12}. The fraction of galaxies with inflows in our sample is  21$\pm$11\%, higher than in other studies. However, this value is based only on three sources, so we can not conclusively state whether the accretion rate in our sample is significantly higher.

Simulations showing the survival of cool clouds traveling through a hot medium find that the clouds undergo hydrodynamical instabilities that create an elongated shape with a wake \citep[e.g.,][]{arm16, gro18}. The coolest and densest gas is typically located inside the head of the cloud, while the ionized gas is found more in the turbulent wake behind the cloud, produced by the mixing between the cool gas ablated from the cloud and the hot medium. One way to investigate this potential structure observationally is to compare the absorption troughs of two transitions of the same species that have different ionization, such as neutral and singly-ionized Mg. 
As mentioned in Section~\ref{fit:mg1}, one of our approaches to fit \mgi \ absorption profiles is to adopt the same kinematic components as \mgii \ and estimate the covering fraction (\cf) that best fits the \mgi \ spectral region. 
An advantage of constraining the kinematics in this way is that we can directly compare the \cf\, of the two different ions as a function of velocity. We find that the shallow troughs of \mgi \ relative to \mgii \ require a lower \cf\, for the former at every velocity, with \cf(\mgi) ranging from $0.16$ to $0.66$ \cf(\mgii), with a median value of \cf(\mgi) = 0.4 \cf(\mgii). The similar kinematics but systematic offset in \cf\, implies that the \mgi \ absorption arises from denser regions within more extended structures traced by \mgii. This finding is in agreement with theoretical models predictions that the coolest and densest gas is typically located in the more internal and self-shielded part of the clouds. 

We note that even if we do not adopt the same kinematics for \mgii \ and \feii, we find a remarkable agreement in most of the velocity components. For these components, we find that \cf(\feii) is systematically lower than \cf(\mgii). This is expected as \feii \ traces denser gas than \mgii, and it is in line with the previous result of \mgi \ being less spatially extended than \mgii.

\subsubsection{Covering Fraction Trends with Outflow Velocity}\label{subsec:cf}
Earlier studies have tried to interpret line profiles of low-ionization ions (e.g. \mgii, \mgi, \feii), and in particular their \cf\ distribution, in the context of driving mechanism models for galactic winds. \citet{mar09} study a sample of five starburst galaxies at z$\sim$0.2 and find that \cf\, decreases as the outflow velocity increases beyond the velocity of minimum intensity, which in their case corresponds to the velocity of maximum \cf. The authors interpret the velocity-dependent \cf\ in their sample as a result of geometric dilution associated with the spherical expansion of a population of absorbers. In the context of this simple physical scenario, their result implies that the high-velocity gas detected in absorption is at a larger radii than the lower velocity (and higher \cf) gas, implying an accelerating wind. This hypothesis is also suggested by \citet{chi16} in a study of the wind-driving mechanisms and distribution of \cf\ in a nearby starburst galaxy. 
This simple accelerating model does not take into account, however, the complexity of the interaction between cool clouds and the hot surrounding wind, such as radiative cooling, cloud compression due to shocks, and effects of shear flow interactions that produce hydrodynamic instabilities. More recent studies have shown that cold clouds are unlikely to survive that kind of acceleration over time \citep{sca15,bru16,sch17,zha17}. 

Figure~\ref{fig:cf} shows the fitted \mgii(\cf) as a function of velocity for the blueshifted (i.e. outflowing) components in our sample. 
\cf\, does not show a unique trend with velocity. Seven galaxies (J0901, J0905, J1219, J1232, J1341, J1613, and J2116) have decreasing \cf\, with increasing velocity, three galaxies (J0826, J1125, and J1558) have increasing \cf\, with increasing velocity, and four galaxies (J0944, J1450, J1506, and J2140) have a constant \cf\ value. This variation in the \cf\ with outflow velocity, along with the variation of the absorption profiles described in Section~\ref{subsec:variation}, may capture the complex morphology of $\sim$kpc scale, inhomogeneously distributed, clumpy gas and the intricacy of the material in the turbulent mixing layers between the cold and the hot phases \citep[e.g.,][]{fie20, nel21}.

Our starburst galaxy sample shows that a model of an outflow accelerating over time is unlikely to be valid. One piece of evidence is the observation of the large scale, multi-phase, and multi-burst wind in the galaxy Makani. \citet{rup19} find that the inner ($R_\mathrm{II}=0-20$~kpc) and younger ($\sim$7 \; Myr ago) wind is faster (with maximum speeds exceeding 2000 \kmps) than the more extended ($R_\mathrm{I} = 20-50$~kpc) and older wind ($\sim$400 \; Myr ago), which has speeds of $\sim$100 \kmps. Rupke et al. (submitted) 
find that the larger wind is consistent with having slowed down in the extended halo and CGM of the galaxy. 

A second piece of evidence against an accelerating model comes from the analysis of the \mgii \ EW and velocity as a function of the light-weighted age of the galaxies in our parent sample.
In this work, we do not find a correlation between the total \mgii \ EW and galaxy light-weighted age. However, the galaxies all have young light-weighted ages, between 13 and 192 Myr, and represent many of the youngest objects in the parent sample studied by our team (Davis et al., (submitted)). 
Using the parent HizEA sample, which has a larger dynamic range in light-weighted age, Davis et al. (in prep) find an inverse correlation between \mgii \ absorption EW and light-weighted age, as well as a complementary inverse correlation between outflow velocity and light-weighted age. These results are consistent with models in which the outflow decelerates with time, which support scenarios in which the cold gas condenses out of a hot flow. In particular, the observed trend of lower outflow velocity with larger light-weighted age matches the analytic models of \citet{loc18} well, where impulsive bursts of star formation driven winds slow down and cool as they expand into the CGM.

Regardless of their physical origin, the large velocity width of the absorption troughs observed here requires contributions from multiple structures along the sightline. It is likely that our sightlines intersect many mixing layers, making it difficult to interpret the \cf\, inferred from our fits in terms of the \cf\, of single clouds that grow over time entrained in the hot wind \citep[e.g.,][]{gro18, fie22}.
To complicate the interpretation of the absorption line profiles further there is the multi-burst nature of the outflows in our sample. It is possible that an initial burst of star formation drives an outflow that consumes most of its energy shock-heating the surrounding ISM and consequently slows down while excavating a hole through which we observe the host galaxy as blue and bright. An outflow driven by a second burst of star formation may have a very different evolution as it can follow the path of minimum density and inertia created by the first burst, retaining more energy and velocity. A single sightline does not provide a full picture of the entire galaxy, and it can intersect multiple outflow episodes, which can overlap in the projected observed velocity space, making it challenging to distinguish the two wind episodes through absorption studies.

\subsubsection{Comparison of MgII and FeII Absorption Troughs}\label{subsec:comparison}

It is currently challenging to use simulations to interpret observed absorption line profiles, as the models do not typically create spectra drawn from realistic sightlines through the simulated halos. They also typically identify the cool wind phase by temperature rather than making predictions for individual ions. However, we can consider their main results and model trends to make comparisons with observed spectra. 

\citet{sch20} present a new simulation part of the Cholla Galactic OutfLow Simulations (CGOLS) project, a series of high-resolution global disk simulations of galaxy outflows. The authors model an M82-like galaxy with a 5 pc resolution in a 10 kpc volume, which is ideally suited to capture both the launching of the wind and the detailed phase structure during the subsequent expansion. Their stellar feedback injection mechanism allows the driving of energetic hot winds from a high surface density galaxy disk that gives rise to a complex, multiphase outflow.
They find that the cold phase is primarily in embedded clouds that are gradually shredded and thereby enhance the mass loading of the hot phase. The authors demonstrate that the mixing between hot and cool gas in the wind is an effective way of transferring momentum from one phase to another, and this occurs at all radii. In cases where the mixed gas has a high enough density to cool, it does so with a higher velocity, leading to a linear relationship between mixed fraction and velocity. 

To test this prediction we can assume that the \feii \ in our spectra primarily traces the entrained cool gas phase component (i.e., gas rich in Fe results from SN Ia explosions), while the \mgii \ primarily traces the cooled hot wind material (i.e., gas ejected from SN II and rich in $\alpha$-elements). As we do not expect the hot wind to be composed entirely of SN II ejecta, it should be $\alpha$-element enriched but quite diluted compared to pure SNe ejecta. Therefore rather than searching for a complete absence of \feii, we can instead examine how the \mgii \ to \feii \ ratio varies as a function of outflow velocity. We expect the EW(\mgii) to EW(\feii) ratio to increase for gas with less mixing (i.e., with less cool gas entrained as the hot wind travels through the ISM); such gas  likely originates from direct cooling of the hot wind phase. 

\begin{figure*}[htp!]
 \centering
 \includegraphics[width=0.9\textwidth]{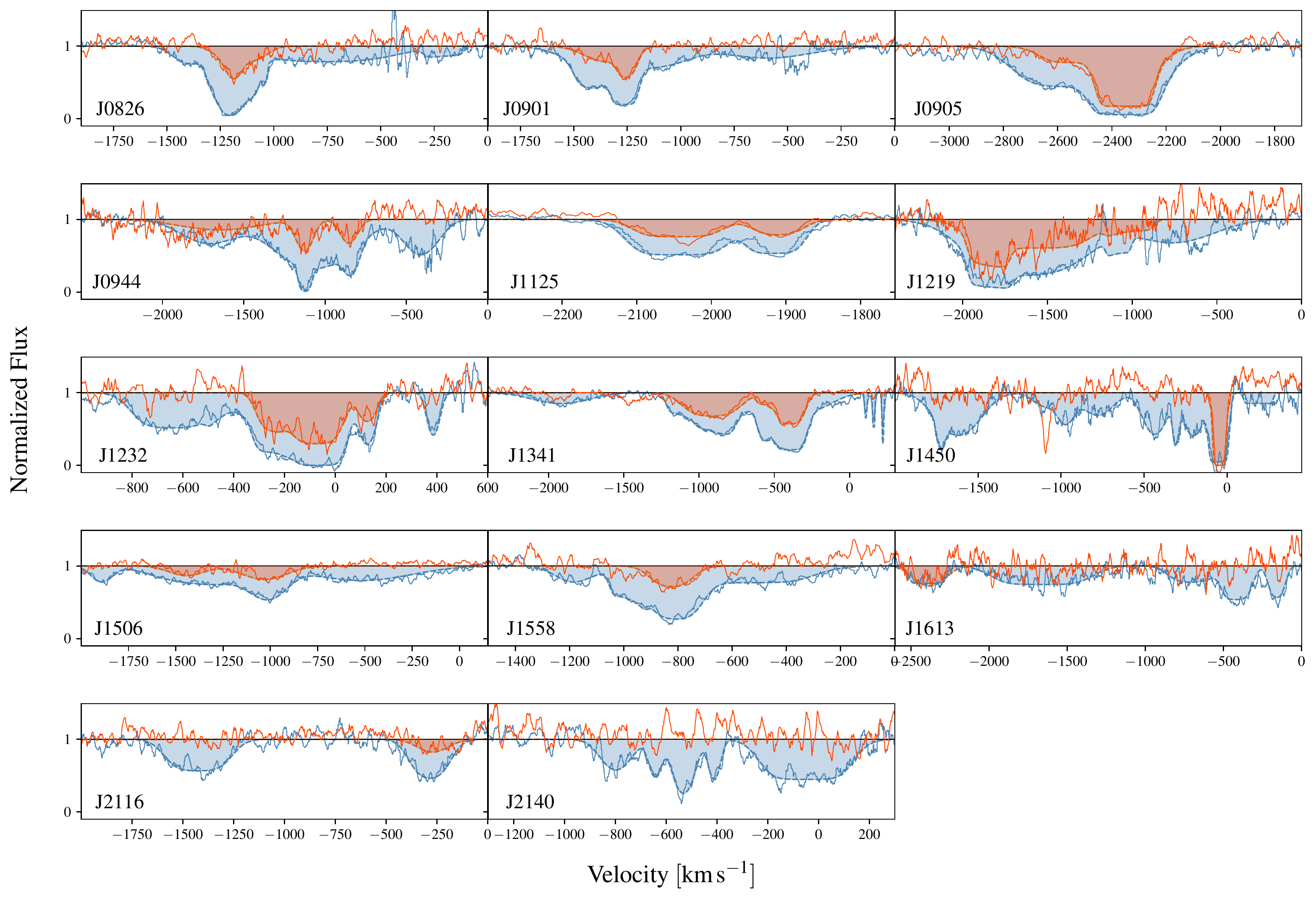}
 \caption{Comparison of the \mgo \ (blue spectrum) and \feii\,$\lambda$2382 (red spectrum) line profiles in the spectra of our galaxies. The shaded blue and red regions represent the \mgo \ and \feii\,$\lambda$2382 best fit profiles, respectively. The \mgo \ troughs generally have higher absorption equivalent width than the \feii\,$\lambda$2382 troughs.}
 \label{fig:overlay}
\end{figure*}

In Figure~\ref{fig:overlay} we overlay the \mgo \ (blue spectrum) and \feii\,$\lambda$2382 (red spectrum) absorption profiles for the galaxies in our sample.
We find that the \mgo\ troughs typically have higher absorption EW. Eight galaxies (J0826, J1232, J1341, J1450, J1506, J1558, J2116, and J2140) exhibit a clear lack of \feii \ at the highest velocities. To verify that this lack of \feii \ is not due to a sensitivity effect, we use Equation~\ref{tau_mgii} to infer the \feii \ optical depth for the velocity components where \feii \ is not detected. For each  component, we use the optical depth of the corresponding \mgii \ and assume \mgii \ and \feii \ to have the same kinematics. Equation~\ref{tau_mgii} takes into account the difference in oscillator strength and dust depletion between \feii \ and \mgii. As \feii \ typically has lower \cf\ compared to \mgii, we adopt a lower \cf(\feii), in line with the other \feii \ absorption lines in the same spectrum. We find that in four of these galaxies (J0826, J1341, J1506, and J1558) we can not exclude that \feii \ absorption may be present but too weak to detect. In the other four galaxies (J1232, J1450, J2116, and J2140) the lack of \feii \ at the highest velocities is real and can be explained by gas cooling directly from the hot wind at the highest velocities. 
We also find that in two galaxies (J0944, J1613) there is a lack of \feii \ at the lowest velocities that is not due to sensitivity effects. Finally, J2140 is the only galaxy in our sample where \feii \ is not detected at any velocity. This is also the only galaxy in our sample that hosts a faint AGN \citep{sel14}, which could result in different feedback mechanisms and spectral features.

\begin{figure*}[htp!]
 \centering
 \includegraphics[width=0.9\textwidth]{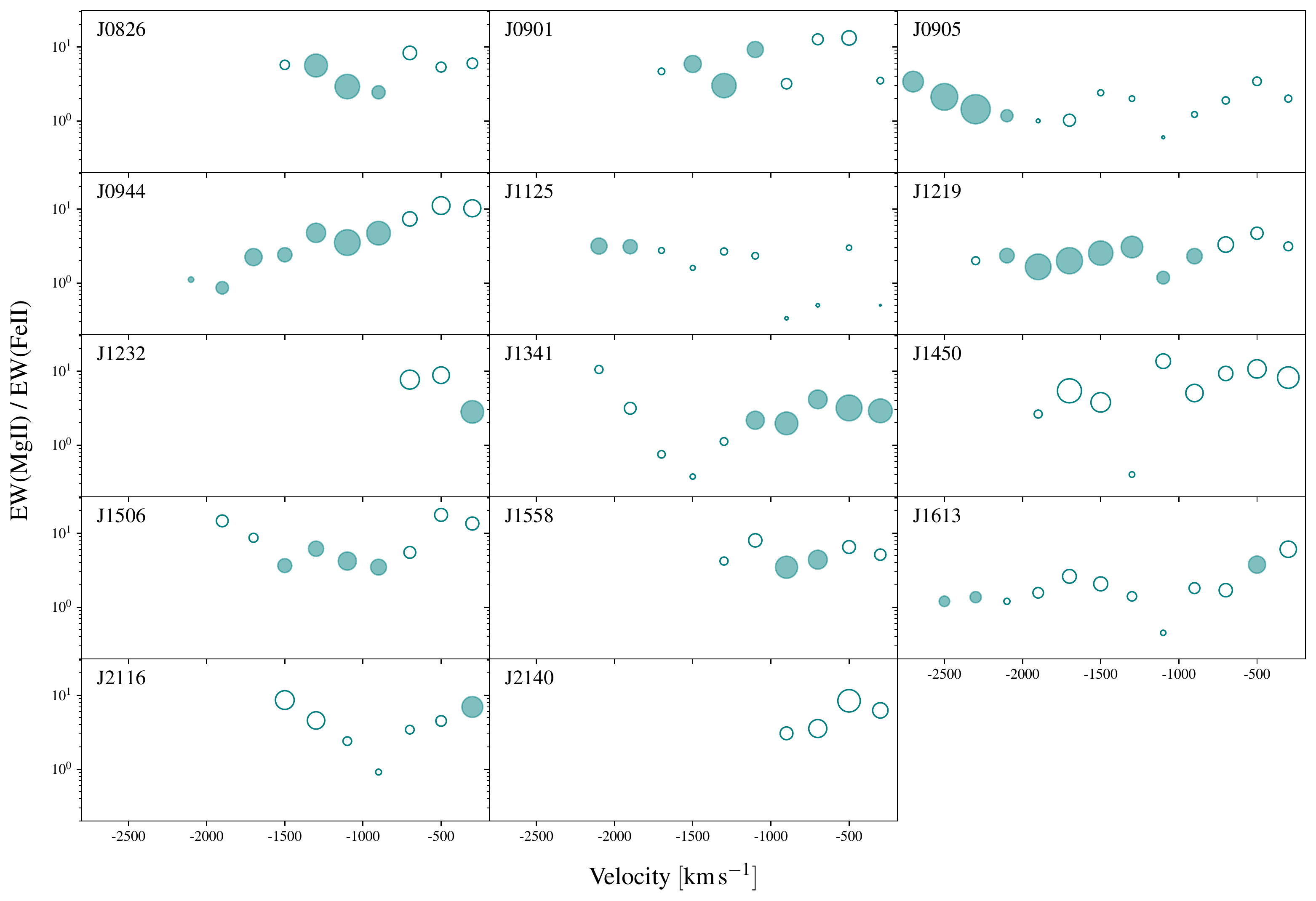}
 \caption{\mgo \ to \feii\,$\lambda$2382 equivalent width (EW) ratio as function of the outflow velocity for the galaxies in our sample. The size of the circles is proportional to the EW(\mgii) strength. Filled circles show velocity bins where we detect \feii, while empty circles show \feii \ non-detections. The EW ratio does not show a unique trend with the outflow velocity. However, for the eight galaxies (J0826, J1232, J1341, J1450, J1506, J1558, J2116, and J2140) that have a clear lack of \feii \ at the highest velocities, the EW ratio is larger than in the velocity bins where we detect \feii, suggesting that we are seeing gas directly condensed out from the hot wind.}
 \label{fig:ratio}
\end{figure*}

In Figure~\ref{fig:ratio} we show the \mgo \ to \feii\,$\lambda$2382 EW ratio as function of outflow velocity. To calculate this ratio we divide the \mgo \ (corrected for \mgd \ absorption) and \feii\,$\lambda$2382 spectral regions in bins of 200 \kmps and integrate the spectra, in order to estimate upper limits where the lines are not detected. The size of the circles in this figure is proportional to the strength of the EW(\mgii). Filled circles show the velocity bins where we detect \feii, while empty circles show \feii \ non-detections. This EW ratio does not have a consistent trend with the outflow velocity. However, for the eight galaxies discussed above that lack \feii \ at the highest velocities, the EW ratio is larger than in the velocity bins where we detect \feii.

In summary, several of the galaxies in our sample do not have \feii \ detected at the highest outflow velocities, implying a lower mixing fraction of entrained cool gas.
While the cold gas outflow phase is most likely produced by a combination of physical mechanisms, these results suggest that the cold gas at the highest velocities could directly condense out of the hot wind phase, as suggested by some theoretical models.

\subsection{Trends of Equivalent Width with Galaxy Properties}\label{sec:trends}

\begin{figure*}[htp!]
 \centering
 \includegraphics[width=0.9\textwidth]{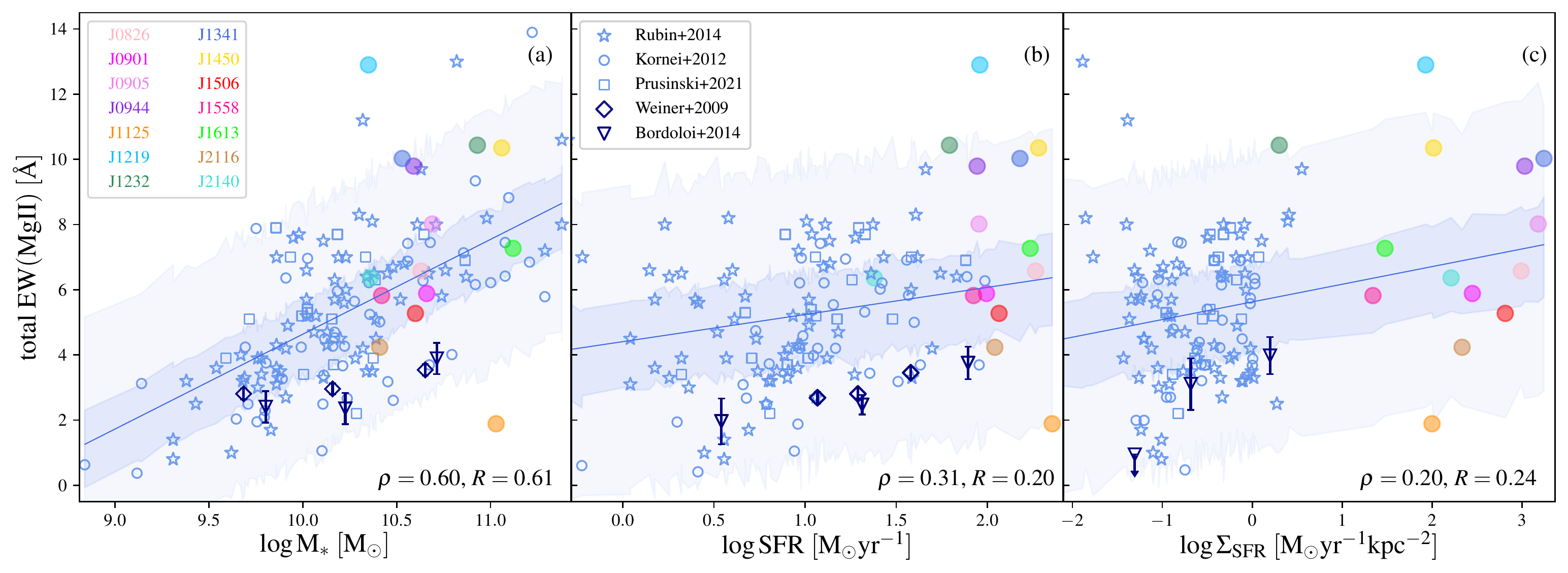}
 \caption{Equivalent width (EW) measured in the \mgb \ transitions versus M$_*$ (a), SFR (b), and $\Sigma_{SFR}$ (c) for the galaxies presented in this work (filled circles; see legend in the upper left corner in the panel (a)) compared with star-forming galaxies with clear wind signatures: 0.3 $< z <$ 1.4 galaxies from \citet{rub14} shown with blue stars; $z\sim1$ galaxies from \citet{kor12} shown with blue circles; $1\leq z\leq 1.5$ galaxies from \citet{pru21} shown with blue rectangles; stacked spectra of 1,406 galaxies at $z \sim1.3-1.5$ from \citet{wei09} shown with dark blue diamonds; coadded spectra of 486 galaxies at $1 < z < 1.5$ from \citet{bor14} shown with dark blue reversed triangles. Error bars are included only for results using stacked data, for clarity. Fit lines are an MCMC-generated linear fit to the data, with 1 and 3$\sigma$ error regions shaded in blue. The Spearman rank order correlation coefficient ($\rho$) and Pearson linear correlation coefficient ($R$) for the relations shown are: for EW vs M$_*$, $\rho$ = 0.60, $R$ = 0.61 (panel (a)); for EW vs SFR, $\rho$ = 0.31, $R$ = 0.20 (panel (b)); for EW vs $\Sigma_{SFR}$, $\rho$ = 0.20, $R$ = 0.24 (panel (c)). The correlation coefficients for EW vs M$_*$ have p-values $\sim$10$^{-16}$, while for EW vs SFR and EW vs $\Sigma_{SFR}$ the correlation coefficients have p-values $\sim$0.02$-$0.0002.}
 \label{fig:ew_gal}
\end{figure*}

We now investigate the relationship between the outflow absorption strength and galaxy properties in our sample. Figure~\ref{fig:ew_gal} compares the total \mgii \ equivalent width (EW) with host galaxy stellar mass (M$_*$), star formation rate (SFR), and star formation rate surface density ($\Sigma_{SFR}$). Since the \mgii \ doublet components are blended together in most of our galaxy spectra, we consider the total EW measured from the \mgb \ transitions.
We compare our sources with data from five representative star-forming galaxy samples in the literature that provide information about both \mgii \ doublet component EWs \citep{wei09, kor12, rub14, bor14, pru21}. 

To quantify the relations presented in Figure~\ref{fig:ew_gal}, we use the python Markov Chain Monte Carlo package PyMC3 \citep{pymc3} to compute a linear fit and the associated uncertainties. In each panel, we show the linear fits with 1 and 3$\sigma$ error regions shaded in blue. We also characterize the correlations using the Spearman rank correlation coefficient ($\rho$) and the Pearson linear correlation coefficient ($R$). These values and their corresponding statistical significance are listed in the figure caption.

 A strong positive correlation between EW(\mgii) and M$_*$ is evident in panel (a) of Figure~\ref{fig:ew_gal}, while EW(\mgii) exhibits a weaker correlation with SFR in panel (b) and $\Sigma_{SFR}$ in panel (c), as has been found in previous work \citep[e.g.,][]{mar12, rub14}. Our galaxies follow the general trends seen in other samples, though with substantial scatter. The EW(\mgii) versus M$_*$ correlation has been explained in these previous works as being due to two factors: 1) the EW of ISM absorption (at the systemic redshift) increases with increasing M$_*$, and 2) emission filling is stronger in lower mass galaxies, preferentially suppressing absorption around the systemic redshift. As a result, the lower emission filling in the spectra of massive galaxies (relative to lower mass galaxies) amplifies the positive correlation between EW(\mgii) and M$_*$. 
 
 As described in Section~\ref{subsection:lack}, our sample of massive galaxies lacks substantial ISM absorption at the systemic velocity, and the large EWs(\mgii) observed here are typically from absorption that is blueshifted more than 200 \kmps, such that emission filling is not substantial. The systematically larger EWs of massive galaxies may be due to a larger number of blueshifted components compared to lower mass galaxies, which would imply a correlation between the maximum velocity of the outflows and M$_*$. This has been
 observed \citep[e.g., Davis et al., submitted,][]{rub14, hec16}, though the correlation with stellar mass does not appear to be as strong as with other galaxy properties. Therefore, the explanation previously proposed for the EW(\mgii) vs. M$_*$ correlation is plausible.

Another potential contributing factor to the EW(\mgii) vs. M$_*$ correlation is that as M$_*$ increases, the reservoir of gas that can potentially be part of the outflow should increase as well, for star-forming galaxies. In this context, the mass and velocity of the outflowing gas are consequences of the star formation conditions in the galaxy. There is growing agreement that $\Sigma_{SFR}$ is one of the most important properties governing the velocities of galactic winds \citep[e.g., Davis et al., submitted,][]{hec15,hec16,pru21}. If we consider two galaxies with similar EW(\mgii), M$_*$, and SFR but different $\Sigma_{SFR}$, we may expect the galaxy with higher $\Sigma_{SFR}$ to have a higher outflow velocity, as the similar energy released by the supernovae explosion is injected into a smaller volume creating a more explosive event. We see this trend if we consider the most massive starburst galaxies from \citet{rub14} and \citet{kor12} with EW(\mgii), M$_*$, and SFR comparable to our sample. The $\sim$2 dex larger $\Sigma_{SFR}$ in our sample is reflected in the substantially higher outflow velocities observed (Davis et al., submitted). A specific example is J1232, which has $\Sigma_{SFR}$ similar to the most massive galaxies from \citet{rub14} and \citet{kor12} and it also has a comparable outflow velocity v$_{98}$ = 761 \kmps (v$_{avg}$ = 217 \kmps), which is the lowest in our sample.
 In this context, considering two galaxies with similar M$_*$, SFR, and $\Sigma_{SFR}$ but very different EW(\mgii), we may expect that the galaxy with lower EW(\mgii) would have a higher outflow velocity, as a comparable amount of energy injected has to accelerate less gas. An example of this is J1125, which has similar M$_*$, SFR, and $\Sigma_{SFR}$ to J1450, but has $\sim$5 times lower EW(\mgii) and a substantially larger outflow velocity of v$_{98}$ = 2244 \kmps (v$_{avg}$ = 1813 \kmps), compared to v$_{98}$ = 1874 \kmps (v$_{avg}$ = 529 \kmps) for J1450.

While the general trends described above apply to many galaxies in Figure~\ref{fig:ew_gal}, there are exceptions, and these relations have substantial scatter. While such scaling relations are not trivial to interpret, they can be useful from a statistical standpoint. However, when we look at the details of individual galaxies, such general trends may not strictly apply, especially given that single sightlines can provide only a partial view of the entire galaxy. Another complication is that EWs measured from saturated absorption lines reflect only a lower limit on the wind absorption strength.

\subsection{Mass Outflow Rates}\label{sec:Mdot}
In this section, we estimate mass outflow rates for the observed winds in these galaxies and discuss the assumptions used to determine these values. We note that the mass outflow rates are confined to the phase we probe with our observations, which is the ionized gas phase, and therefore they are lower limits on the total mass outflow rates which include neutral and molecular gas \citep[e.g.][]{bol13, chi16}. 

Following \citet{rup05b} we assume a simple model for the wind that depends on the physical parameters derived from our line fitting. If we consider a single thin shell wind at radius $r$, with thickness $d \ll r$, the average mass outflow rate is:

\begin{equation}\label{eq:mdot}
\dot{M} = \sum_{i} \Omega_i \mu m_p N_i({\rm H}) v_i r_i
\end{equation}
where $\mu m_p$ is the mean atomic weight ($m_p$ is the proton mass and $\mu$ = 1.4 is the correction for the relative He abundance). We calculate time-averaged outflow rates because they are a more useful quantity than instantaneous values to use when comparing to star formation rates. To average the mass outflow rate over the wind lifetime and obtain Equation~\ref{eq:mdot}, we divide the instantaneous $\dot{M}$ by $t_{wind} = r /v_i$.

The sum in Equation~\ref{eq:mdot} is performed over the individual outflowing (i.e., blueshifted) velocity components in each galaxy: $\Omega_i$ is the solid angle subtended by a given component as seen from the wind's origin, $N_i$(H) is the total hydrogen column density of that component, and $v_i$ is the central velocity of that component. In this model, we split $\Omega$ (the wind's global covering factor) into two parts to account for a potential biconical morphology and a clumpy wind, rather than assuming a smooth shell. We model $\Omega$ in terms of the large-scale covering factor $C_{\Omega}$, related to the wind's opening angle, and the local covering factor \cf, related to the wind's clumpiness. Thus, $\Omega/4\pi = C_{\Omega} C_f$. We adopt $C_{\Omega} = 1$ based on the high outflow detection rate in the parent sample (Davis et al., (submitted)). 
We can then rewrite Equation~\ref{eq:mdot} as

\begin{equation}\label{eq:mdot_final}
\dot{M} = \sum_{i} 4 \pi\, C_{\Omega} C_{fi} \mu m_p N_i({\rm H}) v_i r_i
\end{equation}

An estimate of the total hydrogen column density in the outflow requires knowledge of the ionization state and metallicity of the gas, as well as the amount of dust depletion for the element employed to derive $N$(H). 
Our spectral coverage provides access to a series of strong \feii \ resonance lines which have oscillator strengths spanning a substantial range and therefore can be used to estimate the column density of singly ionized iron. From this one can determine the total hydrogen column density for an assumed metallicity and ionization fraction. 
However, as we detect \mgii \ in a larger number of absorption components for most galaxies in our sample, we begin with estimates of the total hydrogen column derived from \mgii. We describe below how these values might need to be adjusted to provide more accurate estimates using results from \feii \ where they exist.

We adopt $\chi$(\mgii) = 0.7 since, in the case of Mg, the singly-ionized state \mgii \ is likely dominant in photoionized gas at $T \sim$10$^4 K$ \citep{mur07, chu03}. We assume a solar abundance ratio (log Mg/H = $-4.45$; \citealp{asp21}), which is consistent with results from an ensemble of line ratio diagnostic diagrams used to estimate the metallicities of galaxies in our sample \citep{per21}. We assume a dust depletion factor (d(Mg)) of $- 0.5$ dex for Mg, as measured in the local Galactic ISM \citep{jen09}. The total hydrogen column density in the outflow is

\begin{equation}\label{nh}
 N({\rm H}) = \sum_{i} \frac{N_i({\rm MgII)}}{\chi({\rm MgII)} 10^{{\rm log (Mg/H)}} 10^{d({\rm Mg})}}
\end{equation}
where the sum is performed over the individual \mgii \ outflowing velocity components in each galaxy. We list in Table~\ref{table2} the \N(H) values derived using this method. These values are lower limits not only because of our assumed conservative ionization correction, dust depletion, and abundance ratios but also because the absorption in the \mgii \ components is saturated for most of the galaxies in our sample. We emphasize that most of the saturated \mgii \ absorption troughs in our spectra do not appear black which implies that \cf $<$ 1. 

Because the actual hydrogen column densities could be substantially higher than those estimated above from \mgii, we can use bounds on \ta(\mgii) from Equation~\ref{tau_mgii} to obtain a revised \N(\mgii) for all the \mgii \ components that are also detected in \feii. Using these values, we derive updated \N(H) that we report in Table~\ref{table2}. In our galaxy sample the hydrogen column density for the same component obtained from Equation~\ref{nh} using \tmeas\, (i.e. measured directly from \mgii \ fits) and \tinf\, (i.e. inferred using bounds from \feii \ fits) differ from a factor ranging from 0.3 to 19. The total hydrogen column density for each galaxy is obtained as the sum of \N(H) for the individual velocity components. Using the highest \N(H) values in Table~\ref{table2} (i.e., obtained using $\tau_0(\rm MgII)$, or $\tau_0(\rm FeII)$), we obtain total hydrogen outflowing column densities for our galaxies of \N(H)$_{tot}$ = 3 $\times 10^{19}$ $-$ 2 $\times 10^{21}$ \cm, with a median column density of 2 $\times 10^{20}$ \cm. For most galaxies in our sample, these values are still lower limits, as \feii \ provides a bound on \ta\, only for 36\% of the \mgii \ absorption troughs in our sample. Moreover, our estimates are derived assuming a conservative depletion of iron (relative to magnesium) onto dust grains. Dust depletion factors ($d(X)$) measured in the local Galactic ISM fall in the range $-(1.0 - 2.3)$ dex for Fe \citep{jen09}. Here we adopt $d$(Fe) = $-1$ dex; a greater depletion correction would increase \tmeas, and consequently \N(H), by a factor of $\sim$$3-60$. There is therefore an order of magnitude uncertainty due to the dust depletion correction alone. 

Returning to our goal of estimating the mass outflow rates in the winds observed in absorption, we can use the values constrained by our fits for all of the variables in Equation~\ref{eq:mdot_final} except for the spatial extent of the wind ($r$). While the absorption lines are sensitive to absorbing gas at any location along the line of sight to the observed galaxies, they do not provide information on the outflow geometry. Here we choose to adopt a thin shell of uniform radius 5 kpc. This radius is motivated both by observations of star-forming galaxies at similar redshifts \citep[e.g.,][]{bur21, zab21} and by our own integral field spectrograph data of outflows observed in our sources \citep{rup19}. 
\citet{rup19} use KCWI to map the \ion{[O}{2]} and \mgii \ emission in the galaxy Makani. They detect resolved \mgii \ emission on scales $\sim$15 kpc. 
While this galaxy represents an exceptional object with the largest \ion{[O}{2]} nebulae ever observed, our team has collected KCWI observations for an additional 13 sources and uncovered \ion{[O}{2]} and \mgii \ nebulae reaching far beyond the stars in the galaxies, with radial extents of a few tens of kpc (Perrotta et al. in prep). 
Therefore, a physical extent of 5 kpc for the outflows in this paper is used as a reasonable order of magnitude estimate. Determining a more accurate value is challenging as these galaxies show variations object to object and ion to ion. 
Since $\dot{M}$ is directly proportional to $r$, any change in the physical extent of the outflow will proportionally result in a change to $\dot{M}$. For example, if the value of $r$ increases by a factor of two, $\dot{M}$ will increase by a factor of two. Assuming a thick wind instead of a thin wind will decrease the derived outflow rates since the radial factor in Equation~\ref{eq:mdot_final} is the inner radius in the thick wind case, rather than the outer radius. For example, a thick wind extending from 1 to 5 kpc has a mass 5 times lower than the $r = 5$ kpc thin shell that we adopt here. We prefer a shell geometry over a constant velocity wind given the morphology of the ionized outflow observed in Makani that resembles an evacuated and limb-brightened bipolar bubble. Moreover, a constant velocity wind is likely not appropriate for our sample, which has had recent strong star formation bursts due to major merger activity. Finally, the absorption spectra presented here do not support a constant wind velocity model given their multi-component nature and large velocity extent.
We emphasize that while the geometry of the outflow (e.g., physical extent, thickness) is uncertain, this is subdominant to other uncertainties such as the dust depletion factor and optical depth, as discussed above. We may rewrite Equation~\ref{eq:mdot_final} using fiducial values as follows:

\begin{equation*}
\dot{M} = 100 \ {\rm M}_{\odot}\, {\rm yr}^{-1} \ \frac{C_{f}}{0.9} \frac{N({\rm H})}{10^{20}\, \cm} \frac{v}{1600 \, \kmps} \frac{r}{5 \, \rm kpc}
\end{equation*}

The $\dot{M}$ values for our sample derived using Equation~\ref{eq:mdot_final} are reported in Table~\ref{table5}. The $\dot{M}$ values obtained using constraints only from \mgii \ fits are listed in column 1, while column 2 lists $\dot{M}$ calculated utilizing constraints from \mgii \ and, when available, \feii \ fits.
We have robust estimates (i.e., non lower-limits) on the ionized gas column density for only two sources in our sample, J0905 and J1219. For these two objects using the constraints from the \feii \ measurements increases the mass outflow rates values from their lower limits to values that are roughly an order of magnitude larger. This underscores that the lower limits derived for the rest of the galaxies in our sample may underestimate the actual mass outflow rates by a factor of 10.

\begin{figure}[htp!]
 \centering
 \includegraphics[width=1\columnwidth]{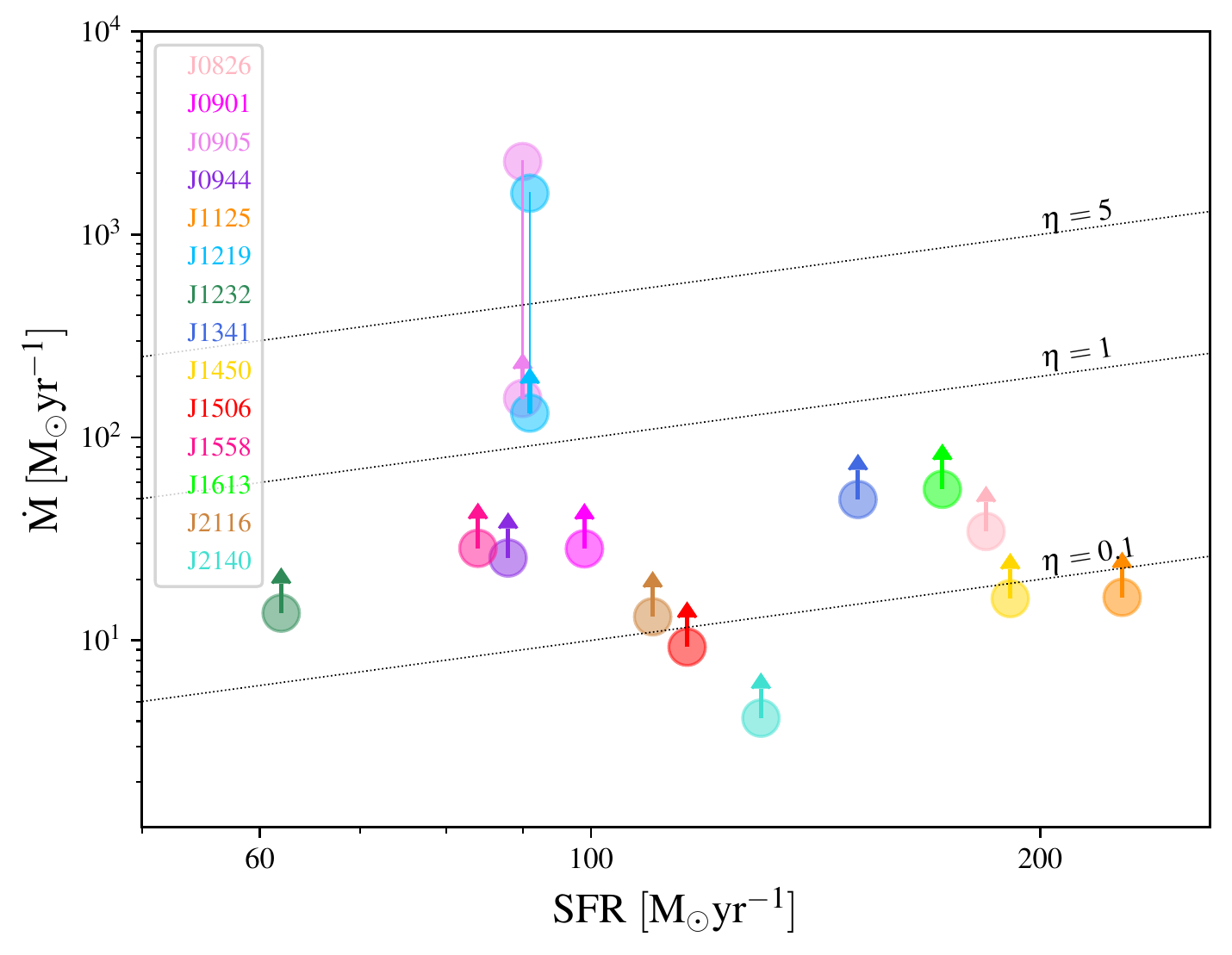}
 \caption{Estimated ionized gas phase mass outflow rates versus star formation rates for our galaxies. These estimates represent lower limits on the actual ionized mass outflow rates for the whole sample, except for J0905 and J1219. For these two galaxies using the constraints from the \feii \ measurements increases the mass outflow rates values from their lower limits to values that are roughly an order of magnitude larger. The tilted lines (from bottom to top) illustrate mass outflow rates that are 0.1, 1, and 5 times the star formation rate. }
 \label{fig:mout}
\end{figure}

As described above, deriving accurate $\dot{M}$ estimates requires precise knowledge of the physical conditions in the outflowing gas, the properties of the interstellar medium where the outflow propagates, and the outflow geometry and kinematics. We discussed above the assumptions adopted here to determine these first-order estimates, and we discussed how they represent lower limits on the actual outflow rates for most of the galaxies in our sample. We stress that the biggest uncertainty in calculating the mass outflow rates is introduced by the uncertainties on the \N(H) estimates. 
In our study, the leading uncertainty on \N(H) arises from the difficulty of breaking the degeneracy between $\tau$ and \cf\, for saturated \mgii \ doublet lines, when additional constraints from the \feii \ multiplet are not available.
We also report in Table~\ref{table5} the mass loading factor, $\eta \equiv \dot{M}$/SFR. $\eta$ compares the amount of gas entrained by the outflow to the amount of gas actively being converted into stars, and it represents a useful way to quantify how the outflow may impact the future evolution of the host galaxy, as it relates to the rate at which gas is being removed from the galaxy to the rate at which it is currently creating stars.

\input{Table5.tex}

Figure~\ref{fig:mout} shows the mass outflow rates as calculated using \mgii \ only constraints (Table~\ref{table5}, column 1) as a function of the star formation rate. 
As discussed these are lower limits on the total ionized gas mass outflow rates. In order to demonstrate the potential magnitude of the systematic underestimate of $\dot{M}$ derived using \mgii \ only, we show the $\dot{M}$ values calculated using bounds on \ta\, from \feii \ fits (Table~\ref{table5}, column 2) for the two sources (J0905 and J1219) in our sample that have the most robust estimates on \ta\, and which are therefore not lower limits. The two $\dot{M}$ estimates for J0905 and J1219 are connected by lines, to facilitate comparison. There is a large correction for these two sources between the \mgii \ only estimates (lower limits) and the more robust estimates (not lower limits), of factors of $\sim$15 and 12, respectively. The rest of the galaxies in the sample might have a similar magnitude correction if better constraints on \N(H) from the \feii \ absorption lines were available. Applying similar corrections to the rest of the sample (i.e. to the $\dot{M}_{\rm MgII}$ values reported in Table~\ref{table5}) as those obtained for J0905 and J1219, we would infer mass outflow rates in the range $\sim$$50-2000$ $\rm M_{\odot}\, yr^{-1}$ for our galaxies.

Figure~\ref{fig:mout} also includes lines of constant mass loading factors, $\eta$. The two galaxies with more robust constraints on $\dot{M}$, J0905 and J1219, exhibit $\eta$ values of the order of $\sim$25 and 17, respectively. Such high values of the mass loading factor indicate that the outflows are ejecting gas at a much higher rate than the production rate of stars. Similar $\eta$ values are found in ultraluminous infrared galaxies (ULIRGs) at $z<0.5$ \citep{rup05c}. While the mass outflow rates are first-order estimates only, they are informative to verify that the observed starburst-driven outflows can clearly affect the evolution of their host galaxies by actively removing substantial amounts of gas. The mass outflow rates for J0905 and J1219 are estimated to be $\sim$2200 and 1500 $\rm M_{\odot}\, yr^{-1}$, which are exceptionally high mass outflow rates. This finding is consistent with results based on observations of molecular gas for two galaxies (J0905 and J1341) in our sample \citep{gea14, gea18}. In particular, \citet{gea14} use the IRAM Plateau de Bure interferometer to study the CO(2-1) emission line, a tracer of the bulk of the cold molecular gas reservoir, in the galaxy J0905. They find that a third of the total molecular gas content appears to have been ejected on a scale of approximately 10 kpc. They estimate that the kinetic energy associated with the outflowing component is consistent with the momentum flux available from stellar radiation pressure, demonstrating that nuclear starbursts are capable of ejecting large amounts of cold gas from the central regions of galaxies.

As discussed in Section~\ref{subsec:comparison}, the cold gas outflow phase is most likely produced by a combination of multiple physical processes. Therefore the high mass loading factors derived here do not necessarily refer to a single production mechanism of the cool gas. However, as a reference, we compare our estimates with predictions from \citet{loc21}, who model hot supersonic winds driven by supernovae energy injection in star-forming galaxies. The authors derive the characteristic momentum rates of hot galactic winds that can undergo single-phase cooling on large scales and produce cold gas as a function of the SFR of the galaxy. Interestingly, the predicted velocities of the wind at the maximum wind momentum rate for SFRs similar to our galaxies are compatible with the highest velocity absorption components in our spectra. However, most of our sample shows higher mass loading factors than the theoretical maximum predicted by their model. This would imply that our observations do not trace only cool outflows that obtained their momentum directly from the free-flowing hot wind, but rather that there may be some mass loading outside the wind-driving region. This supports the hypothesis that the cold gas outflow phase may have multiple origins.

In summary, when robust constraints from the \feii \ multiplet are available we find extreme mass outflow rates ($\rm \sim$$1500-2200 \, M_{\odot}\, yr^{-1}$) and $\eta$ values of $\sim$$17-25$. If similar corrections are applied to the rest of the galaxies, they would have $\dot{M}$ in the range $\rm \sim$$50-2000$ $\rm M_{\odot} \, yr^{-1}$ and $\sim$90\% of the sample would have $\eta > 1$.

\section{Summary and Conclusions}\label{sec:Conclusions}

We use new optical Keck/HIRES spectroscopy of 14 compact starburst galaxies at z $\sim$0.5 
to probe the small-scale structure of the powerful galactic outflows observed in our sample,
gain insights into the role of various physical mechanisms in these extreme feedback episodes, and investigate their potential impact on the evolution of their host galaxies. These galaxies are massive ($\rm M_* \sim$10$^{11} \,M_{\odot}$), extremely compact (half-light radius $\sim$few hundred pc), have very high star formation rates (mean $\rm SFR \sim200 \, M_{\odot} \ yr^{-1}$) and star formation surface densities (mean $\rm \Sigma_{SFR} \sim2000 \,M_{\odot} \ yr^{-1} \ kpc^{-2}$), and host extremely fast (mean maximum velocity $\sim$$-2000$ \kmps) outflows traced by \ion{Mg}{2} absorption lines (\citealp{tre07}; Davis et al., (submitted). The high resolution ($\simeq$ 8 \kmps) spectra presented here cover a suite of Mg and Fe absorption lines (\mgb \ doublet, \mgi$\lambda$2852, and \feii$\lambda$2344, 2374, 2582, 2587, 2600 multiplet). This exquisite data set allows us to directly measure precise column densities and covering fractions (\cf) as a function of velocity for these ions and characterize the kinematics of the cool gas outflow phase (T $\sim$10$^4$ K). Our main conclusions are as follows:

1) Mg II absorption troughs best delineate the outflow kinematics among the ions studied here. We find a substantial variation in the absorption profiles in our data (Section~\ref{subsec:variation} and Figure~\ref{fig:Fits}). In particular, there is a large variation in the minimum number of velocity components required to fit each profile, ranging from two to ten; in their widths, Doppler parameter \bd, ranging from 8 to 344 \kmps \ with a mean of 106 \kmps; and in their \cf, with four galaxies having \cf\, = 1, while the rest of the sample has \cf\, as low as 0.12 and a mean value of 0.57.

2) We investigate if the multiple velocity components observed in our sample are related to the extreme and ``bursty'' star formation episodes in these galaxies. There is not a simple correlation between the number of bursts in the star formation history (SFH) and the \mgii \ absorption profiles (Section~\ref{subsec:variation} and Figure~\ref{fig:SFH}). Galaxies showing similar \mgii \ absorption troughs can have a variety of SFHs and vice-versa. 

3) \mgii \ emission is detected in 9/14 of the galaxies in our sample with velocity shifts of 0 \kmps\ to $+$450 \kmps\ from \zsys. Emission filling is not an issue for most of our galaxies, which have highly blueshifted \mgii \ absorption lines (Section~\ref{subsection:lack} and Figure~\ref{fig:Fits}). Indeed, 11/14 galaxies have less than four percent of the EW(\mgii) within 200 \kmps of \zsys, where emission line filling can be important. We also find a lack of substantial \mgii \ absorption at the systemic velocity, and we present three possible explanations for this: the bulk of the interstellar medium (ISM) is expelled by the powerful outflows, biased observational selection criteria, and/or \mgii\ traces a cold gas phase formed ``in-situ'' within a galactic wind via thermal instabilities and condensation of the fast-moving hot phase. 

4) \mgii, \mgi, and \feii \ exhibit remarkably similar absorption profiles, suggesting these species reside in the same, low-ionization gas structures. However, \mgii \ has on average a higher \cf \ at a given outflow velocity than \feii, and in all cases a higher \cf \ than neutral Mg, implying that the absorbing clouds are not homogeneous (Section~\ref{subsec:dist}). This result is in agreement with theoretical predictions that the coolest and densest gas is typically located in the more internal and self-shielded part of the clouds.

5) We find that \cf\, does not display a unique trend with velocity (Section~\ref{subsec:cf} and Figure~\ref{fig:cf}). This variation, along with the variation of the absorption profiles, may capture the complex morphology of $\sim$kpc scale, inhomogeneously-distributed, clumpy gas, and the intricacy of the material in the turbulent mixing layers between the cold and the hot phases. Moreover, other observations of our galaxies are consistent with models in which the outflows decelerate with time as they expand into the CGM.

6) We consider the possibility of \feii \ as the primary tracer of the entrained cool gas phase component and \mgii \ as the tracer of the cooled hot wind material and examine how their ratio varies as a function of outflow velocity.
Several of the galaxies in our sample do not have \feii \ detected at the highest outflow velocities, indicating a lower mixing fraction of entrained cool gas. This suggests that the cold gas at the highest velocities in these galaxies most likely directly condensed out of the hot wind phase, as suggested by some theoretical models (Section~\ref{subsec:comparison}, Figure~\ref{fig:overlay} and \ref{fig:ratio}).

7) We estimate mass outflow rates for the observed winds in our sample and discuss how these estimates are lower limits on the actual outflow rates for most of the galaxies (Section~\ref{sec:Mdot} and Figure~\ref{fig:mout}). We show that the two galaxies in our sample that have robust constraints from the \feii \ multiplet have extremely high mass outflow rates ($\rm \sim1500-2200\, M_{\odot} \ yr^{-1}$) and mass loading factors ($\eta \sim$17$-$25). If similar corrections are applied to the rest of the galaxies we infer $\dot{M}$ $\sim$50$-$2200 $\rm M_{\odot} \ yr^{-1}$ with a typical value of $\eta \sim$5. 
This demonstrates that starburst galaxies are capable of ejecting very large amounts of cold gas which will substantially impact their future evolution. 

The galaxy sample studied here provides a prime opportunity to investigate star formation and feedback at its most extreme. In a forthcoming paper based on integral field unit Keck/KCWI spectra, we will focus on studying the morphology, physical extent, and kinematics of the outflows in our sample. Such data on these galaxies provides strong observational constraints to theoretical simulations that aim to implement realistic stellar-driven galactic outflows. The comparison of outflow properties between simulations and observations will advance our understanding of galactic feedback, especially from stellar processes, during a critical stage of massive galaxy evolution.

\section*{Acknowledgements}

We thank the referee for her/his time to provide a constructive report. The referee's thoughtful comments have helped to improve the clarity of the manuscript. We acknowledge support from the National Science Foundation (NSF) under a collaborative grant (AST-1813299, 1813365, 1814233, 1813702, and 1814159) and from the Heising-Simons Foundation grant 2019-1659. S.~P. and A.~L.~C. acknowledge support from the Ingrid and Joseph W. Hibben endowed chair at UC San Diego. The data presented herein were obtained at the W. M. Keck Observatory, which is operated as a scientific partnership among the California Institute of Technology, the University of California, and the National Aeronautics and Space Administration. The Observatory was made possible by the generous financial support of the W. M. Keck Foundation.
The authors wish to recognize and acknowledge the very significant cultural role and reverence that the summit of Maunakea has always had within the indigenous Hawaiian community. We are most fortunate to have the opportunity to conduct observations from this mountain.

\clearpage

\appendix

\section{Notes and Fits of Individual Targets}\label{app:A}

\begin{figure*}[htp!]
 \centering
 \includegraphics[width=0.68\textwidth]{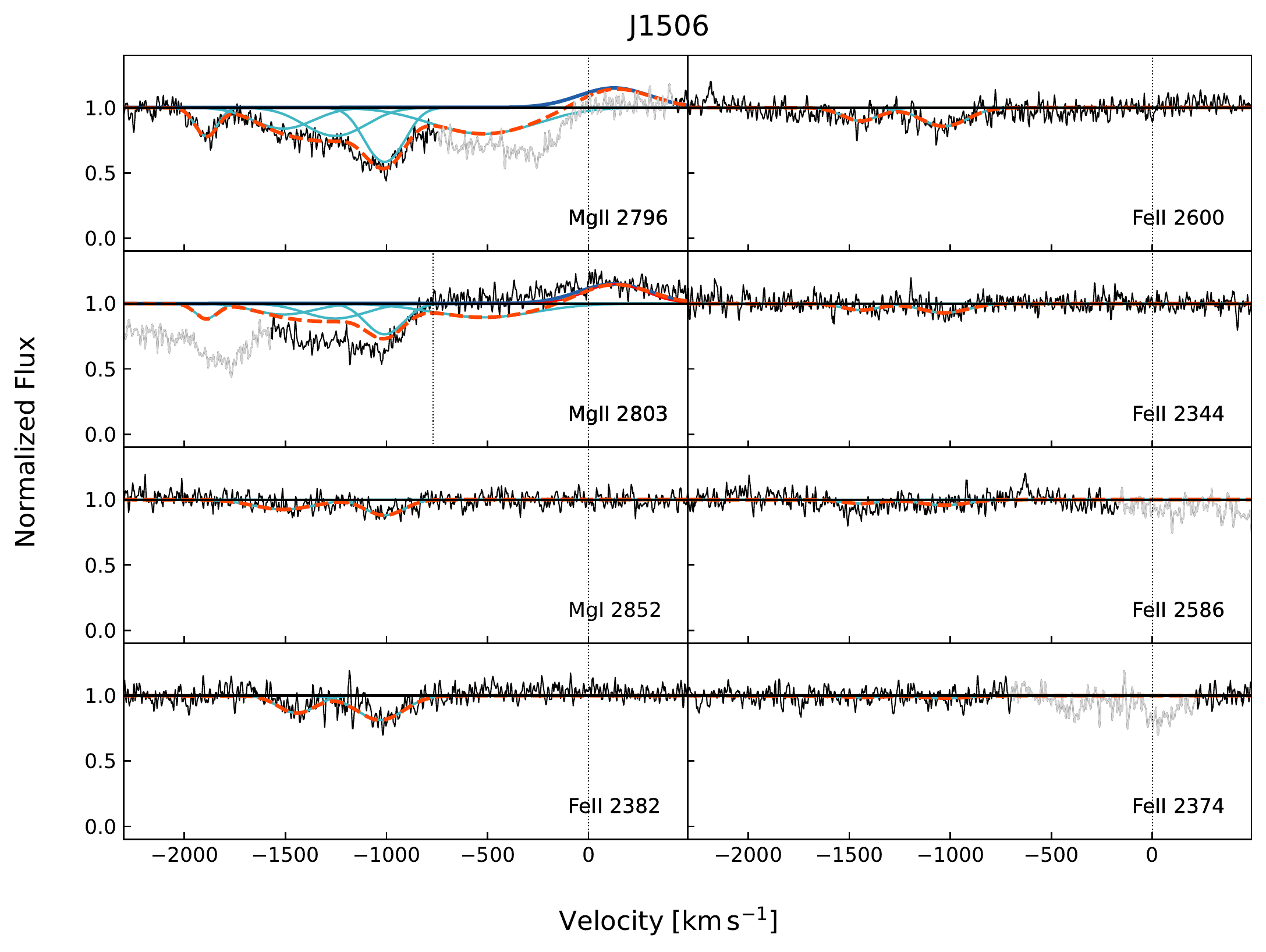}
 \caption{J1506 spectrum, which shows \mgii \ emission within $-300$ and $+550$ \kmps\, of \zsys. In this galaxy, we see strong absorption from \mgii \ falling within $-2030$ and $+150$ \kmps\, of \zsys. The \mgii \ trough is well described by five velocity components. We detect \mgi \ within $-1750$ and $-750$ \kmps\, of \zsys. We detect \feii \ within $-1650$ and $-750$ \kmps\, of \zsys. We model the \mgi \ and \feii \ troughs using two components that show a good qualitative alignment with \mgii.}
 \label{fig:j1506}
\end{figure*}

\begin{figure*}[htp!]
 \centering
 \includegraphics[width=0.68\textwidth]{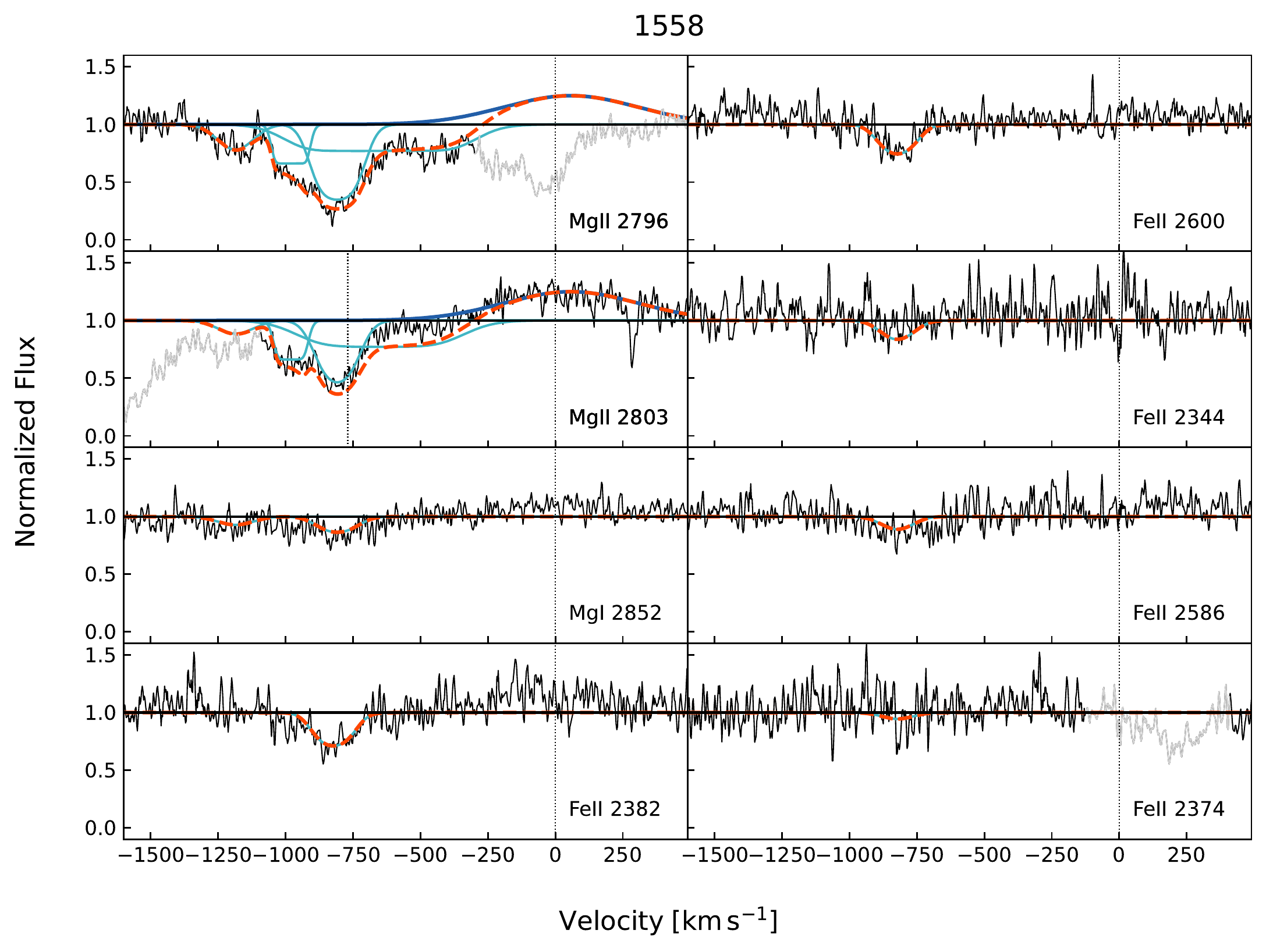}
 \caption{J1558 spectrum, which shows \mgii \ emission within $-650$ and $+700$ \kmps\, of \zsys. In this galaxy, we see strong absorption from \mgii \ falling within $-1350$ and $-100$ \kmps\, of \zsys. The \mgii \ trough is well described by four velocity components. We detect \mgi \ within $-1350$ and $-630$ \kmps\, of \zsys. We model the \mgi \ trough using two components that show a good alignment with \mgii. We detect \feii \ within $-1000$ and $-630$ \kmps\, of \zsys. We model \feii \ using one component.}
 \label{fig:j1558}
\end{figure*}

\begin{figure*}[htp!]
 \centering
 \includegraphics[width=0.68\textwidth]{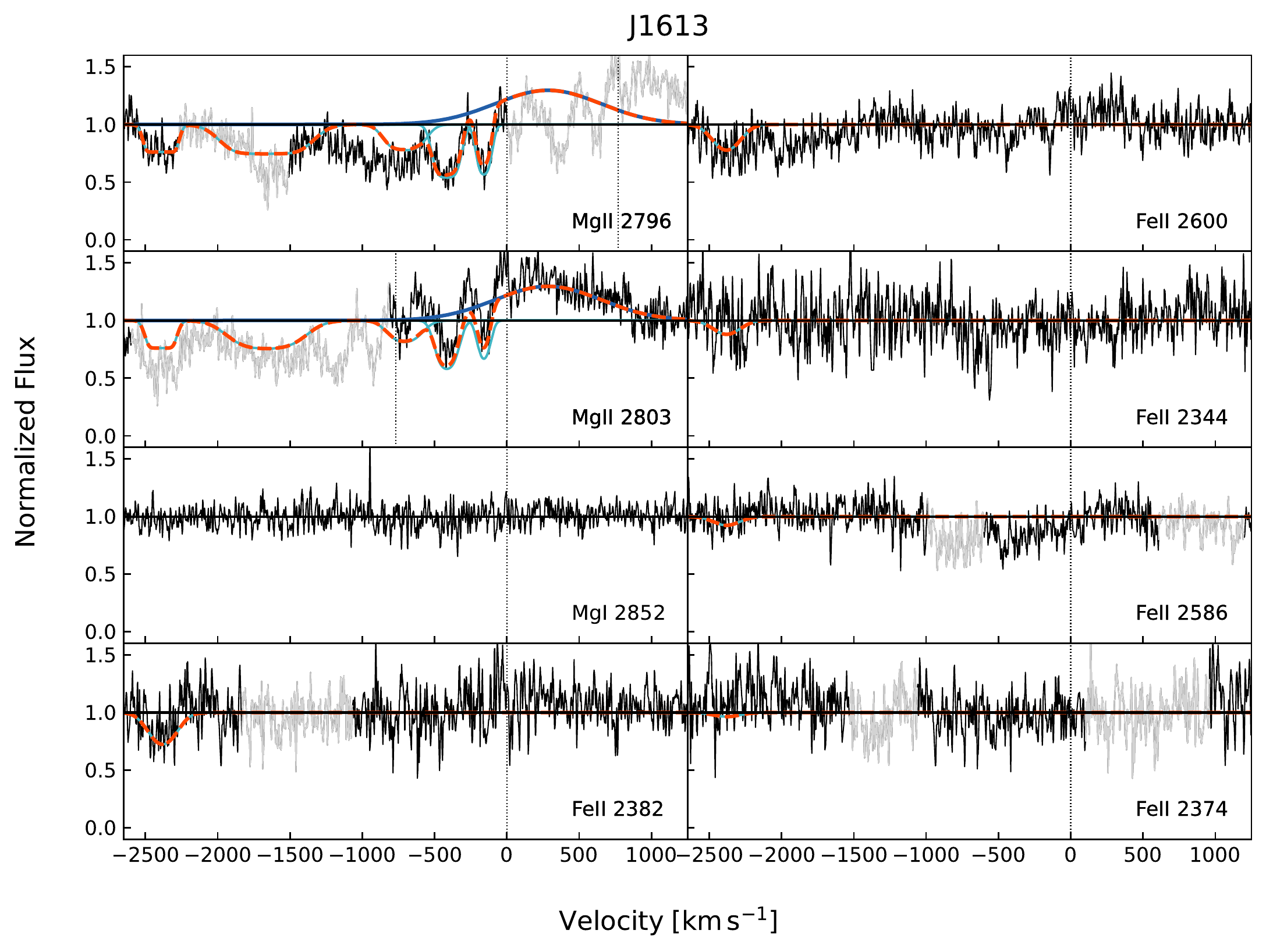}
 \caption{J1613 spectrum, which shows \mgii \ emission within $-700$ and $+1200$ \kmps\, of \zsys. In this galaxy, we see strong absorption from \mgii \ within $-2570$ and $-50$ \kmps\, of \zsys. The \mgii \ trough is well described by five velocity components. We do not detect \mgi. We detect \feii \ at about $-2400$ \kmps\, of \zsys.}
 \label{fig:j1613}
\end{figure*}

Figures~\ref{fig:j0826}$-$\ref{fig:j2140} show the normalized spectra of the other 10 galaxies in our sample centered on the transitions relevant to this study: \mgii$\lambda\lambda$2796,2803\AA, \mgi$\lambda$2803\AA, \feii$\lambda$2382\AA, \feii$\lambda$2600\AA, \feii$\lambda$2344\AA, \feii$\lambda$2586\AA, and \feii$\lambda$2374\AA. For each galaxy, we show the \mgi$\lambda$2853 best fit profile as derived following the constrained approach. Details for each galaxy are given below.

\subsection*{J0826+4305}
Figure~\ref{fig:j0826} shows our best fit to the \mgii, \mgi, and \feii \ absorption lines in the galaxy J0826+4305 (hereafter J0826). J0826 spectrum shows \mgii \ emission within $-400$ and $+400$ \kmps\, of \zsys. In this galaxy, we see strong absorption from \mgii \ falling within $-1560$ and $-75$ \kmps\, of \zsys. The \mgii \ trough is well described by eight velocity components that show complex kinematics. 
The \mgo \ and \mgd \ troughs exhibit comparable intensity at all velocities for seven over eight components. In that case, the \mgii \ traces optically thick gas. As the lines are not black the covering fraction (\cf $ < 1$) determines the shape of the absorption troughs. The most blueshifted component ($v=-1404$ \kmps) has \ta $<$ 1 and \cf = 1. In this galaxy, we note that the \cf\, increases with increasing blueshift from \zsys.
We detect \feii \ and \mgi \ absorption within $-1350$ and $-1000$ \kmps\, of \zsys, which we model using two velocity components. They have quite good kinematic correspondence (within the errors) to two \mgii \ components. However, the \feii \ and \mgi \ (independent) fit find the least blueshifted component to be substantially broader than the corresponding \mgii \ (20 and 55\%, respectively). We find both \feii \ components to trace optically thin gas and have lower \cf\, than \mgii \ (32 and 45\% of \cf(\mgii), respectively). Based on comparison of \tmeas\, and \tinf, we conclude that the \mgii \ fit gives a lower limit estimate of the N(\mgii) for the two components we have detection of both \mgii \ and \feii. The actual \ta(\mgii) is $\sim$3 times larger than the measured one. As for previous targets we favor the \mgi \ constrained fit solution because the neutral Mg fraction must be less than one percent to produce an optically thin \mgi$\lambda$2853 trough with \cf=1 as found by the independent fit. The constrained fit finds for both \mgi \ components a smaller \cf than \mgii \ (28 and 33\% of \cf(\mgii), respectively).

\subsection*{J0901+0314}
\begin{figure*}[htp!]
 \centering
 \includegraphics[width=0.68\textwidth]{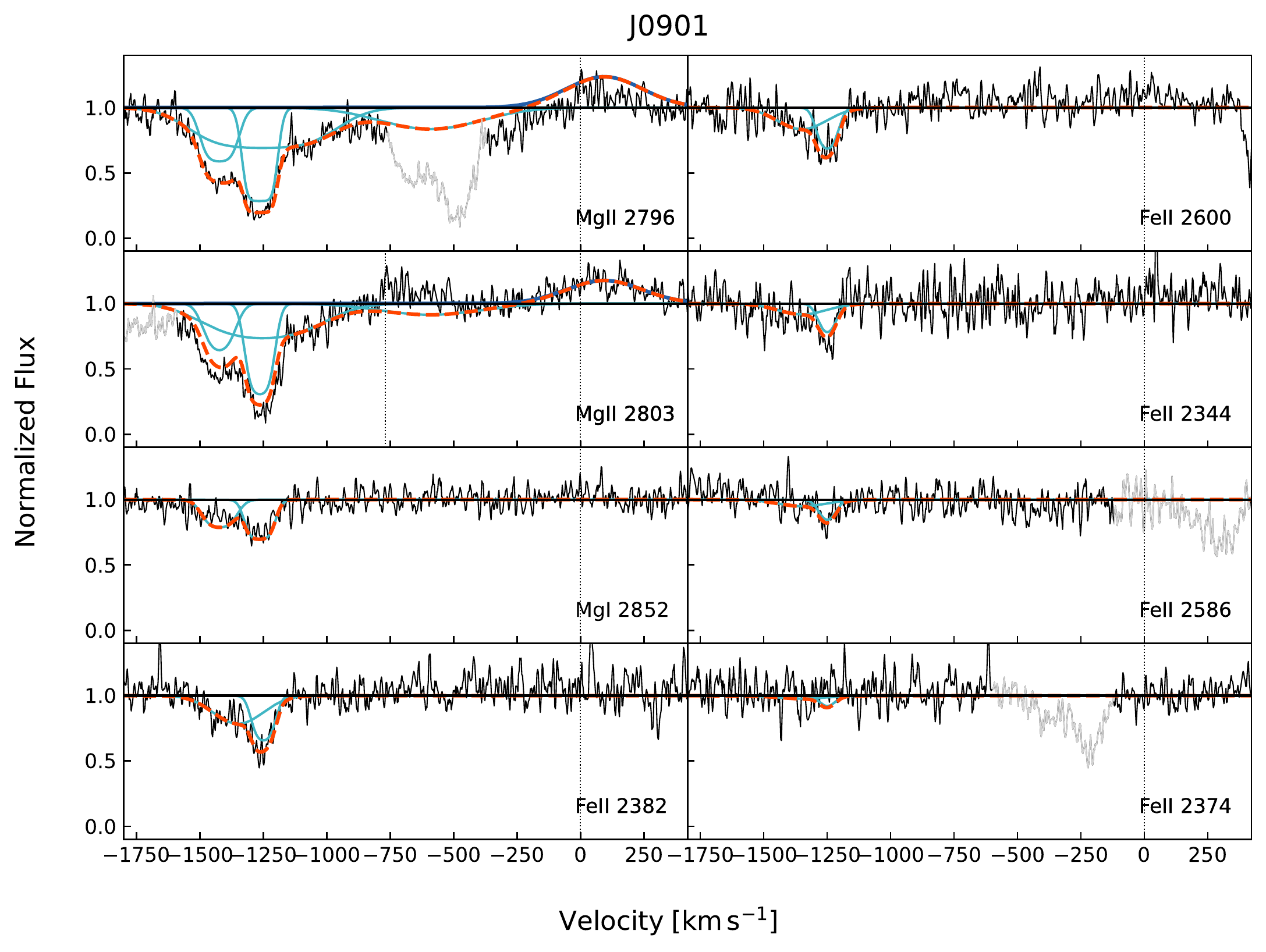}
 \caption{J0901 spectrum, which exhibits \mgii \ emission within $-300$ and $+500$ \kmps\, of \zsys. In this galaxy, we see strong absorption from \mgii \ falling within $-1750$ and $-100$ \kmps\, of \zsys. The \mgii \ trough is well described by four velocity components. We detect \feii \ and \mgi \ absorption within $-1550$ and $-1000$ \kmps\, of \zsys, which we model using two velocity components.}
 \label{fig:j0901}
\end{figure*}

Figure~\ref{fig:j0901} shows our best fit to the \mgii, \mgi, and \feii \ absorption lines in the galaxy J0901+0314 (hereafter J0901). J0901 spectrum exhibits \mgii \ redshifted emission ($+93$ \kmps) within $-300$ and $+500$ \kmps\, of \zsys. In this galaxy, we detect strong absorption from \mgii \ falling within $-1750$ and $-100$ \kmps\, of \zsys. The \mgii \ trough is well characterized by four velocity components that exhibit complex kinematics. One \mgo \ component blends together with three \mgd \ ones. Our \mgii \ best fit includes one optically thin component (at $v=-591$ \kmps) with \cf=1, and three more blueshifted components. The latters trace optically thick gas, with \mgo \ and \mgd \ showing nearly identical intensity at all velocities. Their absorption profiles are not black and their shape is determined by \cf$< 1$. We detect \feii \ and \mgi \ absorption within $-1550$ and $-1000$ \kmps\, of \zsys, which we model using two velocity components. We find that one of these two components has a good kinematics correspondence (comparable $v$ and $b$ to within the error bars) to one \mgii \ component. \cf\, for both \feii \ and \mgi \ (from the constrained fit), is found to be lower than \cf(\mgii), 52\% and 43\% of \cf(\mgii), respectively. The values of \tmeas\, and \tinf\, for the component detected at $-1266$ \kmps represent a lower limit (the actual \ta(\mgii) is $\sim$3 times larger), while agreeing well within the errors for the component detected at $-1426$ \kmps.
The independent \mgi \ fit shows kinematics in line with the constrained fit. However, to be a valid solution, the neutral Mg fraction must be less than one percent. The second and more blueshifted \feii \ and \mgi \ (from the independent fit) component in our model show a substantially broader absorption profile (46\% and 72\% of \bd(\mgii), respectively). Based on comparison of \tmeas\, and \tinf\, we infer that \mgii \ should be optically thin, but that is inconsistent with the nearly identical intensity of \mgo \ and \mgd. We attribute this discrepancy to the low SNR in the \feii \ and \mgi \ profiles blueward of $-1350$ \kmps that is inadequate to definitively constrain the kinematics of the absorption troughs.

\subsection*{J0944+0930}
\begin{figure*}[htp!]
 \centering
 \includegraphics[width=0.68\textwidth]{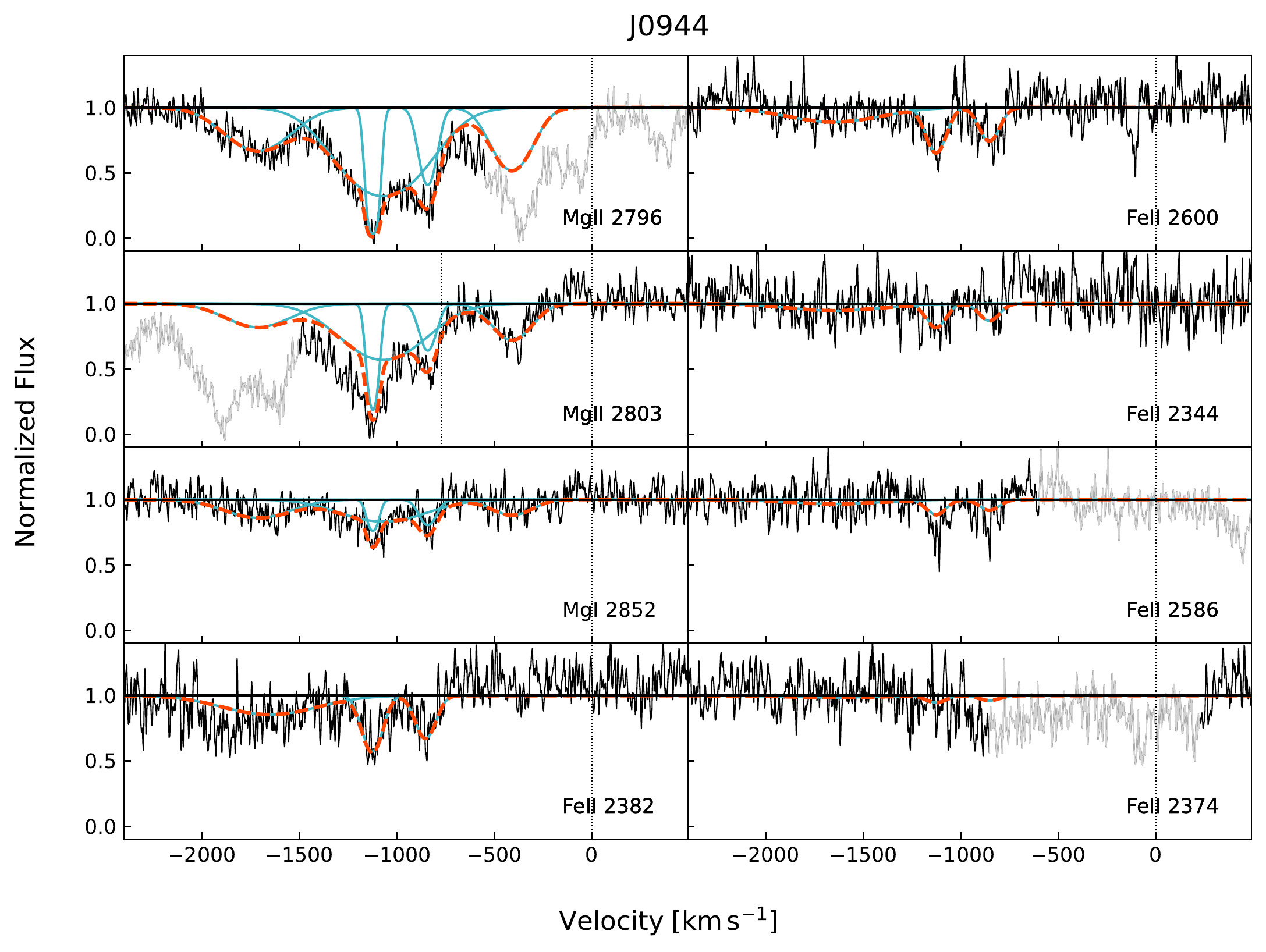}
 \caption{J0944 spectrum, which does not exhibit \mgii \ emission. In this galaxy, we see strong absorption from \mgii \ falling within $-2100$ and $-120$ \kmps\, of \zsys. The \mgii \ trough is well described by five velocity components. The detected \mgi \ trough has a remarkable alignment with \mgii. We model the \mgi \ absorption using five components. We detect \feii \ within $-1350$ and $-700$ \kmps\, of \zsys, which we model using three velocity components.}
 \label{fig:j0944}
\end{figure*}

Figure~\ref{fig:j0944} shows our best fit to the \mgii, \mgi, and \feii \ absorption lines in the galaxy J0944+0930 (hereafter J0944). J0944 spectrum does not show \mgii \ emission. In this galaxy, we see strong absorption from \mgii \ falling within $-2100$ and $-120$ \kmps\, of \zsys. The \mgii \ trough is well described by five velocity components that show convoluted kinematics. Most of them (four \mgo, and four \mgd) blend together. We find that the profile shape for all \mgii \ components is determined by the optical depth, rather than the \cf, that is unity. Our \mgii \ best fit includes only one optically thick component (at $v=-1125$ \kmps), the remaining components trace optically thin gas.
The \mgi \ trough has a remarkable alignment with \mgii. We model the \mgi \ absorption using five components, as well. In the constrained fit the five \mgi \ components are described by lower \cf\, compared to \mgii, with values in the range of $32-56\%$ of \cf(\mgii). The independent \mgi \ fit finds a great agreement in terms of kinematics with the corresponding \mgii \ components. However, to be a valid solution, the neutral Mg fraction must be less than one percent. Therefore, we favor the constrained fit result.
We detect \feii \ within $-1350$ and $-700$ \kmps\, of \zsys, that we model using three velocity components. They have good alignment with three \mgii \ components, but they are characterized by broader line profiles (in the range of 30 to 50 \%). Based on a comparison of \tmeas\, and \tinf\, we conclude that the \mgii \ fit provides a decent estimate of the N(\mgii). However, we note that the difference in the kinematics of the two absorption troughs makes the comparison of the \ta\, not accurate.

\subsection*{J1125$-$0145}
\begin{figure*}[htp!]
 \centering
 \includegraphics[width=0.68\textwidth]{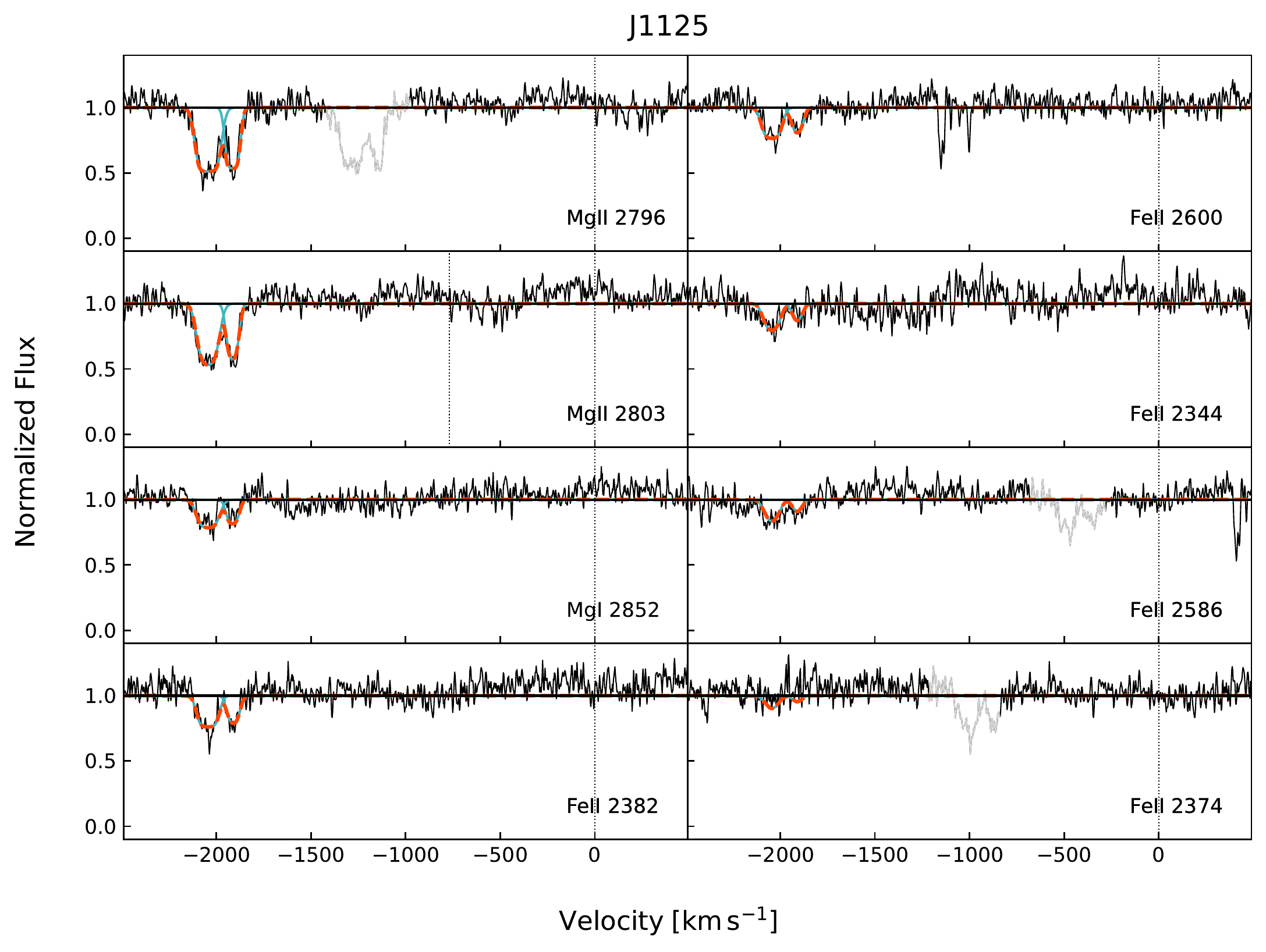}
 \caption{J1125 spectrum, which does not exhibit \mgii \ emission. In this galaxy, we see absorption from \mgii \ falling within $-2170$ and $-1830$ \kmps\, of \zsys. The \mgii \ trough is well described by two velocity components. The detected \mgi \ and \feii \ troughs have a remarkable alignment with \mgii. We model the \mgi \ and \feii \ absorption using two components.}
 \label{fig:j1125}
\end{figure*}

Figure~\ref{fig:j1125} shows our best fit to the \mgii, \mgi, and \feii \ absorption lines in the galaxy J1125$-$0145 (hereafter J1125). J1125 spectrum does not exhibit \mgii \ emission. In this galaxy, we see absorption from \mgii \ falling within $-2170$ and $-1830$ \kmps\, of \zsys. The \mgii \ trough is well described by two optically thick velocity components. The identical intensity at all velocities of the \mgii \ absorption lines indicates that their profile shapes are determined by \cf $<1$. We detect \mgi \ and \feii \ within the same velocity range as \mgii. We model the \mgi \ and \feii \ troughs using two components and find a remarkable kinematic correspondence (comparable $v$ and \bd\, to within the errors) to the \mgii \ fit. Both, \mgi \ (from constrained fit) and \feii \ absorption profiles are described well by lower \cf\, than \mgii ($\sim$50 and $\sim$43 \% of \cf(\mgii), respectively). Comparing \tmeas\, and \tinf\, we argue that the \mgii \ fit provides a lower limit estimate of the N(\mgii) and that the actual central optical depth for the two \mgd \ components is closer to 9, 22 (rather than 2 and 3 as measured from the \mgii \ fit).

\subsection*{J1219+0336}
Figure~\ref{fig:j1219} shows our best fit to the \mgii, \mgi, and \feii \ absorption lines in the galaxy J1219+0336 (hereafter J1219).
J1219 spectrum very clearly shows \mgd \ emission within $-600$ and $+600$ \kmps\, of \zsys. The corresponding \mgo \ line is not obvious because of \mgd \ absorption at the same wavelengths. The inclusion of an emission component in the model improves the overall fit to the \mgii \ absorption trough. In this galaxy, we see strong absorption from \mgii \ falling within $-2400$ and $-150$ \kmps\, of \zsys. The \mgii \ trough is well described by four velocity components that blend together producing extremely intricate kinematics. Our \mgii \ best fit includes one optically thin component (at $v=-755$\kmps) with \cf=1. The remaining three components trace optically thick gas, and their profiles are shaped by \cf$<1$. We detect \mgi \ within $-2400$ and $-950$ \kmps\, of \zsys. We model the \mgi \ trough using three components that show a good qualitative alignment with \mgii. The constrained \mgi \ fit finds lower \cf than \mgii, in the range $29 - 56$ \% of \cf(\mgii). In our independent \mgi \ fit model, we find for the most blueshifted component an excellent kinematic agreement to the corresponding \mgii \ one ($v$ and \bd\, within the errors). The other two components show kinematics substantially different from \mgii. We attribute this difference to the low SNR in the \mgi \ profile redward of $-1400$ \kmps which is not sufficient to conclusively constrain the kinematics of the absorption troughs. We detect \feii \ within $-2400$ and $-750$ \kmps\, of \zsys. Before performing the \feii \ fit, we identify and model the \ion{Mn}{2} $\lambda\lambda\lambda$ 2576, 2594, and 2606 triplet (yellow solid lines). We model the \mgi \ trough using three components that show extremely good kinematic agreement with \mgii. We find \feii \ to have lower \cf than \mgii \ (53, 46, and 55 \% of \cf(\mgii), respectively). Based on a comparison of \tmeas\, and \tinf, we conclude that the \mgii \ fit gives only a lower limit to the N(\mgii), and that the actual \ta\, for the two \mgd \ components, is closer to 90, 21, and 91 (than 10, 1.2, and 10).

\subsection*{J1506+5402}
Figure~\ref{fig:j1219} shows our best fit to the \mgii, \mgi, and \feii \ absorption lines in the galaxy J1506+5402 (hereafter J1506). J1506 spectrum very clearly shows redshifted ($\sim$126 \kmps) \mgd \ emission from $-300$ to $+550$ \kmps\, of \zsys. The corresponding \mgo \ line is not obvious because of \mgd \ absorption at the same wavelengths. The inclusion of an emission component in the model improves the overall fit to the \mgii \ absorption trough. In this galaxy, we see strong absorption from \mgii \ falling within $-2030$ and $+150$ \kmps\, of \zsys. The \mgii \ trough is well characterized by five velocity components that blend together creating complex kinematics. We find that all \mgii \ components trace optically thin gas, and their absorption profiles are shaped by the optical depth, rather than the \cf, that is unity.
We detect \mgi \ within $-1750$ and $-750$ \kmps\, of \zsys. We model the \mgi \ trough using two components that show a good alignment with \mgii. The constrained fit finds lower \cf(\mgi) than \mgii \ (45 and 60\% of \cf(\mgii), respectively). Adopting the independent \mgi \ fit we find the two \mgi \ components to be slightly narrower (7 and 20\%) than the corresponding \mgii. In both cases, the \mgi \ traces optically thin gas as \mgii. However, we favor the constrained fit solution since to have \cf(\mgi) = 1 the neutral Mg fraction must be less than one percent. We detect \feii \ within $-1650$ and $-750$ \kmps\, of \zsys. We model the \feii \ troughs using two components that show a good alignment with \mgii. However, we note that \bd(\feii) is 38\% narrower than \bd(\mgii) for the most blueshifted component, and 15\% broader for the other one. Despite the discrepancy between their kinematics, we find a very good agreement between \tmeas\, and \tinf within the errors.

\subsection*{J1558+3957}
Figure~\ref{fig:j1219} shows our best fit to the \mgii, \mgi, and \feii \ absorption lines in the galaxy J1558+3957 (hereafter J1558). J1558 spectrum distinctly shows slightly redshifted (57 \kmps) \mgd \ emission within $-650$ and $+700$ \kmps\, of \zsys. The corresponding \mgo \ line is not apparent because of \mgd \ absorption at the same wavelengths. The inclusion of an emission component in the model improves the fit to the \mgii \ absorption trough. In this galaxy, we see strong absorption from \mgii \ falling within $-1350$ and $-100$ \kmps\, of \zsys. The \mgii \ trough is well described by four velocity components that exhibit intricate kinematics. Our \mgii \ best fit finds the most blueshifted component to be optically thing and with \cf = 1. For the other three components, the measured equivalent width ratio for \mgo \ and \mgd \ is not consistent ($\sim$30 to 65\% lower) with the optically thin limit. They trace optically thick gas and their absorption profiles are shaped by \cf $<$ 1. We detect weak \mgi \ within $-1350$ and $-630$ \kmps\, of \zsys. We model the \mgi \ trough using two components. The constrained fit finds \mgi \ to be characterized by lower \cf\, than \mgii \ (31 and 50\% of \cf(\mgii), respectively). The independent \mgi \ fit finds two components that show a good qualitative alignment with \mgii, but fairly broader absorption profiles (56 and 18\% respectively). We prefer the constrained fit solution since to have optically thin \mgi \ for the least blueshifted component and \cf(\mgi) = 1 the neutral Mg fraction must be less than one percent. We detect \feii \ within $-1000$ and $-630$ \kmps\, of \zsys. We model \feii \ using only one component with a good kinematic correspondence to one \mgii \ component. We find that \feii \ traces optically thin gas and the profile is well described by a lower \cf than \mgii (51 \% of \cf(\mgii)). Based on comparison of \tmeas\, and \tinf, we conclude that the \mgii \ fit gives a lower limit estimate of the N(\mgii) for the component we have detection of both \mgii \ and \feii. The actual \ta(\mgii) is $\sim$4 times larger than the measured one.

\subsection*{J1613+2834}
\begin{figure*}[htp!]
 \centering
 \includegraphics[width=0.68\textwidth]{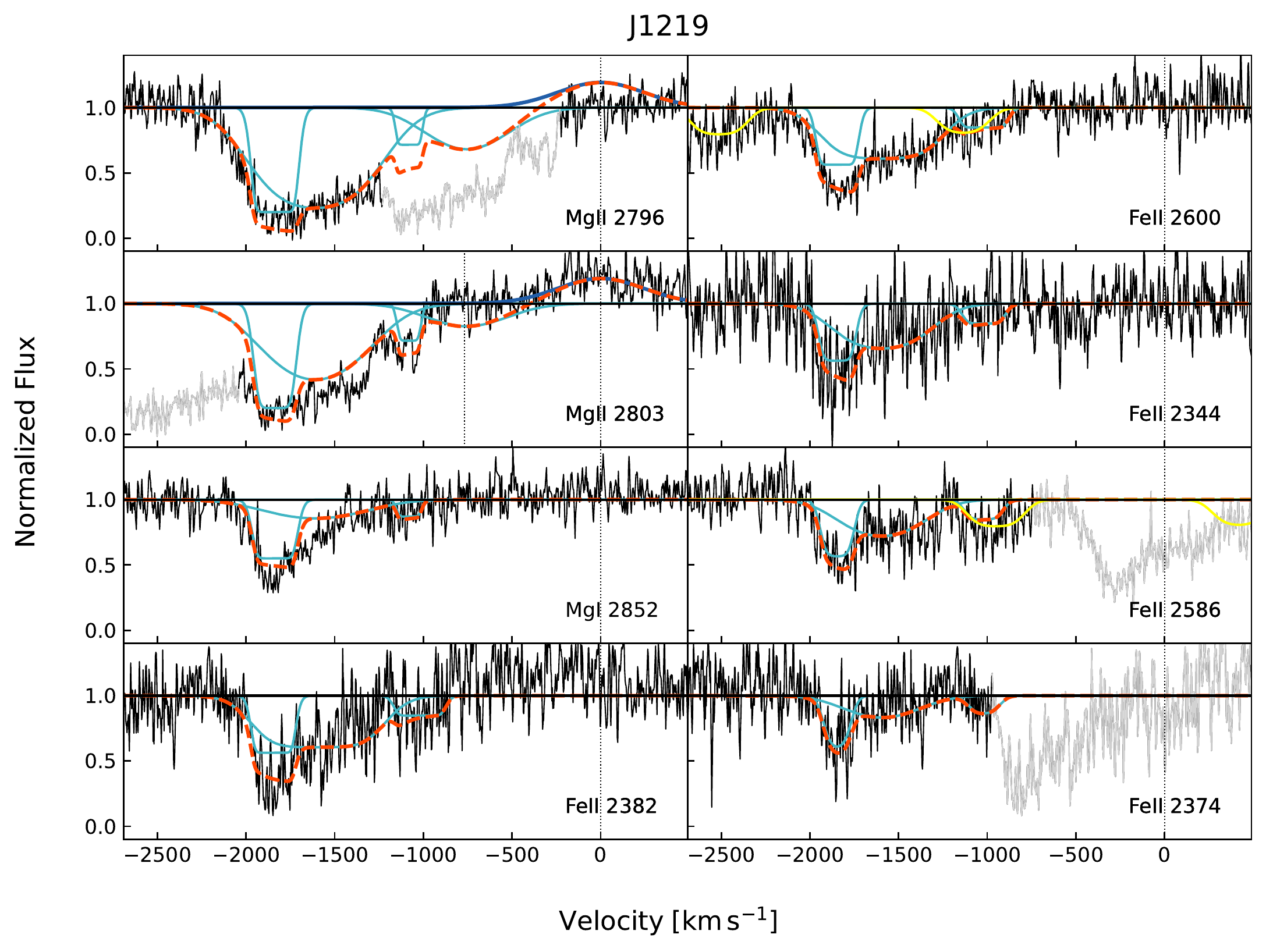}
 \caption{J1219 spectrum, which shows \mgii \ emission within $-600$ and $+600$ \kmps\, of \zsys. In this galaxy, we see strong absorption from \mgii \ falling within $-2400$ and $-150$ \kmps\, of \zsys. The \mgii \ trough is well described by four velocity components. We detect \mgi \ within $-2400$ and $-950$ \kmps\, of \zsys. We model the \mgi \ trough using three components that show a good qualitative alignment with \mgii. We detect \feii \ within $-2400$ and $-750$ \kmps\, of \zsys. We model the \feii \ trough using three components that show good kinematic agreement with \mgii.}
 \label{fig:j1219}
\end{figure*}

Figure~\ref{fig:j1219} shows our best fit to the \mgii, \mgi, and \feii \ absorption lines in the galaxy J1613+2834 (hereafter J1613). J1613 spectrum clearly exhibits redshifted (284 \kmps) \mgd \ emission within $-700$ and $+1200$ \kmps\, of \zsys. The corresponding \mgo \ line is not obvious because of \mgd \ absorption at the same wavelengths. The inclusion of an emission component in the model improves the fit to the \mgii \ absorption trough. In this galaxy, we see strong absorption from \mgii \ falling within $-2570$ and $-50$ \kmps\, of \zsys. The \mgii \ trough is well described by five velocity components that blend together and produce complicated kinematics. We find that \mgii \ traces optically thick gas. The measured equivalent width ratio for \mgo \ and \mgd \ is not consistent ($\sim$25 to 45\% lower) with the optically thin limit. The absorption profile of all five components is well described by \cf$<$ 1. We do not identify any significant absorption trough in the \mgi \ spectral region. We detect weak \feii \ in correspondence with the most blueshifted \mgii \ component. We model the \feii \ trough using only one component. Despite the \feii \ component showing a good alignment with \mgii \ (similar $v$ within the errors), our best fit finds a significantly (44\%) broader absorption profile. We attribute this to the low SNR in the \feii \ spectral region. According to our best fit \feii \ traces optically thin gas. We infer from the \feii \ fit that the corresponding \mgii \ component should be less optically thick, with \tinf = 2.7 and \tmeas = 10. This is probably due to a difference in the kinematics (\bd(\mgii) = 57\% \bd(\feii)) of the two absorption lines that makes the comparison of the \ta\, not accurate, underestimating \tinf. 

\subsection*{J2116$-$0624}
\begin{figure*}[htp!]
 \centering
 \includegraphics[width=0.68\textwidth]{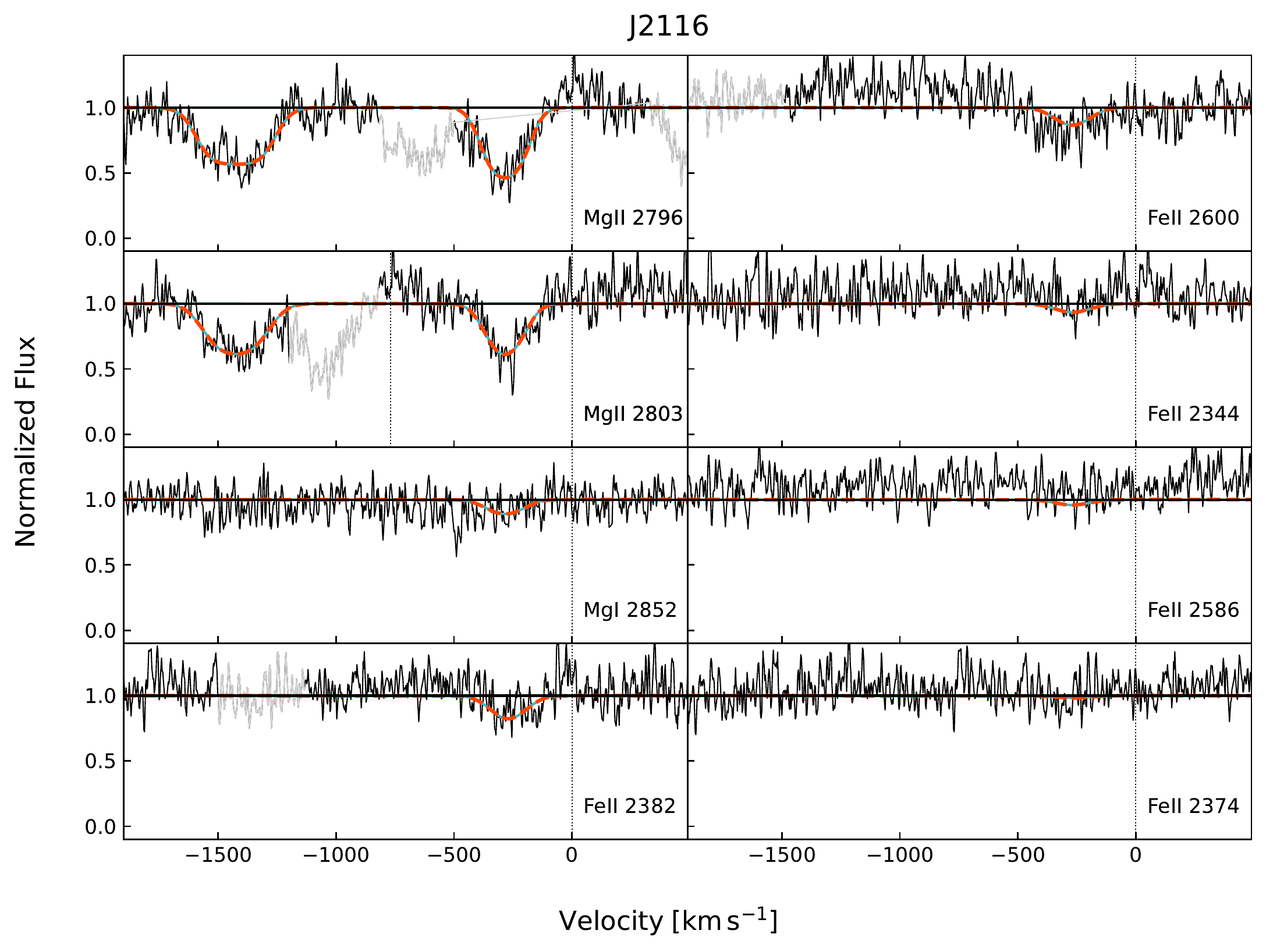}
 \caption{J2116 spectrum, which does not exhibit \mgii \ emission. In this galaxy, we identify two strong \mgii \ absorption troughs at around $-280$ and $-1430$ \kmps\, of \zsys, respectively. We detect weak \mgi \ and \feii \ only in correspondence to the least blueshifted \mgii \ component.}
 \label{fig:j2116}
\end{figure*}

Figure~\ref{fig:j2116} shows our best fit to the \mgii, \mgi, and \feii \ absorption lines in the galaxy J2116$-$0624 (hereafter J2116). J2116 spectrum does not exhibit \mgii \ emission. In this galaxy, we identify two \mgii \ absorption troughs at around $-280$ and $-1430$ \kmps\, of \zsys, respectively. We find that the most blueshifted \mgii \ component ($v=-1428$ \kmps) traces optically thick gas and its profile shape is characterized by \cf$<$1. The least blueshifted \mgii \ component ($v=-284$ \kmps) is close to the transition between optically thin and thick gas, with \ta=0.97. The measured equivalent width ratio for \mgo \ and \mgd \ is $\sim$25\% lower than the theoretical value. Therefore, we conclude that the absorption profile shape is determined by the \cf$<$1 rather than the optical depth. We detect weak \mgi \ and \feii \ in correspondence with the least blueshifted \mgii \ component. We model \mgi \ and \feii \ using only one component. The independent \mgi \ fit finds a substantially broader ($\sim$60\%) absorption profile than the corresponding \mgii. We attribute this discrepancy to the difficulties of identifying such a weak absorption trough. As for previous targets we favor the \mgi \ constrained fit solution because the neutral Mg fraction must be less than one percent to produce an optically thin \mgi$\lambda$2853 trough with \cf=1 as found by the independent fit. The constrained fit finds \mgi \ to be characterized by a smaller \cf than \mgii \ (32\% of \cf(\mgii)). The \feii \ fit shows a very good kinematic agreement with \mgii \ (similar $v$ and \bd within the errors). The \feii \ absorption traces optically thin gas with \cf=1. We find a good agreement between \tmeas\, and \tinf within the errors.

\subsection*{J2140+1209}
\begin{figure*}[htp!]
 \centering
 \includegraphics[width=0.68\textwidth]{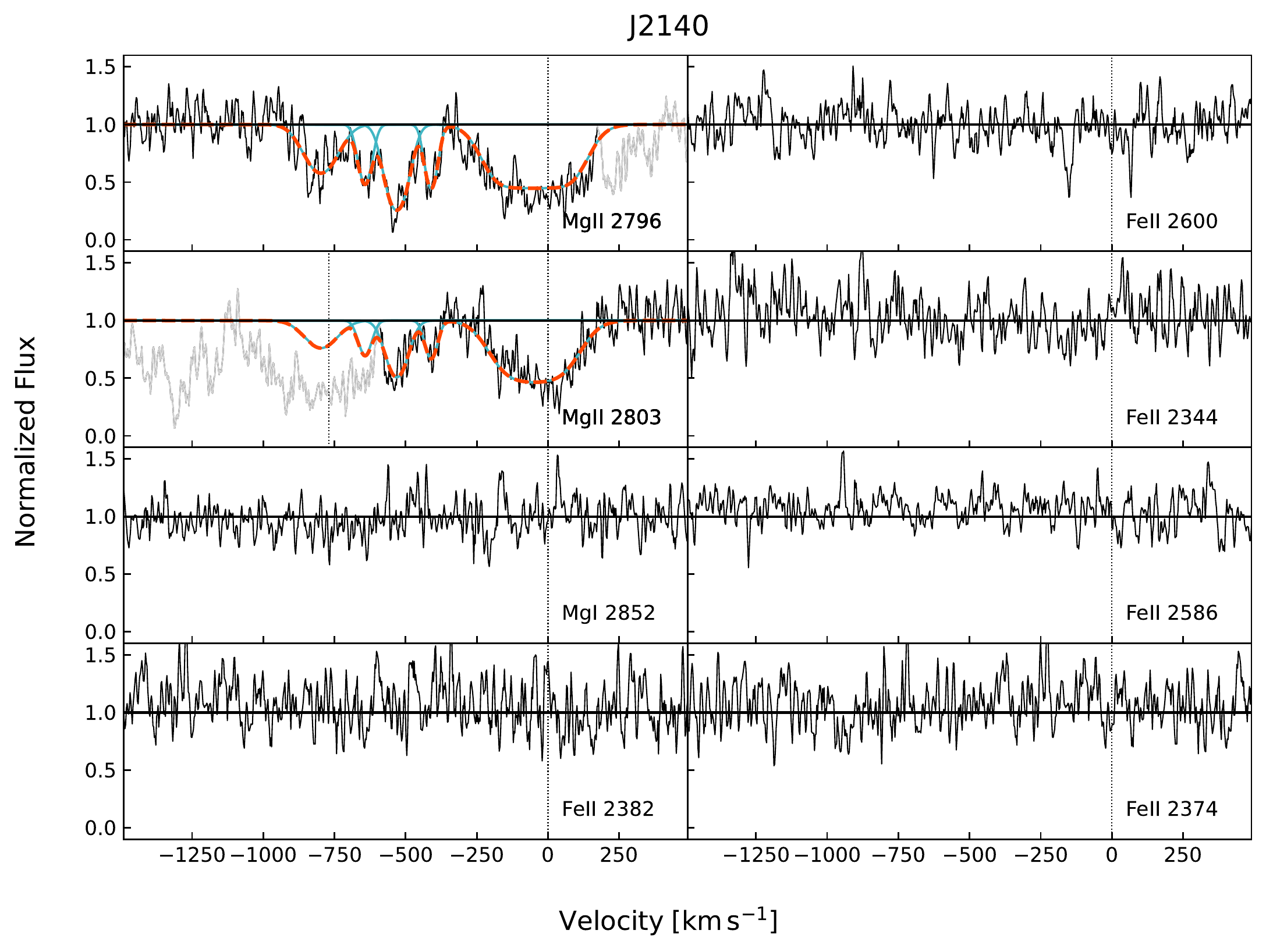}
 \caption{J2140 spectrum, which does not exhibit \mgii \ emission. In this galaxy, we see strong absorption from \mgii \ falling within $-950$ and $+250$ \kmps\, of \zsys. The \mgii \ trough is well described by five velocity components that marginally blend together. We do not detect any \mgi \ and \feii \ absorption lines.}
 \label{fig:j2140}
\end{figure*}

Figure~\ref{fig:j2140} shows our best fit to the \mgii, \mgi, and \feii \ absorption lines in the galaxy J2140+1209 (hereafter J2140). J2140 spectrum does not exhibit \mgii \ emission. In this galaxy, we identify strong \mgii \ absorption troughs falling within $-950$ and $+250$ \kmps\, of \zsys. The \mgii \ trough is well described by five velocity components that marginally blend together (one \mgo \ and three \mgd \ components). We find that \mgii \ traces optically thick gas for the least blueshifted component ($v=-48$ \kmps). The measured equivalent width ratio for \mgo \ and \mgd \ is not consistent ($\sim$40\% lower) with the optically thin limit. The absorption profile of this component is shaped by \cf$<$ 1. The remaining four components trace optically thin gas and their profile shape is well described by the optical depth. We do not identify any significant absorption trough in the \mgi \ and \feii \ spectral regions.

\begin{figure*}[htp!]
 \centering
 \includegraphics[width=0.68\textwidth]{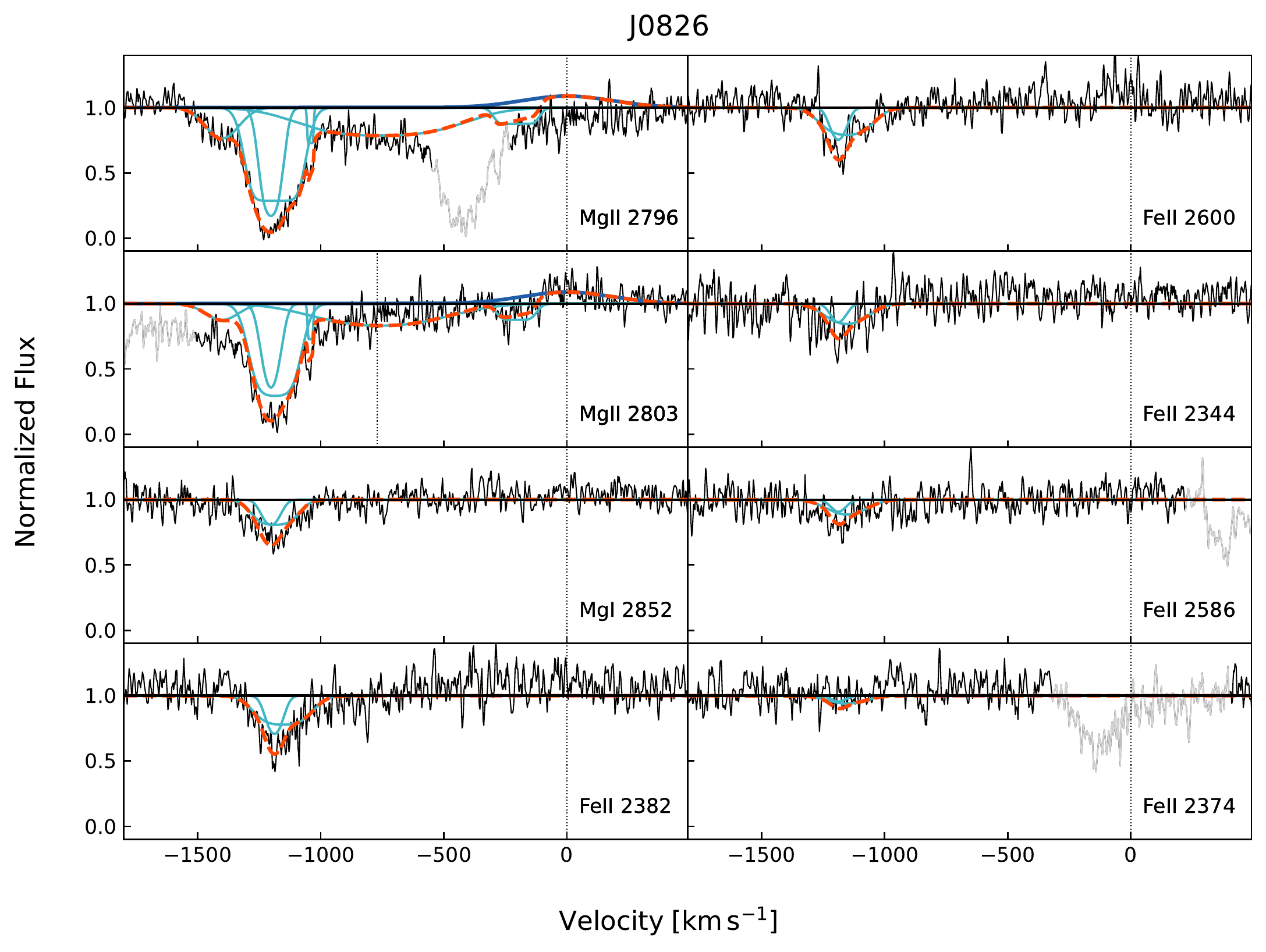}
 \caption{J0826 spectrum, which exhibits \mgii \ emission within $-400$ and $+400$ \kmps\, of \zsys. In this galaxy, we see strong absorption from \mgii \ falling within $-1560$ and $-75$ \kmps\, of \zsys. The \mgii \ trough is well described by eight velocity components which show complex kinematics. We detect \feii \ and \mgi \ absorption within $-1350$ and $-1000$ \kmps\, of \zsys, which we model using two velocity components.}
 \label{fig:j0826}
\end{figure*}

\bibliography{biblio}{}
\bibliographystyle{aasjournal}

\end{document}

%% file: Table1.tex
\begin{deluxetable*}{lrrrccccc}[htp!]
\tabletypesize{\small}
\tablecaption{Sample properties\label{table1}}
\tablehead{
\colhead{ID} & \colhead{\zsys} & 
\colhead{RA} & \colhead{Dec} & 
\colhead{log(M$_*$/M$_\odot$)} & \colhead{r$_e$} & 
\colhead{SFR} & \colhead{$\Sigma_{SFR}$} & \colhead{LW Age}\\
\colhead{} & \colhead{} & \colhead{J2000} & \colhead{J2000} &
\colhead{} & \colhead{kpc} & \colhead{M$_\odot$ yr$^{-1}$} &
\colhead{M$_\odot$ yr$^{-1}$ kpc$^{-2}$} & \colhead{Myr} \\
\colhead{(1)} & \colhead{(2)} & \colhead{(3)} &
\colhead{(4)} & \colhead{(5)} & \colhead{(6)} &
\colhead{(7)} & \colhead{(8)}& \colhead{(9)} }
\startdata
J0826+4305   & 0.603 & 126.66006 & 43.091498 & 10.63$^{+0.2}_{-0.2}$  & 0.173$^{+0.075}_{-0.053}$  & 184$^{+53}_{-41}$  &  981 &  22$^{+11}_{-5}$  \\
J0901+0314   & 0.459 & 135.38926 &  3.236799 & 10.66$^{+0.2}_{-0.2}$  & 0.237$^{+0.144}_{-0.088}$  &  99$^{+39}_{-26}$  &  281 &  54$^{+22}_{-13}$   \\
J0905+5759   & 0.711 & 136.34832 & 57.986791 & 10.69$^{+0.3}_{-0.3}$  & 0.097$^{+0.044}_{-0.033}$  &  90$^{+23}_{-20}$  & 1519 &  17$^{+10}_{-4}$   \\
J0944+0930   & 0.514 & 146.07437 &  9.505385 & 10.59$^{+0.2}_{-0.2}$  & 0.114$^{+0.067}_{-0.047}$  &  88$^{+26}_{-21}$  & 1074 &  88$^{+48}_{-43}$   \\
J1125$-$0145 & 0.519 & 171.32874 & -1.759006 & 11.03$^{+0.2}_{-0.2}$  & 0.600$^{+0.300}_{-0.18}$  & 227$^{+104}_{-68}$  &  100 & 102$^{+27}_{-25}$   \\
J1219+0336   & 0.451 & 184.98241 &  3.604417 & 10.35$^{+0.3}_{-0.2}$  & 0.412$^{+0.194}_{-0.124}$  &  91$^{+28}_{-23}$  &   85 &  22$^{+14}_{-5}$   \\
J1232+0723   & 0.400 & 188.06593 &  7.389085 & 10.93$^{+0.1}_{-0.1}$  & 2.200$^{+1.100}_{-0.660}$  &  62$^{+16}_{-13}$  &    2 &  79$^{+30}_{-27}$   \\
J1341$-$0321 & 0.661 & 205.40333 & -3.357019 & 10.53$^{+0.2}_{-0.1}$  & 0.117$^{+0.040}_{-0.032}$  & 151$^{+34}_{-23}$  & 1755 &  14$^{+6}_{-3}$   \\
J1450+4621   & 0.782 & 222.62024 & 46.360473 & 11.06$^{+0.1}_{-0.1}$  & 0.540$^{+0.270}_{-0.162}$  & 191$^{+146}_{-70}$  &  104 & 107$^{+21}_{-13}$  \\
J1506+5402   & 0.608 & 226.65124 & 54.039095 & 10.60$^{+0.2}_{-0.2}$  & 0.168$^{+0.076}_{-0.054}$  & 116$^{+32}_{-25}$  &  652 &  13$^{+6}_{-2}$   \\
J1558+3957   & 0.402 & 239.54683 & 39.955787 & 10.42$^{+0.3}_{-0.3}$  & 0.778$^{+0.383}_{-0.244}$  &  84$^{+16}_{-15}$  &   22 &  44$^{+14}_{-10}$   \\
J1613+2834   & 0.449 & 243.38552 & 28.570772 & 11.12$^{+0.2}_{-0.2}$  & 0.949$^{+0.274}_{-0.207}$  & 172$^{+36}_{-36}$  &   30 &  72$^{+33}_{-26}$   \\
J2116$-$0624 & 0.728 & 319.10479 & -6.579113 & 10.41$^{+0.2}_{-0.2}$  & 0.284$^{+0.131}_{-0.092}$  & 110$^{+55}_{-27}$  &  216 &  21$^{+13}_{-4}$   \\
J2140+1209$^*$   & 0.752 & 325.00205 & 12.154051 & 10.36$^{+0.2}_{-0.2}$  &  0.153$^{+0.092}_{-0.064}$  &  24$^{+44}_{-10}$  &  163  & 192$^{+6}_{-2}$        
\enddata
\tablecomments{ Column 2: Galaxy systemic redshift. Column 5: Stellar mass from Prospector. Column 6: Effective radii from HST. Column 7: SFRs from Prospector. Column 8: SFR surface densities estimated using columns (6) and (7). Column 9: Light-weighted
ages of the stellar populations younger than 1 Gyr.}
\tablenotetext{*}{While this target has confirmed Type I AGN, its bolometric luminosity is not AGN-dominated. For this galaxy we take the SFR and light-weighted age measurements from our analysis of the UV-optical spectrum with the pPXF software described in Section~\ref{subsection:properties}. See Davis et al., (submitted) for more details.} 
\end{deluxetable*}

%% file: Table2.tex
\begin{deluxetable*}{ccccccccccc}
\tablecaption{\mgii\, and \feii\, best fit \,parameters\label{table2}}
\tabletypesize{\small}
\tablehead{
\multicolumn{1}{c}{}&
\multicolumn{4}{c}{\mgii}&
\multicolumn{1}{c}{$N(\rm H)$}&
\multicolumn{1}{c}{$N(\rm H)$}&
\multicolumn{4}{c}{\feii}\\
 \cline{2-5}  \cline{8-11}
\colhead{ID} & \colhead{$v$} & \colhead{\cf} & \colhead{\bd} & \colhead{\N} &\colhead{$\tau_0(\rm MgII)$}&\colhead{$\tau_0(\rm FeII)$} & \colhead{$v$} &\colhead{\cf} & \colhead{\bd} & \colhead{\N} \\
\colhead{} & \colhead{\footnotesize\kmps} &  & \colhead{\footnotesize\kmps} & \colhead{\footnotesize\cm}& \colhead{\footnotesize\cm} & \colhead{\footnotesize\cm} & \colhead{\footnotesize\kmps} &  & \colhead{\footnotesize\kmps} & \colhead{\footnotesize\cm}}
\startdata
J0826   & $-205^{+7}_{-7}$  &  0.12$^{+0.31}_{-0.21}$  &   55$^{+8}_{-21}$  & 14.50$^{+0.07}_{-0.17}$&        19.60$^{+0.07}_{-0.17}$ &\nodata&   \nodata     & \nodata            & \nodata & \nodata   \\
        & $-757^{+21}_{-19}$  &  0.23$^{+0.04}_{-0.04}$ &  293$^{+48}_{-65}$ & 14.48$^{+0.09}_{-0.11}$&      19.58$^{+0.09}_{-0.11}$ &\nodata&  \nodata      & \nodata            & \nodata & \nodata \\
        & $-1044^{+2}_{-1}$ &  0.27$^{+0.65}_{-0.66}$  &    8$^{+2}_{-4}$   & 13.77$^{+0.13}_{-0.22}$&        18.87$^{+0.13}_{-0.22}$ &\nodata&   \nodata     & \nodata            & \nodata & \nodata   \\
        & $-1187^{+2}_{-2}$ &  0.71$^{+0.39}_{-0.61}$  &   80$^{+6}_{-7}$   & 14.46$^{+0.04}_{-0.04}$&        19.56$^{+0.04}_{-0.04}$ &20.00&   $-1156^{+78}_{-14}$  & 0.23$^{+0.02}_{-0.03}$ &  100$^{+13}_{-45}$ & 14.40$^{+0.60}_{-0.31}$ \\
        & $-1204^{+2}_{-2}$ &  0.90$^{+0.13}_{-0.23}$  &   44$^{+6}_{-6}$   & 13.63$^{+0.08}_{-0.09}$&        18.73$^{+0.08}_{-0.09}$ &19.30&    $-1190^{+6}_{-5}$  & 0.44$^{+0.02}_{-0.02}$ &  41$^{+16}_{-7}$   & 13.57$^{+0.36}_{-0.22}$   \\
        & $-1404^{+8}_{-12}$ &  1.00$^{+0.01}_{-0.04}$  &   92$^{+14}_{-20}$ & 12.98$^{+0.26}_{-2.30}$&       18.08$^{+0.26}_{-2.30}$ &\nodata&  \nodata      & \nodata             & \nodata & \nodata   \\
\midrule             
J0901   &  $-591^{+31}_{-22}$  &  1.00$^{+0.01}_{-0.04}$ & 271$^{+55}_{-95}$ & 13.28$^{+1.18}_{-4.72}$&       18.38$^{+1.18}_{-4.72}$ &\nodata&   \nodata    & \nodata             & \nodata &  \nodata  \\
        & $-1254^{+10}_{-9}$ &  0.32$^{+0.08}_{-0.14}$  &  229$^{+19}_{-19}$ & 14.51$^{+0.05}_{-0.06}$&       19.61$^{+0.05}_{-0.06}$ &\nodata&   \nodata    & \nodata             & \nodata &  \nodata  \\
        & $-1266^{+1}_{-1}$ &  0.72$^{+0.21}_{-0.33}$ &   48$^{+3}_{-3}$    & 14.11$^{+0.03}_{-0.03}$ &       19.21$^{+0.03}_{-0.03}$ &19.68 &    $-1252^{+3}_{-3}$  & 0.38$^{+0.02}_{-0.02}$ & 42$^{+6}_{-7}$    & 13.92$^{+0.15}_{-0.14}$ \\
        & $-1426^{+3}_{-3}$ &  0.42$^{+0.11}_{-0.19}$  &   61$^{+6}_{-6}$   & 13.95$^{+0.06}_{-0.06}$ &       19.05$^{+0.06}_{-0.06}$ &18.80 &    $-1352^{+29}_{-21}$  & 0.97$^{+0.01}_{-0.01}$ & 132$^{+32}_{-18}$ & 13.43$^{+0.51}_{-4.71}$ \\
\midrule
J0905   & $-2362^{+1}_{-1}$ &  0.87$^{+0.38}_{-0.50}$  &   82$^{+3}_{-3}$    & 14.59$^{+0.02}_{-0.02}$&       19.69$^{+0.02}_{-0.02}$ &20.81&   $-2360^{+1}_{-1}$  & 0.78$^{+0.10}_{-0.11}$ &  63$^{+1}_{-1}$ & 15.01$^{+0.01}_{-0.01}$ \\
        & $-2480^{+3}_{-3}$ &  0.56$^{+0.13}_{-0.16}$  &  196$^{+4}_{-4}$    & 14.82$^{+0.01}_{-0.01}$&       19.92$^{+0.01}_{-0.01}$ &21.12&   $-2436^{+4}_{-4}$  & 0.23$^{+0.07}_{-0.08}$ & 157$^{+5}_{-5}$ & 15.29$^{+0.02}_{-0.02}$ \\
\midrule
J0944   &  $-410^{+4}_{-4}$  &  1.00$^{+0.01}_{-0.01}$  &  135$^{+7}_{-8}$   & 13.54$^{+0.04}_{-0.04}$&       18.64$^{+0.04}_{-0.04}$&\nodata&  \nodata    & \nodata              & \nodata &  \nodata  \\
        &  $-843^{+3}_{-2}$  &  1.00$^{+0.02}_{-0.02}$  &   58$^{+6}_{-6}$   & 13.31$^{+0.06}_{-0.06}$&       18.41$^{+0.06}_{-0.06}$& 18.96 &   $-854^{+4}_{-4}$  & 1.00$^{+0.01}_{-0.01}$ &  70$^{+6}_{-6}$     & 13.35$^{+0.18}_{-0.18}$ \\
        & $-1072^{+1}_{-2}$ &  1.00$^{+0.13}_{-0.17}$  &  282$^{+10}_{-10}$  & 14.09$^{+0.02}_{-0.02}$&       18.77$^{+0.04}_{-0.04}$&\nodata&  \nodata    & \nodata              & \nodata &  \nodata  \\
        & $-1125^{+6}_{-6}$ &  1.00$^{+0.01}_{-0.01}$  &   35$^{+3}_{-3}$    & 13.67$^{+0.04}_{-0.04}$&       19.19$^{+0.02}_{-0.02}$& 19.80 &   $-1128^{+3}_{-3}$  & 1.00$^{+0.01}_{-0.01}$ &  69$^{+4}_{-5}$    & 13.49$^{+0.10}_{-0.10}$ \\
        & $-1718^{+7}_{-8}$ &  1.00$^{+0.01}_{-0.01}$  &  225$^{+13}_{-17}$  & 13.55$^{+0.05}_{-0.05}$&       18.65$^{+0.05}_{-0.05}$& 19.16 &   $-1655^{+24}_{-24}$  & 1.00$^{+0.01}_{-0.01}$ & 331$^{+32}_{-43}$  & 13.63$^{+0.56}_{-0.51}$ \\
\midrule
J1125   & $-1920^{+2}_{-2}$ &  0.47$^{+0.10}_{-0.16}$  &   31$^{+2}_{-2}$    & 13.77$^{+0.04}_{-0.04}$&       18.87$^{+0.04}_{-0.04}$ & 19.45&   $-1917^{+2}_{-2}$  & 0.24$^{+0.01}_{-0.01}$ & 32$^{+3}_{-3}$ &  13.76$^{+0.21}_{-0.14}$ \\
        & $-2051^{+2}_{-2}$ &  0.49$^{+0.12}_{-0.17}$  &   49$^{+3}_{-2}$    & 14.10$^{+0.03}_{-0.02}$&       19.20$^{+0.03}_{-0.02}$ & 20.03&   $-2052^{+2}_{-2}$  & 0.24$^{+0.02}_{-0.02}$ & 44$^{+2}_{-2}$ &  14.28$^{+0.03}_{-0.04}$ \\
\midrule
J1219   &  $-755^{+14}_{-15}$ &  1.00$^{+0.01}_{-0.04}$&  307$^{+41}_{-42}$ & 13.66$^{+0.33}_{-1.23}$ &      18.76$^{+0.33}_{-1.23}$&\nodata&   \nodata    & \nodata             & \nodata & \nodata   \\
        & $-1083^{+3}_{-3}$ &  0.28$^{+0.68}_{-0.21}$  &   43$^{+4}_{-8}$   & 14.53$^{+0.04}_{-0.08}$ &      19.63$^{+0.04}_{-0.08}$& 20.59 &  $-1022^{+9}_{-8}$  & 0.15$^{+0.18}_{-0.31}$ & 81$^{+11}_{-11}$ & 15.17$^{+0.05}_{-0.07}$ \\
        & $-1613^{+6}_{-6}$ &  0.86$^{+0.05}_{-0.08}$  &  344$^{+14}_{-14}$ & 14.49$^{+0.03}_{-0.04}$ &      19.59$^{+0.03}_{-0.04}$& 20.89 &  $-1598^{+12}_{-13}$  & 0.40$^{+0.02}_{-0.03}$ & 267$^{+11}_{-11}$ & 15.09$^{+0.03}_{-0.04}$ \\
        & $-1842^{+1}_{-1}$ &  0.80$^{+0.01}_{-0.01}$  &   76$^{+5}_{-5}$   & 14.78$^{+0.10}_{-0.10}$ &      19.88$^{+0.00}_{-0.00}$& 20.84 &  $-1852^{+3}_{-3}$  & 0.44$^{+0.24}_{-0.37}$ &  70$^{+4}_{-5}$   & 15.11$^{+0.03}_{-0.04}$ \\
\midrule
J1232   &  $386^{+1}_{-1}$  &  1.00$^{+0.02}_{-0.04}$  &  31$^{+3}_{-3}$  & 13.02$^{+0.07}_{-0.18}$ &        18.12$^{+0.07}_{-0.18}$ & \nodata&   \nodata    & \nodata           & \nodata &  \nodata  \\
        &  $128^{+1}_{-1}$  &  1.00$^{+0.03}_{-0.04}$  &  40$^{+3}_{-3}$  & 13.28$^{+0.05}_{-0.09}$ &        18.38$^{+0.05}_{-0.09}$ & 19.49  &  $128^{+3}_{-3}$    & 0.46$^{+0.03}_{-0.04}$ &  37$^{+5}_{-5}$ & 13.75$^{+0.46}_{-0.28}$ \\
        &  $-67^{+2}_{-2}$  &  1.00$^{+0.20}_{-0.24}$  &  80$^{+4}_{-4}$  & 14.41$^{+0.02}_{-0.02}$ &        19.51$^{+0.02}_{-0.02}$ & 20.19  &  $-74^{+3}_{-4}$    & 0.71$^{+0.03}_{-0.04}$ &  82$^{+4}_{-4}$ & 14.51$^{+0.03}_{-0.03}$ \\
        & $-247^{+3}_{-3}$  &  0.82$^{+0.15}_{-0.22}$  &  64$^{+6}_{-7}$  & 14.17$^{+0.04}_{-0.05}$ &        19.27$^{+0.04}_{-0.05}$ & 19.80  &  $-249^{+5}_{-5}$   & 0.56$^{+}0.03_{-0.03}$ &  54$^{+6}_{-7}$ & 14.04$^{+0.14}_{-0.13}$ \\
        & $-431^{+7}_{-7}$  &  0.23$^{+0.62}_{-0.24}$  &  45$^{+6}_{-13}$ & 14.54$^{+0.06}_{-0.12}$ &        19.64$^{+0.06}_{-0.12}$ & \nodata&   \nodata    & \nodata           & \nodata &  \nodata  \\
        & $-642^{+4}_{-4}$  &  0.49$^{+0.11}_{-0.14}$  & 119$^{+8}_{-9}$  & 14.34$^{+0.03}_{-0.04}$ &        19.44$^{+0.03}_{-0.04}$ & \nodata&   \nodata    & \nodata           & \nodata &  \nodata  \\
\midrule            
J1341   & $  221^{+1}_{-1}$ &  0.98$^{+0.05}_{-0.08}$  &   8$^{+1}_{-1}$  & 12.49$^{+0.08}_{-0.14}$ &        17.59$^{+0.08}_{-0.14}$ & \nodata&   \nodata    & \nodata            & \nodata &  \nodata  \\
        & $  160^{+1}_{-1}$ &  0.48$^{+0.14}_{-0.32}$  &   8$^{+1}_{-1}$  & 13.03$^{+0.07}_{-0.08}$ &        18.13$^{+0.07}_{-0.08}$ & \nodata&   \nodata    & \nodata            & \nodata &  \nodata  \\
        & $  108^{+1}_{-1}$ &  0.48$^{+0.04}_{-0.09}$  &   8$^{+1}_{-2}$  & 12.61$^{+0.29}_{-0.25}$ &        17.71$^{+0.29}_{-0.25}$ & \nodata&   \nodata    & \nodata            & \nodata &  \nodata  \\
        & $ -395^{+2}_{-8}$ &  0.71$^{+0.12}_{-0.21}$  &  77$^{+2}_{-2}$  & 14.37$^{0.01+}_{-0.01}$ &        20.09$^{+0.03}_{-0.03}$ & 20.68  &  $ -395^{+1}_{-1}$ & 0.44$^{+0.01}_{-0.01}$ &  78$^{+2}_{-2}$   & 14.45$^{+0.01}_{-0.01}$ \\ 
        & $ -525^{+1}_{-1}$ &  0.25$^{+0.12}_{-0.14}$  & 263$^{+20}_{-19}$& 14.99$^{+0.03}_{-0.03}$ &        19.47$^{+0.01}_{-0.01}$ & \nodata&   \nodata    & \nodata  & \nodata &  \nodata  \\
        & $ -566^{+1}_{-1}$ &  0.58$^{+0.04}_{-0.04}$  &  45$^{+3}_{-3}$  & 13.59$^{+0.04}_{-0.04}$ &        18.69$^{+0.04}_{-0.04}$ & 19.98  &  $ -566^{+3}_{-3}$ & 0.14$^{+0.01}_{-0.01}$ &  39$^{+3}_{-3}$   & 14.22$^{+0.05}_{-0.05}$ \\
        & $ -903^{+2}_{-3}$ &  0.44$^{+0.16}_{-0.21}$  &  93$^{+5}_{-5}$  & 14.59$^{+0.02}_{-0.02}$ &        19.69$^{+0.02}_{-0.02}$ & 20.08   &  $ -916^{+2}_{-2}$ & 0.34$^{+0.01}_{-0.01}$ & 163$^{+3}_{-3}$   & 14.62$^{+0.02}_{-0.02}$ \\ 
        & $-1138^{+4}_{-6}$ &  0.29$^{+0.10}_{-0.18}$  &  62$^{+7}_{-8}$  & 14.06$^{+0.06}_{-0.06}$ &        19.16$^{+0.06}_{-0.06}$ & \nodata&  \nodata    & \nodata             & \nodata &  \nodata  \\
        & $-1922^{+7}_{-6}$ &  0.30$^{+0.01}_{-0.02}$  & 234$^{+17}_{-18}$& 13.79$^{+0.84}_{-0.39}$ &        18.89$^{+0.84}_{-0.39}$ & \nodata&   \nodata    & \nodata            & \nodata &  \nodata  \\
\midrule
J1450   & $  189^{+6}_{-5}$ &  0.15$^{+0.26}_{-0.01}$ &  41$^{+5}_{-12}$   & 15.20$^{+0.05}_{-0.13}$ &      20.30$^{+0.05}_{-0.13}$ & \nodata&   \nodata    & \nodata           & \nodata &  \nodata  \\
        & $  -50^{+1}_{-1}$ &  0.95$^{+0.98}_{-1.33}$ &  38$^{+2}_{-2}$    & 14.36$^{+0.03}_{-0.03}$ &      19.46$^{+0.03}_{-0.03}$ & 20.44  &   $-45^{+1}_{-1}$  & 1.00$^{+0.18}_{-0.20}$ & 28$^{+1}_{-1}$  & 14.56$^{+0.01}_{-0.01}$ \\
        & $ -201^{+5}_{-4}$ &  1.00$^{+0.02}_{-0.02}$ &  81$^{+11}_{-16}$  & 13.45$^{+0.07}_{-0.10}$ &      18.55$^{+0.07}_{-0.10}$ & \nodata&   \nodata    & \nodata           & \nodata &  \nodata  \\ 
        & $ -309^{+2}_{-2}$ &  1.00$^{+0.04}_{-0.04}$ &  21$^{+4}_{-5}$    & 12.91$^{+0.10}_{-0.11}$ &      18.01$^{+0.10}_{-0.11}$ & \nodata&   \nodata    & \nodata           & \nodata &  \nodata  \\
        & $ -435^{+4}_{-4}$ &  1.00$^{+0.01}_{-0.01}$ &  90$^{+7}_{-7}$    & 13.47$^{+0.05}_{-0.05}$ &      18.57$^{+0.05}_{-0.05}$ & \nodata&   \nodata    & \nodata           & \nodata &  \nodata  \\
        & $ -780^{+18}_{-16}$ & 1.00$^{+0.01}_{-0.02}$ & 121$^{+24}_{-29}$ & 13.24$^{+0.20}_{-0.19}$ &      18.34$^{+0.20}_{-0.19}$ & \nodata&   \nodata    & \nodata           & \nodata &  \nodata  \\
        & $ -973^{+11}_{-10}$ & 1.00$^{+0.03}_{-0.03}$ &  68$^{+28}_{-33}$ & 13.14$^{+0.25}_{-0.26}$ &      18.24$^{+0.25}_{-0.26}$ & \nodata&   \nodata    & \nodata           & \nodata &  \nodata  \\
        & $-1103^{+30}_{-61}$ & 1.00$^{+0.02}_{-0.02}$ &  93$^{+38}_{-60}$ & 12.87$^{+0.69}_{-0.67}$ &      17.97$^{+0.69}_{-0.67}$ & \nodata&   \nodata    & \nodata           & \nodata &  \nodata  \\
        & $-1624^{+4}_{-4}$ &  1.00$^{+0.01}_{-0.01}$ & 164$^{+6}_{-6}$    & 13.75$^{+0.03}_{-0.03}$ &      18.85$^{+0.03}_{-0.03}$ & \nodata&   \nodata    & \nodata           & \nodata &  \nodata  \\
        & $-1728^{+2}_{-2}$ &  1.00$^{+0.04}_{-0.04}$ &  30$^{+4}_{-5}$    & 13.04$^{+0.09}_{-0.10}$ &      18.14$^{+0.09}_{-0.10}$ & \nodata&   \nodata    & \nodata           & \nodata &  \nodata  \\
\midrule
J1506   & $ -519^{+9}_{-9}$ &  1.00$^{+0.01}_{-0.01}$  & 351$^{+18}_{-20}$  & 13.49$^{+0.08}_{-0.07}$ &      18.59$^{+0.08}_{-0.07}$ & \nodata &   \nodata    & \nodata           & \nodata &  \nodata  \\
        & $-1011^{+4}_{-5}$ &  1.00$^{+0.02}_{-0.01}$  & 127$^{+12}_{-9}$   & 13.42$^{+0.10}_{-0.08}$ &      18.52$^{+0.10}_{-0.08}$ & 19.02   &  $-1033^{+6}_{-6}$ & 1.00$^{+0.01}_{-0.01}$ & 159$^{+8}_{-9}$  & 13.42$^{+0.19}_{-0.17}$ \\
        & $-1256^{+40}_{-47}$ &  1.00$^{+0.03}_{-0.04}$& 206$^{+82}_{-126}$ & 13.29$^{+1.43}_{-0.82}$ &      18.39$^{+1.43}_{-0.82}$ & \nodata &   \nodata    & \nodata           & \nodata &  \nodata  \\
        & $-1514^{+89}_{-40}$ &  1.00$^{+0.03}_{-0.01}$& 203$^{+91}_{-50}$  & 13.15$^{+1.21}_{-0.20}$ &      18.25$^{+1.21}_{-0.20}$ & 19.07   &  $-1447^{+7}_{-8}$ & 1.00$^{+0.01}_{-0.01}$ & 126$^{+10}_{-11}$& 13.17$^{+0.44}_{-0.43}$ \\
        & $-1897^{+3}_{-3}$ &  1.00$^{+0.01}_{-0.01}$  &  76$^{+5}_{-6}$    & 12.86$^{+0.10}_{-0.11}$ &      17.96$^{+0.10}_{-0.11}$ & \nodata &   \nodata    & \nodata           & \nodata &  \nodata  \\
\midrule      
J1558   & $ -650^{+14}_{-35}$ &  0.23$^{+0.17}_{-0.26}$ &  219$^{+30}_{-23}$ & 14.98$^{+0.06}_{-0.05}$&      20.08$^{+0.06}_{-0.05}$ & \nodata &   \nodata    & \nodata                  & \nodata &  \nodata  \\
        & $ -809^{+3}_{-2}$ &  0.68$^{+0.05}_{-0.09}$ &   84$^{+3}_{-3}$    & 14.00$^{+0.02}_{-0.03}$ &      19.10$^{+0.02}_{-0.03}$ & 20.19   &  $-824^{+4}_{-4}$  & 0.35$^{+0.01}_{-0.02}$ & 78$^{+6}_{-74}$ & 14.05$^{+0.24}_{-0.19}$ \\
        & $ -982^{+2}_{-2}$ &  0.34$^{+0.52}_{-0.22}$ &   41$^{+2}_{-4}$    & 14.50$^{+0.02}_{-0.04}$ &      19.60$^{+0.02}_{-0.04}$ & \nodata &   \nodata    & \nodata                  & \nodata &  \nodata  \\
        & $-1188^{+5}_{-5}$ &  1.00$^{+0.01}_{-0.02}$ &   90$^{+11}_{-12}$  & 12.93$^{+0.29}_{-2.14}$ &      18.03$^{+0.29}_{-2.14}$ & \nodata &   \nodata    & \nodata                  & \nodata &  \nodata  \\
\midrule
J1613   & $ -159^{+2}_{-2}$ &  0.48$^{+0.05}_{-0.06}$  &   51$^{+4}_{-4}$    & 13.67$^{+0.07}_{-0.06}$&       18.77$^{+0.07}_{-0.06}$ &\nodata & \nodata    & \nodata              & \nodata &  \nodata\\  
        & $ -415^{+2}_{-2}$ &  0.47$^{+0.08}_{-0.13}$  &   77$^{+5}_{-5}$    & 14.14$^{+0.03}_{-0.04}$&       19.24$^{+0.03}_{-0.04}$ &\nodata & \nodata    & \nodata              & \nodata &  \nodata \\ 
        & $ -707^{+7}_{-8}$ &  0.23$^{+0.07}_{-0.25}$  &  122$^{+24}_{-23}$  & 14.19$^{+0.12}_{-0.14}$&       19.29$^{+0.12}_{-0.14}$ &\nodata & \nodata    & \nodata              & \nodata &  \nodata  \\
        & $-1667^{+8}_{-8}$ &  0.26$^{+0.12}_{-0.21}$  &  227$^{+19}_{-19}$  & 14.73$^{+0.04}_{-0.05}$&       19.83$^{+0.04}_{-0.05}$ &\nodata & \nodata    & \nodata              & \nodata &  \nodata  \\
        & $-2402^{+3}_{-3}$ &  0.24$^{+0.18}_{-0.13}$  &   71$^{+4}_{-9}$    & 14.75$^{+0.03}_{-0.06}$&       19.85$^{+0.03}_{-0.06}$ & 19.29  & $-2390^{+8}_{-8}$  & 0.56$^{+0.01}_{-0.02}$ & 125$^{+12}_{-13}$ & 13.84$^{+0.98}_{-0.97}$ \\
\midrule
J2116   & $ -284^{+2}_{-2}$ &  0.63$^{+0.04}_{-0.04}$  &   96$^{+5}_{-5}$  & 13.86$^{+0.06}_{-0.05}$  &       18.96$^{+0.06}_{-0.05}$ & 18.88  &  $ -270^{+11}_{-10}$ & 1.00$^{+0.01}_{-0.01}$ & 103$^{+14}_{-17}$ & 13.21$^{+2.75}_{-6.47}$ \\
        & $-1428^{+3}_{-3}$ &  0.44$^{+0.06}_{-0.09}$  &  126$^{+6}_{-5}$  & 14.32$^{+0.03}_{-0.03}$  &       19.42$^{+0.03}_{-0.03}$ & \nodata&   \nodata    & \nodata           & \nodata &  \nodata  \\
\midrule
J2140   & $ -48^{+3}_{-3}$ &  0.55$^{+0.19}_{-0.28}$  &  126$^{+6}_{-7}$  & 14.52$^{+0.03}_{-0.03}$  &        19.62$^{+0.03}_{-0.03}$ &\nodata &   \nodata    & \nodata  & \nodata &  \nodata  \\
        & $-411^{+2}_{-2}$ &  1.00$^{+0.03}_{-0.07}$  &   29$^{+3}_{-4}$  & 12.96$^{+0.11}_{-0.32}$  &        18.06$^{+0.11}_{-0.32}$ &\nodata &   \nodata    & \nodata  & \nodata &  \nodata  \\
        & $-532^{+2}_{-2}$ &  1.00$^{+0.03}_{-0.06}$  &   50$^{+4}_{-4}$  & 13.43$^{+0.04}_{-0.09}$  &        18.53$^{+0.04}_{-0.09}$ &\nodata &   \nodata    & \nodata  & \nodata &  \nodata  \\
        & $-643^{+2}_{-2}$ &  1.00$^{+0.04}_{-0.14}$  &   32$^{+5}_{-6}$  & 12.94$^{+0.16}_{-0.60}$  &        18.04$^{+0.16}_{-0.60}$ &\nodata &   \nodata    & \nodata  & \nodata &  \nodata  \\
        & $-798^{+4}_{-4}$ &  1.00$^{+0.02}_{-0.09}$  &   78$^{+10}_{-9}$ & 13.22$^{+0.14}_{-0.83}$  &        18.32$^{+0.14}_{-0.83}$ &\nodata &   \nodata    & \nodata  & \nodata &  \nodata  \\
\enddata
\tablecomments{ $v$: line centroid; \cf: covering fraction; \bd: Doppler parameter; \N: column density; N(H)$\tau_0(\rm MgII)$: Hydrogen column density from Eq.~\ref{nh}, i.e. using \ta(\mgii); N(H)$\tau_0(\rm FeII)$: Hydrogen column density inferred using bounds on \ta(\mgii) from Eq.~\ref{tau_mgii}.}
\end{deluxetable*}

%% file: Table3.tex
\begin{deluxetable*}{cccc|ccc}[htp!]
\tabletypesize{\small}
\tablecaption{\mgi \ best fit parameters\label{table3}}
\tablehead{
\colhead{ID} & \colhead{$v$ (\mgii)} & \colhead{\cf(\mgi)} & \colhead{\N(\mgi)}&
\colhead{$v$ (\mgi)} & \colhead{\bd (\mgi)} & \colhead{\N(\mgi)}  \\
\colhead{} & \colhead{\footnotesize\kmps}  & \colhead{} & \colhead{\footnotesize\cm}& \colhead{\footnotesize\kmps} &
\colhead{\footnotesize\kmps} & \colhead{\footnotesize\cm} \\
\colhead{(1)} & \colhead{(2)} & \colhead{(3)} &
\colhead{ (4)} & \colhead{(5)} & \colhead{(6)}& \colhead{(7)}  }
\startdata
J0826+4305   & $-1187$ &  0.20$^{+0.01}_{-0.01}$   & 13.53$^{+0.04}_{-0.04}$ &  $-1177^{+5}_{-3}$  &  $179^{+42}_{-42}$  & $12.72^{+0.05}_{-0.05}$ \\
             & $-1204$ &  0.30$^{+0.03}_{-0.03}$    & 12.78$^{+0.05}_{-0.05}$ &  $-1212^{+4}_{-4}$  &  $40^{+13}_{-13}$  & $12.01^{+0.21}_{-0.21}$ \\
\midrule
J0901+0314   & $-1266$ &  0.31$^{+0.02}_{-0.02}$   & 13.37$^{+0.05}_{-0.05}$ &  $-1267^{+3}_{-2}$  &  $60^{+12}_{-12}$  & $12.22^{+0.12}_{-0.12}$ \\
             & $-1426$ &  0.23$^{+0.03}_{-0.03}$   & 13.31$^{+0.03}_{-0.03}$ &  $-1421^{+7}_{-7}$ &  $220^{+42}_{-42}$ & $12.64^{+0.07}_{-0.07}$ \\
\midrule
J0905+5759   & $-2362$ &  0.58$^{+0.02}_{-0.02}$   & 14.05$^{+0.03}_{-0.03}$ &   $-2311^{+2}_{-2} $  &  $94^{+6}_{-6}$   & $13.23^{+0.042}_{-0.04}$ \\
             & $-2480$ &  0.22$^{+0.01}_{-0.01}$   & 14.04$^{+0.03}_{-0.03}$ &   $-2435^{+3}_{-2} $ &  $235^{+32}_{-32}$ & $12.91^{+0.093}_{-0.09}$ \\
\midrule
J0944+0930   &  $ -410$ &  0.41$^{+0.03}_{-0.03}$   & 12.97$^{+0.04}_{-0.04}$ &  $-436^{+3}_{-2} $  &  $154^{+34}_{-34}$ & $12.36^{+0.08}_{-0.08}$ \\
             &  $ -843$ &  0.47$^{+0.03}_{-0.03}$   & 12.86$^{+0.06}_{-0.06}$ &  $-863^{+2}_{-4} $   &  $39^{+7}_{-7}$  & $12.17^{+0.08}_{-0.08}$ \\
             & $-1125$ &  0.32$^{+0.01}_{-0.01}$  & 12.77$^{+0.05}_{-0.05}$ & $-1143^{+5}_{-4} $ &  $43^{+11}_{-11}$  & $12.02^{+0.16}_{-0.16}$ \\   
             & $-1072$ &  0.39$^{+0.02}_{-0.02}$  & 13.20$^{+0.04}_{-0.04}$ & $-1173^{+8}_{-6} $ &  $248^{+37}_{-37}$ & $12.89^{+0.04}_{-0.04}$ \\  
             & $-1718$ &  0.56$^{+0.02}_{-0.02}$   & 13.01$^{+0.04}_{-0.04}$ & $-1711^{+12}_{-14} $  &  $213^{+37}_{-37}$  & $12.61^{+0.07}_{-0.07}$ \\
\midrule
J1125$-$0145 & $-1920$ &  0.20$^{+0.01}_{-0.01}$   & 12.92$^{+0.03}_{-0.03}$ & $-1910^{+1}_{-2} $   &  $ 35^{+6}_{-6}$  & $11.97^{+0.07}_{-0.07}$ \\
             & $-2051$ &  0.23$^{+0.01}_{-0.01}$   & 13.30$^{+0.02}_{-0.02}$ & $-2048^{+2}_{-2} $   &  $ 65^{+8}_{-8}$  & $12.37^{+0.04}_{-0.04}$ \\
\midrule
J1219+0336   & $-1083$ &  0.13$^{+0.02}_{-0.02}$   & 13.81$^{+0.01}_{-0.01}$ & $-1156^{+2}_{-2} $  &  $203^{+49}_{-49}$ & $12.62^{+0.08}_{-0.08}$ \\
             & $-1613$ &  0.25$^{+0.02}_{-0.02}$   & 13.57$^{+0.02}_{-0.02}$ & $-1761^{+6}_{-7} $  &  $283^{+2}_{-2}$  & $13.34^{+0.02}_{-0.02}$ \\
             & $-1842$ &  0.45$^{+0.03}_{-0.03}$   & 14.15$^{+0.02}_{-0.02}$ & $-1863^{+3}_{-3} $   &  $71^{+10}_{-10}$ & $12.73^{+0.07}_{-0.07}$ \\
\midrule
J1232+0723   &   $128$ &  0.33$^{+0.03}_{-0.03}$   & 12.42$^{+0.04}_{-0.04}$ &  $137^{+2}_{-3} $  &  $ 70^{+6}_{-6} $ & $12.03^{+0.10}_{-0.10}$ \\
             &   $-67$ &  0.16$^{+0.02}_{-0.02}$   & 13.24$^{+0.03}_{-0.03}$ &  $-84^{+3}_{-3} $  &  $ 64^{+18}_{-18}$  & $12.33^{+0.10}_{-0.10}$ \\
             &   $-247$ &  0.24$^{+0.08}_{-0.08}$  & 13.26$^{+0.03}_{-0.03}$ &  $-263^{+3}_{-4}$  &  $ 86^{+19}_{-19}$ & $12.50^{+0.07}_{-0.07}$ \\
\midrule
J1341$-$0321 & $ -395$ &  0.18$^{+0.04}_{-0.04}$   & 13.39$^{+0.03}_{-0.03}$ & $-443^{+2}_{-2} $  &  $ 80^{+8}_{-8}$ & $12.33^{+0.04}_{-0.04}$ \\
             & $ -566$ &  0.19$^{+0.04}_{-0.04}$   & 12.72$^{+0.04}_{-0.04}$ & $-574^{+2}_{-2} $  &  $ 26^{+12}_{-12}$ & $11.41^{+0.17}_{-0.17}$ \\
             & $ -903$ &  0.18$^{+0.05}_{-0.05}$   & 13.82$^{+0.04}_{-0.04}$ & $-901^{+3}_{-3} $  &  $130^{+28}_{-28}$ & $12.33^{+0.05}_{-0.05}$ \\
             & $-1138$ &  0.13$^{+0.05}_{-0.05}$   & 13.34$^{+0.03}_{-0.03}$ & $-1183^{+3}_{-2}$  &  $ 86^{+45}_{-45}$ & $11.88^{+0.20}_{-0.20}$ \\
\midrule
J1450+4621   & $  -50$ &  0.30$^{+0.09}_{-0.09}$   & 13.48$^{+0.03}_{-0.03}$ & $  -37^{+3}_{-3} $  &  $ 34^{+4}_{-4}$  & $12.44^{+0.05}_{-0.05}$ \\
             & $-1624$ &  0.30$^{+0.03}_{-0.03}$   & 12.85$^{+0.03}_{-0.03}$ & $-1702^{+6}_{-6} $ &  $56^{+44}_{-44}$ & $11.91^{+0.49}_{-0.49}$ \\
             & $-1728$ &  0.40$^{+0.03}_{-0.03}$   & 12.27$^{+0.06}_{-0.06}$ & $-1753^{+5}_{-6} $ &  $283^{+147}_{-147}$ & $12.70^{+0.12}_{-0.12}$ \\
\midrule
J1506+5402   & $-1011$ &  0.45$^{+0.01}_{-0.01}$   & 12.70$^{+0.06}_{-0.06}$ & $-1014^{+12}_{-10} $ &  $140^{+16}_{-16}$ & $12.30^{+0.04}_{-0.04}$ \\
             & $-1514$ &  0.60$^{+0.01}_{-0.01}$   & 12.55$^{+0.07}_{-0.07}$ & $-1447^{+11}_{-11} $ &  $165^{+30}_{-30}$ & $12.19^{+0.06}_{-0.06}$ \\
\midrule
J1558+3957   & $ -809$ &  0.20$^{+0.03}_{-0.03}$   & 13.09$^{+0.04}_{-0.04}$ & $-799^{+4}_{-4} $  &  $191^{+27}_{-27}$ & $12.68^{+0.05}_{-0.05}$ \\
             & $-1188$ &  0.50$^{+0.02}_{-0.02}$   & 12.26$^{+0.08}_{-0.08}$ & $-1243^{+10}_{-9}$  &  $ 110^{+50}_{-50}$ & $12.16^{+0.17}_{-0.17}$ \\
\midrule
J1613+2834   & \nodata&  \nodata   & \nodata &        \nodata     &        \nodata     &        \nodata       \\
\midrule
J2116$-$0624 & $ -276$ &  0.20$^{+0.02}_{-0.02}$   & 13.11$^{+0.03}_{-0.03}$ & $-336^{+4}_{-4} $ & $230 \pm 4$ & $12.63 \pm 0.06$ \\
             & $-1421$ &  0.18$^{+0.02}_{-0.02}$   & 13.58$^{+0.04}_{-0.04}$ & $-1436^{+5}_{-4}$ & $330 \pm 37$ & $12.75 \pm 0.06$ \\
\midrule
J2140+1209    & \nodata&  \nodata   & \nodata &        \nodata     &        \nodata     &        \nodata       \\
\enddata
\tablecomments{ Col 2: \mgii\, line centroid; Col 3: \mgi \,covering fraction from constrained fit; Col 4: \mgi\, column density from constrained fit; Col 5: \mgi \, line centroid from independent fit; Col 6: \mgi \,Doppler parameter from independent fit; Col 7: \mgi \,column density from independent fit.}
\end{deluxetable*}   

%% file: Table4.tex
\begin{deluxetable}{ccccc}[htp!]
\tabletypesize{\small}
\tablecaption{\mn \ best fit parameters\label{table4}}
\tablehead{
\colhead{ID} & \colhead{$v$(\mn)} & \colhead{\cf(\mn)} & \colhead{\bd(\mn)} & \colhead{\N(\mn)} \\
\colhead{} & \colhead{\footnotesize\kmps}  & \colhead{} & \colhead{\footnotesize\kmps}& \colhead{\footnotesize\cm}\\
\colhead{(1)} & \colhead{(2)} & \colhead{(3)} &
\colhead{ (4)} & \colhead{(5)}  
}
\startdata
J0905   & $-2356^{+6}_{-5}$ &  0.12$^{+0.26}_{-0.07}$   & $59^{+5}_{-9}$  &  $14.63^{+0.04}_{-0.04}$  \\
\midrule
J1219   & $-1865^{+9}_{-8}$ &  0.21$^{+0.14}_{-0.35}$   & $130^{+6}_{-8}$ &  $14.62^{+0.04}_{-0.04}$  \\
\enddata
\tablecomments{ Col 2: \mn\, line centroid; Col 3: \mn\, covering fraction; Col 4: \mn\, Doppler parameter; Col 5: \mn\, column density}
\end{deluxetable}

%% file: Table5.tex
\begin{deluxetable}{ccccc}[htp!]
\tabletypesize{\small}
\tablecaption{Derived Parameters\label{table5}}
\tablehead{
\colhead{ID} & \colhead{$\dot{M}_{\rm MgII}$\tablenotemark{1}}  & \colhead{$\dot{M}_{\rm FeII}$\tablenotemark{2}}  & \colhead{$\eta$\tablenotemark{3}} & \colhead{$\eta_{corr}$\tablenotemark{4}}\\
\colhead{} & \colhead{[$M_{\odot} yr^{-1}$]}  & \colhead{[$M_{\odot} yr^{-1}$]} & \colhead{}
}
\startdata
J0826   &    $>$ 34  & $>$ 85  &  $>$  0.46 & 5.52 \\
J0901   &    $>$ 28  & $>$ 49  &   $>$ 0.49 & 5.93 \\
J0905   &    $>$ 156 & 2288$^*$ &   25.43   & 25.43 \\
J0944   &    $>$ 25  & $>$ 80  &   $>$ 0.91 & 10.94 \\
J1125   &    $>$ 16  & $>$ 97  &   $>$ 0.43 & 5.10 \\
J1219   &    $>$ 132 & 1598$^*$ &   17.56   & 17.56 \\
J1232   &    $>$ 14  & $>$ 26  &  $>$  0.42 & 5.04 \\
J1341   &    $>$ 49  & $>$ 162  & $>$  1.08 & 12.91 \\
J1450   &    $>$ 16  & $>$ 25  &  $>$  0.13 & 1.54 \\
J1506   &    $>$  9  & $>$ 25  &  $>$  0.22 & 2.63 \\
J1558   &    $>$ 28  & $>$  85  & $>$  1.01 & 12.11 \\
J1613   &    55      & \nodata &  $>$  0.32 & 3.87 \\
J2116   &    13      & \nodata &  $>$  0.12 & 1.42 \\
J2140   &    4       & \nodata &  $>$  0.03 & 0.38 \\
\enddata
\tablenotetext{1}{Mass outflow rate calculated using Equation~\ref{eq:mdot_final} and the blueshifted hydrogen column density reported in Table~\ref{table2}, column 6.}
\tablenotetext{2}{Mass outflow rate calculated using Equation~\ref{eq:mdot_final} and the highest value between column 6 and 7 of Table~\ref{table2} (only blueshifted components).}
\tablenotetext{3}{Mass loading factor, i.e. $\eta \equiv \dot{M}/SFR$, calculated using $\dot{M}_{\rm FeII}$ as $\dot{M}$.}
\tablenotetext{4}{Mass loading factor calculated using $\dot{M}_{\rm MgII}$ as $\dot{M}$ and applying a factor of 12 correction to the upper limit values.}
\tablenotetext{*}{These galaxies have the best constraints from \feii absorption lines, the values for the rest of the sample should be considered as lower limits.}
\end{deluxetable}